\documentclass[12pt]{elsarticle}
\makeatletter
\def\ps@pprintTitle{%
 \let\@oddhead\@empty
 \let\@evenhead\@empty
 \def\@oddfoot{\centerline{\thepage}}%
 \def\@evenfoot{\thepage}
 \let\@evenfoot\@oddfoot}
\makeatother

\usepackage[final]{changes}
\definecolor{C0}{HTML}{1F77B4} 
\definechangesauthor[color=C0]{R1}

\usepackage{hyperref}

\usepackage[T1]{fontenc}
\usepackage{lmodern}
\usepackage{dirtytalk}
\usepackage{microtype}
\usepackage{comment}
\usepackage[margin=1in]{geometry}
\usepackage[]{algorithm2e}
\usepackage{amsmath,amssymb,amsthm,bm}
\usepackage{mathtools}
\usepackage{graphicx}
 \usepackage{algpseudocode} 
\usepackage{subcaption}
\biboptions{sort&compress}
\usepackage{epstopdf}
\usepackage{float}

\usepackage[utf8x]{inputenc}
\usepackage{xcolor}
\AppendGraphicsExtensions{.tif}

\begin{document}

\title{Multiscale graph neural networks with adaptive mesh refinement for accelerating mesh-based simulations}
\author[auburn]{Roberto Perera}
\author[auburn]{Vinamra Agrawal}
\cortext[cor1]{Corresponding author: vza0013@auburn.edu}
\address[auburn]{Department of Aerospace Engineering, Auburn University, Auburn, AL, USA}

\begin{abstract}
    Mesh-based Graph Neural Networks (GNNs) have recently shown capabilities to simulate complex multiphysics problems with accelerated performance times.
    However, mesh-based GNNs require a large number of message-passing (MP) steps and suffer from over-smoothing for problems involving very fine mesh.
    In this work, we develop a multiscale mesh-based GNN framework mimicking a conventional iterative multigrid solver, coupled with adaptive mesh refinement (AMR), to mitigate challenges with conventional mesh-based GNNs.
    We use the framework to accelerate phase field (PF) fracture problems involving coupled partial differential equations with a near-singular operator due to near-zero modulus inside the crack.
    We define the initial graph representation using all mesh resolution levels.
    We perform a series of downsampling steps using Transformer MP GNNs to reach the coarsest graph followed by upsampling steps to reach the original graph.
    We use skip connectors from the generated embedding during coarsening to prevent over-smoothing.
    We use Transfer Learning (TL) to significantly reduce the size of training datasets needed to simulate different crack configurations and loading conditions.
    The trained framework showed accelerated simulation times, while maintaining high accuracy for all cases compared to physics-based PF fracture model.
    Finally, this work provides a new approach to accelerate a variety of mesh-based engineering multiphysics problems. 
\end{abstract}

\begin{keyword}
    Machine Learning; Phase Field Model; Multiscale; Mesh-based; Graph Neural Network; Algebraic Multigrid Scheme, Transfer Learning; Crack Propagation; Displacement Fields
\end{keyword}

\maketitle    

\section{Introduction} \label{sec:introduction}
    Computational models play an important role across many engineering fields to simulate the behavior of complex physics phenomena without performing experiments. 
    These models commonly rely on solving coupled system of partial differential equations (PDEs) that define the underlying physics of the problem.
    A favored approach to solve these PDEs and propagate the physics in time has been to discretize the problem domain into a mesh, where the solutions of the PDEs are then approximated.
    While these methods have proven to generate accurate and reliable results in the past, they quickly become computationally expensive as problem complexity is increased.

    Fracture is one of the common means of failure in engineered materials and, therefore, widely studied through computational techniques, including the phase field (PF) method \cite{ambati2015review, ambati2016phase, ernesti2020fast, goswami2020adaptive, zhang2022mixed}.
    The PF approach to fracture regularizes the discontinuous crack using a continuous damage field $\phi \in \{0,1\}$ and uses it to formulate an energy functional, $\Pi$ \cite{francfort1998revisiting, egger2019discrete, ribot2019new,runnels2020phase}. 
    The PF fracture approach has been used to study crack propagation, nucleation, and branching in brittle and ductile materials, composite materials with anisotropy, and even biological systems \cite{xu2022phase,vajari2022thermodynamically, li2023adaptive,han2022variational,agrawal2023robust,agrawal2021block,chen2023parallel,brach2019phase, gultekin2018numerical, marulli2022combined, yin2020anisotropic}. 
    Despite their robustness and ease of implementation, PF methods require fine mesh resolution near the crack tip to adequately capture the crack interface, increasing computational requirements.
    As such, PF fracture methods typically employ adaptive mesh refinement (AMR) to use fine mesh only near the crack surfaces. 
    Despite AMR, PF fracture methods are still computationally demanding, often requiring many CPU hours on high-performance computing clusters.

    Machine Learning (ML) techniques provide a reduced-order modeling approach to mitigate the computational costs of current models.
    The rapid increase in popularity of ML has led to the development of various works for accelerating simulation models \cite{hunter2019reduced,lew2021deep,euser2019simulation, montes2021accelerating,yang2021deep,sharma2021polygonal,zhang2022prediction,wang2021stressnet}. 
    Graph neural networks (GNNs) combine graph theory with neural networks and represent the model input as a graph using connecting nodes and edges (i.e., similar to a mesh).
    While neural networks update the model's weights through stochastic backward gradient descent \cite{amari1993backpropagation,bottou2010large,ketkar2017stochastic}, GNN models define the weights using the resulting graph representation of nodes and edges.  
    Recently, GNNs have been used to predict finite element (FE) convergence of elastostatic problems \cite{black2022learning}, molecular dynamics of translationally-invariant and rotationally-covariant local atomic environments \cite{park2021accurate}, predict the graph structures of 3D boundaries for granular flow processes \cite{mayr2023boundary}, and predict the dynamics of various solid mechanics and fluid mechanics problems \cite{vlassis2020geometric,vlassis2022geometric,perera2022graph,li2022graph,bhattoo2022learning}. 
    
    The graph representation approach is also ideal for mesh-based problems where the simulation mesh can be used as the model’s graph representation.
    Recently, mesh-based GNNs have been used for engineering problems, such as simulating FE stresses and displacements \cite{Jin2023Leveraging, JIANG2023106370, 10022022}, crack propagation in PF models with AMR \cite{perera2023dynamic}, and flow over cylinders and airfoils \cite{SHAO2023110056, pfaff2020learning}.
    However, mesh-based GNNs struggle when working with problems involving very fine mesh.
    GNN models use message-passing (MP) blocks \cite{gasteiger2020directional} to transfer information and learn relations between the nodes and edges \cite{scarselli2008graph,gilmer2017neural}.
    For fine meshes, GNN-based frameworks require a large number of MP blocks to transfer and capture the relations between nodes that are far apart from each other.
    This is unfavorable for most mesh-based problems where information must be shared between nodes beyond the local neighborhoods to achieve high accuracies.
    Increasing the MP blocks also results in over-smoothing \cite{Li_Han_Wu_2018, rusch2023survey, cai2020note}.
    This challenge has led to the development of Multiscale GNNs \cite{fortunato2022multiscale}, which mimic the conventional iterative multigrid scheme.
    Multiscale GNNs perform multiple graph coarsening operations obtaining smaller meshes at each level, which are passed through MP blocks \cite{stuben2001introduction,xu2017algebraic,eliasof2020diffgcn}.
    The smaller mesh levels provide new connections between distant nodes, thus, transferring information beyond the previous local neighborhood and reducing the required number of MP steps.   
    This technique has shown higher accuracies compared to conventional mesh-based GNNs in dynamic problems involving flow field predictions \cite{yang2022amgnet}, time-independent PDEs in unstructured meshes and sparse linear systems \cite{gladstone2023gnn,liu2021multi,luz2020learning}, and fluid motion over rigid bodies of varying shape \cite{lino2021simulating,lino2022towards}.
    While these techniques provide a new avenue for simulating various physics problems while avoiding over-smoothing and errors due to long-range interactions, there are some limitations.  
    First, these techniques were designed for a fixed initial graph and employed different graph-pooling techniques for downsampling steps \cite{cao2023efficient, lino2021simulating, gao2019graph, barwey2023multiscale}.
    These techniques can add to the overall computational cost and have yet to be adapted to problems involving AMR.

\begin{figure}            
		\centering 
	        \includegraphics[width=1\linewidth]{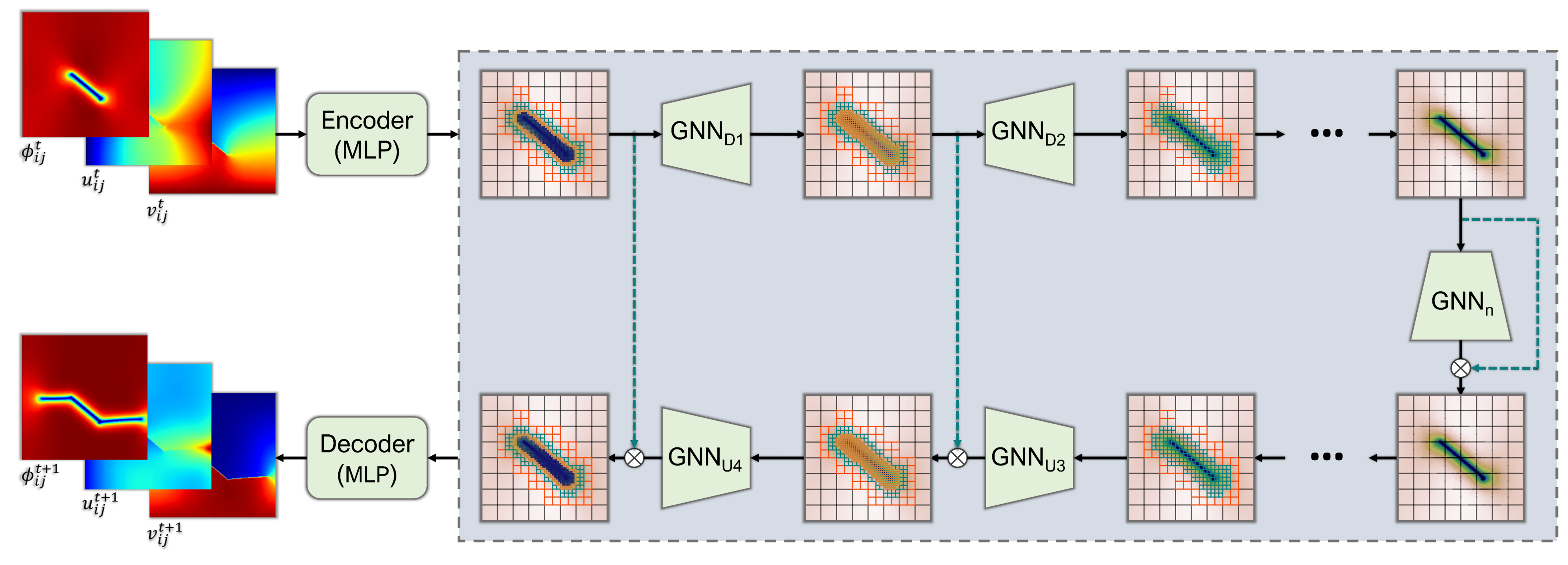}
	        \caption{Architecture of the multiscale GNN framework for the Four-Stage Refinement architecture. The model first uses MLP networks to encode the input graph representation. The feature embedding is then passed through MP GNN blocks followed by mesh coarsening operations denoted by $GNN_{D1}$ and $GNN_{D2}$. $GNN_{n}$ involves an additional MP GNN block to operate on the coarsest mesh. The resulting coarsened embedding is then reconstructed through MP GNN blocks followed by mesh upscaling operations denoted by $GNN_{U3}$ and $GNN_{U4}$. The dashed green lines represent skip connectors. The final reconstructed embedding is passed through a decoder MLP network to predict the crack field and displacement fields at future times.}
	        \label{fig:GNN_flowchart}
	\end{figure}

	In this work, we develop a new multiscale GNN coupled with block-structured AMR (BSAMR) to study time-evolving problems with near-singular operators, such as PF fracture. 
	The framework introduces a new simple and computationally efficient downsampling/upscaling approach for mesh-based AMR problems by exploiting the various mesh resolution levels.
    In the developed downsampling/upscaling approach the framework reduces the size of the mesh by eliminating the highest level of mesh refinement at each coarsening operation, as shown in Figure \ref{fig:GNN_flowchart}. 
	We study PF fracture problems with five AMR levels and compare the accuracy and computational time of three coarsening approaches from the finest to the coarsest mesh: (i) four coarsening/downscale operations, (ii) two coarsening/downscale operations, and (iii) one coarsening/downscale operation.
    We begin by developing the framework for Mode-I fracture problems with a single crack on the left edge of the domain.
    We employ transfer learning (TL) \cite{yosinski2014transferable, perera2023generalized} to extend the framework for other configurations such as center cracks, right-edge cracks, and shear loading.
    We then analyze the framework's accuracy for each problem configuration.
	Lastly, we compare the computational time of the developed multiscale and adaptive GNN framework against the high-fidelity PF model.

\section{Methods}

    \subsection{Physics based PF fracture model}\label{subsec:methods_phase_field_model} 
        In this work, we develop the multiscale mesh-based GNN framework with BSAMR for PF fracture problems. 
        PF fracture problems are computationally demanding due to a) near singular operator due to near-zero modulus inside the crack field and b) the requirement of high mesh resolution near the crack tip. 
        In the PF formulation of fracture, the sharp crack is regularized using a smooth scalar field $\phi(\bm{x})$ ranging from 0 to 1. 
        An energy functional $\Pi$ is constructed using $\phi(\bm{x})$ which is then minimized to obtain a set of coupled partial differential equations, a vector equation corresponding to elastic equilibrium, and a scalar equation for crack equilibrium.
        The reader is referred to the extensive literature on PF fracture for details on the formulation \cite{francfort2022variational}.
        In this work, we use the second-order energy functional as below \cite{goswami2020adaptive}.
        \begin{flalign}
            \Pi = \int_{\mathcal{V}} \left[ W\left( \varepsilon (\textbf{u},\phi) \right) + \frac{G_{c}}{2} \left( \frac{\phi^{2}}{d} + d| \nabla \phi |^{2} \right) \right]d\mathcal{V}.
            \label{eq:energy_functional}
        \end{flalign}
        Here, $\bm{u}$ is the displacement field, $\bm{\varepsilon}$ is the strain, $W$ is the strain energy density, $G_{c}$ is the fracture energy constant, and $d$ is the regularization length scale of the crack.
        Here, $\phi$ takes the value $0$ inside the crack, and the value $1$ in the bulk material. 
        In this work, we use the PF open-source high-fidelity model with BSAMR capabilities from \cite{goswami2020adaptive} to simulate various crack propagation problems.	
        Following \cite{perera2023dynamic}, we chose a linear elastic isotropic brittle material $(0.5m\times 0.5m)$, with Young's modulus $E = 210 GPa$, Poisson's ratio $\nu=0.3$, fracture energy density $G_c = 2.7 N/m^2$, and $d=0.0125m$.
        We fixed the bottom edge and applied displacement to the top edge in $y$-direction for tensile load and $x$-direction for shear load. 
        We varied initial crack length, $C_{L}$, edge position, $C_{P}$, and crack angle, $C_{\theta}$ to gather the datasets of unique initial conditions.

	\subsection{Graph Neural Network}\label{subsec:methods_graph_network}

		\begin{figure}
			\centering
			\begin{subfigure}[t]{0.32\textwidth}
				\centering
				\includegraphics[width=1\linewidth]{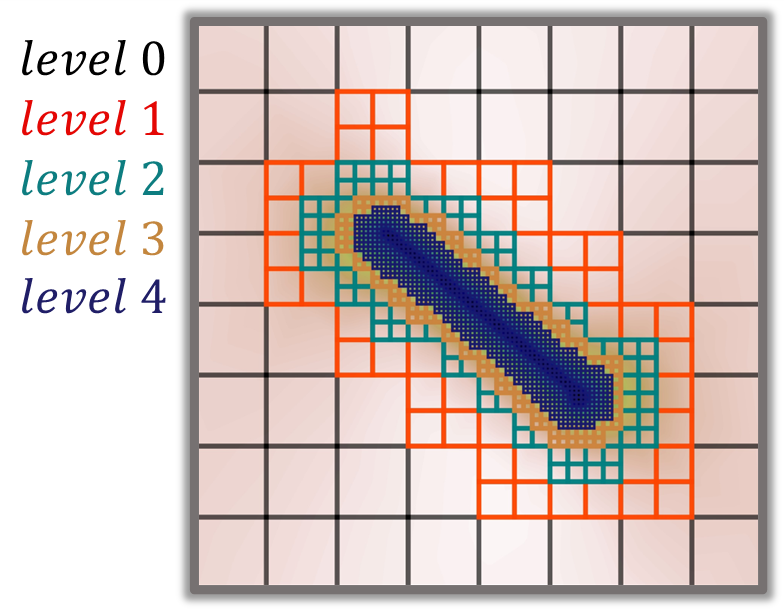}
				\caption{Initial refined graph: $\mathcal{M}^{ref}_{4}$}
				\label{subfig:graph_4}
			\end{subfigure}
			\centering
			\begin{subfigure}[t]{0.32\textwidth}
				\centering
				\includegraphics[width=1\linewidth]{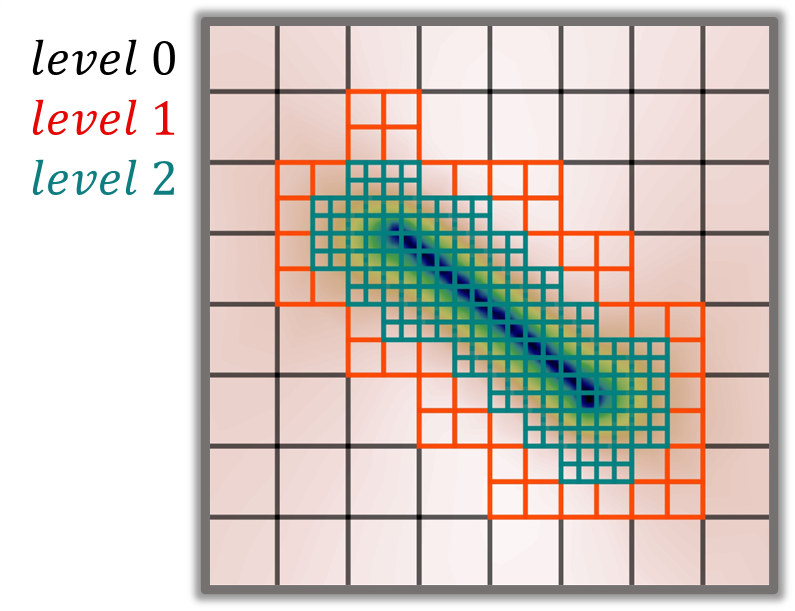}
				\caption{Second downscale graph: $\mathcal{M}^{ref}_{2}$}
				\label{subfig:graph_2}
			\end{subfigure}
			\centering
			\begin{subfigure}[t]{0.32\textwidth}
				\centering
				\includegraphics[width=1\linewidth]{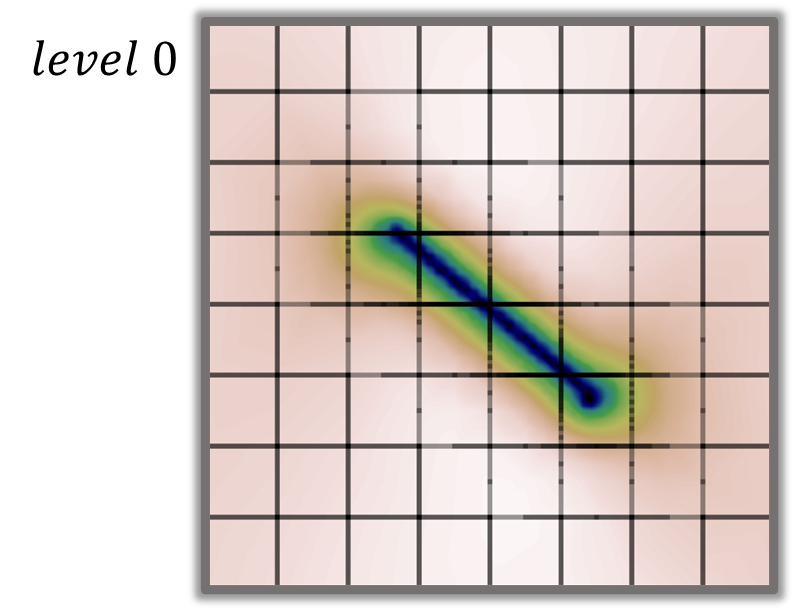}
				\centering
				\caption[r]{Fourth downscale graph: $\mathcal{M}^{ref}_{0}$}
				\label{subfig:graph_0}
			\end{subfigure}
			\caption{Representation of the instantaneous refined mesh graphs involving a) refinement levels 0-4, b) refinement levels 0-2, and c) refinement level 0 (i.e., coarsest mesh).}
			\label{fig:graph_mesh_representation}
		\end{figure}	

		As shown in Figure \ref{fig:graph_mesh_representation}, we formulated the graph representation using the instantaneous refined mesh, {$\mathcal{M}^{ref} : \langle {\mathcal{\mathbf{U}}},\mathbf{E} \rangle$}, where $\mathcal{\mathbf{U}}$ includes the mesh vertices, and $\mathbf{E}$ the resulting mesh edges. 
        The node features, $\xi_{s}$, include the positions of the mesh vertices, $\hat{P}_{s}$, their x- and y-displacement values, $\hat{D}_{s}$, their crack field values, ${\phi}_{s}$, and the applied displacement loading along the x- and y-directions, $\{ {u}_{0_{s}}, {v}_{0_{s}} \}$.
        {\begin{flalign}
        && \hat{P}_{s} = \{ \left(x_{s}, y_{s} \right)\} && \{s \in \mathcal{\mathbf{U}}\}, \{\mathbf{U} \in \mathcal{M}^{ref}\},  \nonumber\\
        && \hat{D}_{s} = \{ \left(u_{s}, v_{s} \right) \} && \{s \in \mathcal{\mathbf{U}}\}, \{\mathbf{U} \in \mathcal{M}^{ref}\},  \nonumber\\
        && \{\xi_{s}\} = \{\hat{P}_{s}, \hat{D}_{s}, \phi_{s}, {u}_{0_{s}}, {v}_{0_{s}}\} && \{s \in \mathcal{\mathbf{U}}\}, \{\mathbf{U} \in \mathcal{M}^{ref}\}. \label{eq:vertex_states}
        \end{flalign}}
        The applied displacement loading node feature, $\{ {u}_{0_{s}}, {v}_{0_{s}} \}$, allows the framework to distinguish between cases subjected to tensile loading versus cases under shear loading.

		The edge features, $e_{sr}$, are defined using the binary connectivity value, $b_{sr} \in \{ 0,1 \}$, where $s$ and $r$ indicate the ``sender'' and ``receiver'' nodes, respectively.
		This means that $b_{sr}=1$ for nodes that share an edge and for cases where $s=r$.
		{\begin{flalign}
		&& \{e_{sr}\} = \{\xi_{s}, \xi_{r}, b_{sr}\} && \{(s,r,b_{sr}) \in \mathcal{\mathbf{E}}\}, \{\mathbf{E} \in \mathcal{M}^{ref}\}. \label{eq:edge_states}
		\end{flalign}}

		Lastly, we note that for each time-step in a simulation, we construct the initial graph from the instantaneous refined mesh containing all five levels of mesh resolution, $\mathcal{M}^{ref}$, shown in Figure \ref{subfig:graph_4}.
		However, as shown in Figure \ref{fig:GNN_flowchart}, the framework removes each refinement level iteratively until obtaining the coarsest mesh at level 0 resolution.
		We define the corresponding graphs as (i) $\mathcal{M}^{ref}_{4}$ the initial instantaneous refined mesh with resolution levels 0-4 (shown in Figure \ref{subfig:graph_4}), (ii) $\mathcal{M}^{ref}_{3}$ the first downscaled mesh with resolution levels 0-3, (iii) $\mathcal{M}^{ref}_{2}$ the second downscaled mesh with resolution levels 0-2 (shown in Figure \ref{subfig:graph_2}), (iv) $\mathcal{M}^{ref}_{1}$ the third downscaled mesh with resolution levels 0-1, and (v) $\mathcal{M}^{ref}_{0}$ the coarsest downscaled mesh with resolution level 0 only (shown in Figure \ref{subfig:graph_0}). 		 

	\subsection{Multiscale GNN framework}\label{subsec:methods_framework}
		The developed GNN framework integrates AMR to the multigrid approach where the initial refined mesh is coarsened by one mesh level at each step.
		The resulting coarsened mesh resolutions provide new connections between distant nodes to transfer information beyond the previous local neighborhoods.
		This approach reduces the required number of MP steps, thus, avoiding over-smoothing and significantly reducing computational costs while maintaining high accuracy. 

		The architecture of the developed multiscale GNN framework is depicted in Figure \ref{fig:GNN_flowchart}.
		The first step of the framework is to input the graph representation defined in Section \ref{subsec:methods_graph_network} and shown in Figure \ref{subfig:graph_4}, at a given time ``$t$'', into an encoder network.
		We use an MLP model as the encoder network, denoted by $MLP_{\text{in}}$.
		{\begin{flalign}
			&& \{\xi_{s}^{'}\} \leftarrow MLP_{\text{in}}\{\xi_{s}^{t}, e_{sr}^{t}\} && \{s \in \mathcal{\mathbf{U}}\},  \{(s,r,b_{sr}) \in \mathcal{\mathbf{E}}\}, \{(\mathbf{U},\mathbf{E}) \in \mathcal{M}^{ref}\}. \label{eq:encoder}
		\end{flalign}}
		We input the generated latent-space embedding, $\xi_{s}^{'}$, to a MP network, $GNN_{1}$, to learn relations within the local neighborhoods.
		We employ Graph Transformers as our MP models in this work \cite{shi2020masked}.
		Graph Transformers were first introduced for the tasks of sequence modeling and language translation \cite{vaswani2017attention}. 
		Recently, this technique was extended for graph-based tasks by applying the multi-head and self-attention mechanisms to the vertex and edge features of neighboring vertices, thus, creating additional attention embeddings \cite{shi2020masked}.  
		The Transformer MP models in the developed framework follow the same architecture, involving four attention heads and a hidden node dimension of 128.
		
		The ``{GNN}$_{D1}$'' block depicted in Figure \ref{fig:GNN_flowchart} includes the MP network, $GNN_{1}$, and the first Downscale operation, $Down_{1}$.
		Each Downscale operation removes the current finest level of mesh refinement.
		For instance, the first Downscale operation in Figure \ref{fig:GNN_flowchart} removes refinement level 4 from graph $\mathcal{M}^{ref}_{4}$ (Figure \ref{subfig:graph_4}), resulting in graph $\mathcal{M}^{ref}_{3}$. 
		{\begin{flalign}
			&& \{\xi_{s}^{4}\} \leftarrow GNN_{1}\{\xi_{s}^{'}, e_{sr}^{t}\} && \{s \in \mathcal{\mathbf{U}}\},  \{(s,r,b_{sr}) \in \mathcal{\mathbf{E}}\}, \{(\mathbf{U},\mathbf{E}) \in \mathcal{M}^{ref}\}, \nonumber \\
			&& \{\xi_{s}^{3}\} \leftarrow Down_{1}\{\xi_{s}^{4}, e_{sr}^{t}, e_{sr}^{3}\} && \{s \in \mathbf{U}^{3}\},  \{(s,r,b_{sr}) \in \mathbf{E}^{3}\}, \{(\mathbf{U}^{3},\mathbf{E}^{3}) \in \mathcal{M}^{ref}_{3}\}.
			\label{eq:downscale_1}
		\end{flalign}}
		Here, $\xi_{s}^{4}$ is the resulting node embedding in refined mesh $\mathcal{M}^{ref}$ from the MP block GNN$_{1}$, $e_{sr}^{3}$ is the edge feature vector for the downscaled mesh $\mathcal{M}^{ref}_{3}$, and $\xi_{s}^{3}$ is the resulting downscaled node embedding from $\mathcal{M}^{ref}$ to $\mathcal{M}^{ref}_{3}$.       
		Similarly, ``{GNN}$_{D2}$'' from Figure \ref{fig:GNN_flowchart} includes a MP network, $GNN_{2}$, and the second Downscale operation, $Down_{2}$, which removes refinement level 3 from graph $\mathcal{M}^{ref}_{3}$, resulting in graph $\mathcal{M}^{ref}_{2}$ shown in Figure \ref{subfig:graph_2}.
		{\begin{flalign}
			&& \{\xi_{s}^{3}\} \leftarrow GNN_{2}\{\xi_{s}^{3}, e_{sr}^{3}\} && \{s \in \mathcal{\mathbf{U}^{3}}\},  \{(s,r,b_{sr}) \in \mathcal{\mathbf{E}^{3}}\}, \{(\mathbf{U}^{3},\mathbf{E}^{3}) \in \mathcal{M}^{ref}_{3}\}, \nonumber \\
			&& \{\xi_{s}^{2}\} \leftarrow Down_{2}\{\xi_{s}^{3}, e_{sr}^{3}, e_{sr}^{2}\} && \{s \in \mathbf{U}^{2}\},  \{(s,r,b_{sr}) \in \mathbf{E}^{2}\}, \{(\mathbf{U}^{2},\mathbf{E}^{2}) \in \mathcal{M}^{ref}_{2}\},
			\label{eq:GNN_D2}
		\end{flalign}}
		where $e_{sr}^{2}$ is the edge feature vector for the downscaled mesh $\mathcal{M}^{ref}_{2}$, and $\xi_{s}^{2}$ is the resulting downscaled node embedding from $\mathcal{M}^{ref}_{3}$ to $\mathcal{M}^{ref}_{2}$.  
		We repeat this process using two additional MP blocks each followed by their coarsening operations to obtain the downscaled node embedding (from $\mathcal{M}^{ref}_{1}$ to $\mathcal{M}^{ref}_{0}$) for the coarsest mesh, $\xi_{s}^{0}$, shown in Figure \ref{subfig:graph_0}.  
		The framework then includes an additional MP network (denoted by ``{GNN}$_{n}$'' in Figure \ref{fig:GNN_flowchart}) which operates on the resulting coarsest mesh node embedding. 
		{\begin{flalign}
			&& \{\xi_{s}^{0^{'}}\} \leftarrow Agg\left[\xi_{s}^{0}, {GNN}_{n}\{\xi_{s}^{0}, e_{sr}^{0}\}\right] && \{s \in \mathbf{U}^{0}\},  \{(s,r,b_{sr}) \in \mathbf{E}^{0}\}, \{(\mathbf{U}^{0},\mathbf{E}^{0}) \in \mathcal{M}^{ref}_{0}\}. \label{eq:GNN_n}
		\end{flalign}}
		Here, $e_{sr}^{0}$ is the edge feature vector for the downscaled mesh $\mathcal{M}^{ref}_{0}$ shown in Figure \ref{subfig:graph_0}, $Agg$ is an aggregation operation for the skip connectors (shown as dashed green lines in Figure \ref{fig:GNN_flowchart}), and $\xi_{s}^{0^{'}}$ is the resulting node embedding for mesh $\mathcal{M}^{ref}_{0}$ from the aggregation of the MP block GNN$_{n}$ and the skip connector.

		Next, we perform a series of upscale steps to reconstruct the original refined mesh.
		For instance, ``{GNN}$_{U1}$'' includes the first Upscale operation, $Up_{1}$, to reconstruct mesh level 1 ($\mathcal{M}^{ref}_{1}$), followed by the skip connector from $\xi_{s}^{1}$ to the MP GNN, $GNN_{4}$. 
		{\begin{flalign}
			&& \{\xi_{s}^{1^{'}}\} \leftarrow Up_{1}\{\xi_{s}^{0^{'}}, e_{sr}^{0}, e_{sr}^{1}\} && \{s \in \mathcal{\mathbf{U}}^{1}\},  \{(s,r,b_{sr}) \in \mathcal{\mathbf{E}}^{1}\}, \{(\mathbf{U}^{1},\mathbf{E}^{1}) \in \mathcal{M}^{ref}_{1}\}, \nonumber \\ 
			&& \{\xi_{s}^{1^{''}}\} \leftarrow Agg\left[\xi_{s}^{1}, {GNN}_{4}\{\xi_{s}^{1^{'}}, e_{sr}^{1}\}\right] && \{s \in \mathcal{\mathbf{U}}^{1}\},  \{(s,r,b_{sr}) \in \mathcal{\mathbf{E}}^{1}\}, \{(\mathbf{U}^{1},\mathbf{E}^{1}) \in \mathcal{M}^{ref}_{1}\}. \label{eq:GNN_U1}
		\end{flalign}}
		In equation (\ref{eq:GNN_U1}), $\xi_{s}^{1^{'}}$ defines the resulting reconstructed node embedding from the upscale operation, $Up_{1}$, (i.e., from $\mathcal{M}^{ref}_{0}$ to $\mathcal{M}^{ref}_{1}$), and $\xi_{s}^{1^{''}}$ denotes the resulting node embedding for mesh $\mathcal{M}^{ref}_{1}$ from the aggregation of the MP block $GNN_{4}$ and the skip connector from $\xi_{s}^{1}$.
		Next, ``${GNN}_{U2}$'' applies the second Upscale operation, $Up_{2}$, to regenerate mesh level 2 ($\mathcal{M}^{ref}_{2}$ from Figure \ref{subfig:graph_2}), followed by the skip connector from $\xi_{s}^{2}$ to the MP GNN, ${GNN}_{3}$.
		{\begin{flalign}
			&& \{\xi_{s}^{2^{'}}\} \leftarrow Up_{2}\{\xi_{s}^{1^{'}}, e_{sr}^{1}, e_{sr}^{2}\} && \{s \in \mathcal{\mathbf{U}}^{2}\},  \{(s,r,b_{sr}) \in \mathcal{\mathbf{E}}^{2}\}, \{(\mathbf{U}^{2},\mathbf{E}^{2}) \in \mathcal{M}^{ref}_{2}\}, \nonumber \\ 
			&& \{\xi_{s}^{2^{''}}\} \leftarrow Agg\left[\xi_{s}^{2}, {GNN}_{3}\{\xi_{s}^{2^{'}}, e_{sr}^{2}\}\right] && \{s \in \mathcal{\mathbf{U}}^{2}\},  \{(s,r,b_{sr}) \in \mathcal{\mathbf{E}}^{2}\}, \{(\mathbf{U}^{2},\mathbf{E}^{2}) \in \mathcal{M}^{ref}_{2}\}, \label{eq:GNN_U2}
		\end{flalign}}
		where $\xi_{s}^{2^{'}}$ defines the resulting reconstructed node embedding from the upscale operation, $Up_{2}$, (i.e., from $\mathcal{M}^{ref}_{1}$ to $\mathcal{M}^{ref}_{2}$), and $\xi_{s}^{2^{''}}$ denotes the resulting node embedding for mesh $\mathcal{M}^{ref}_{2}$ from the aggregation of the MP block $GNN_{3}$ and the skip connector from $\xi_{s}^{2}$.
		We repeat this process using two additional upscale blocks (i.e., $GNN_{U3}$ and $GNN_{U4}$ in Figure \ref{fig:GNN_flowchart}) until we reconstruct the initial refined mesh $\mathcal{M}^{ref}$ with mesh levels 0-4 and obtain the resulting aggregated node embedding $\xi_{s}^{4^{''}}$. 
		Lastly, we use the final generated reconstructed embedding, $\xi_{s}^{4''}$, as input to a Decoder MLP network, $MLP_{\text{out}}$, to transfer the embedding from the latent-space to the real-space and predict the displacements and scalar damage field at the next time-step, ``$t+1$''. 
		{\begin{flalign}
			&& \{\xi_{s}^{t+1}\} \leftarrow MLP_{\text{out}}\{\xi_{s}^{4''}, e_{sr}^{t}\} && \{s \in \mathcal{\mathbf{U}}\},  \{(s,r,b_{sr}) \in \mathcal{\mathbf{E}}\}, \{(\mathbf{U},\mathbf{E}) \in \mathcal{M}^{ref}\}. \label{eq:decoder}
		\end{flalign}}

	\subsection{Single-Stage, Two-Stage, and Four-Stage Refinement GNNs}\label{subsec:methods_singledoubledownscale}
		
		While a higher number of downscale/upscale operations and MP GNNs may provide the framework with higher prediction accuracy, it comes with an increased computational cost. 
		We evaluated the prediction accuracy and computational cost of three different architectures following the procedure described in the previous Section \ref{subsec:methods_framework}.
		For each architecture, we reduced the number of MP GNNs, upscale operations, and downscale operations. 		
		The first framework was a Four-Stage Refinement (FSR) GNN involving four downscale and four upscale operations as shown in Figure \ref{fig:GNN_flowchart}.
		The FSR framework architecture is described in detail in Section \ref{subsec:methods_framework}.
		
		The second framework was a Two-Stage Refinement (TSR) GNN involving two downscale and two upscale operations.
		In the TSR framework, a single downscale/upscale operation accounts for two mesh resolution levels.
		For instance, given an initial instantaneous refined mesh $\mathcal{M}^{ref}_{4}$ (Figure \ref{subfig:graph_4}), the first downscale operation in TSR, $GNN_{D1}$, removes mesh resolution levels 4 and 3 to obtain the new mesh graph $\mathcal{M}^{ref}_{2}$ (i.e., Figure \ref{subfig:graph_2}), along with the resulting downscaled node embedding $\xi_{s}^{2}$.
		By removing four MP GNNs, and two downscale and upscale operations, we expect the TSR framework to be faster than FSR. 
		
		Lastly, the third framework was a Single-Stage GNN (SSR) involving a single downscale/upscale operation.
		In the SSR framework, we removed six MP GNNs and three downscale and upscale operations.
		For instance, SSR involves a single downscale operation, $GNN_{D1}$, which removes mesh resolution levels 1-4 directly from the initial node embedding, $\xi_{s}^{4}$, to obtain the coarsest mesh level shown in Figure \ref{subfig:graph_0}.
		The SSR framework is significantly less computationally expensive compared to FSR and TSR as it does not require computing new graphs for intermediate mesh resolution levels (1-4), nor additional MP Transformer GNNs.    
		
	\subsection{Transfer learning}\label{subsec:methods_transferlearning}

		\begin{figure}
			\centering
			\begin{subfigure}[t]{0.24\textwidth}
				\centering
				\includegraphics[width=1\linewidth]{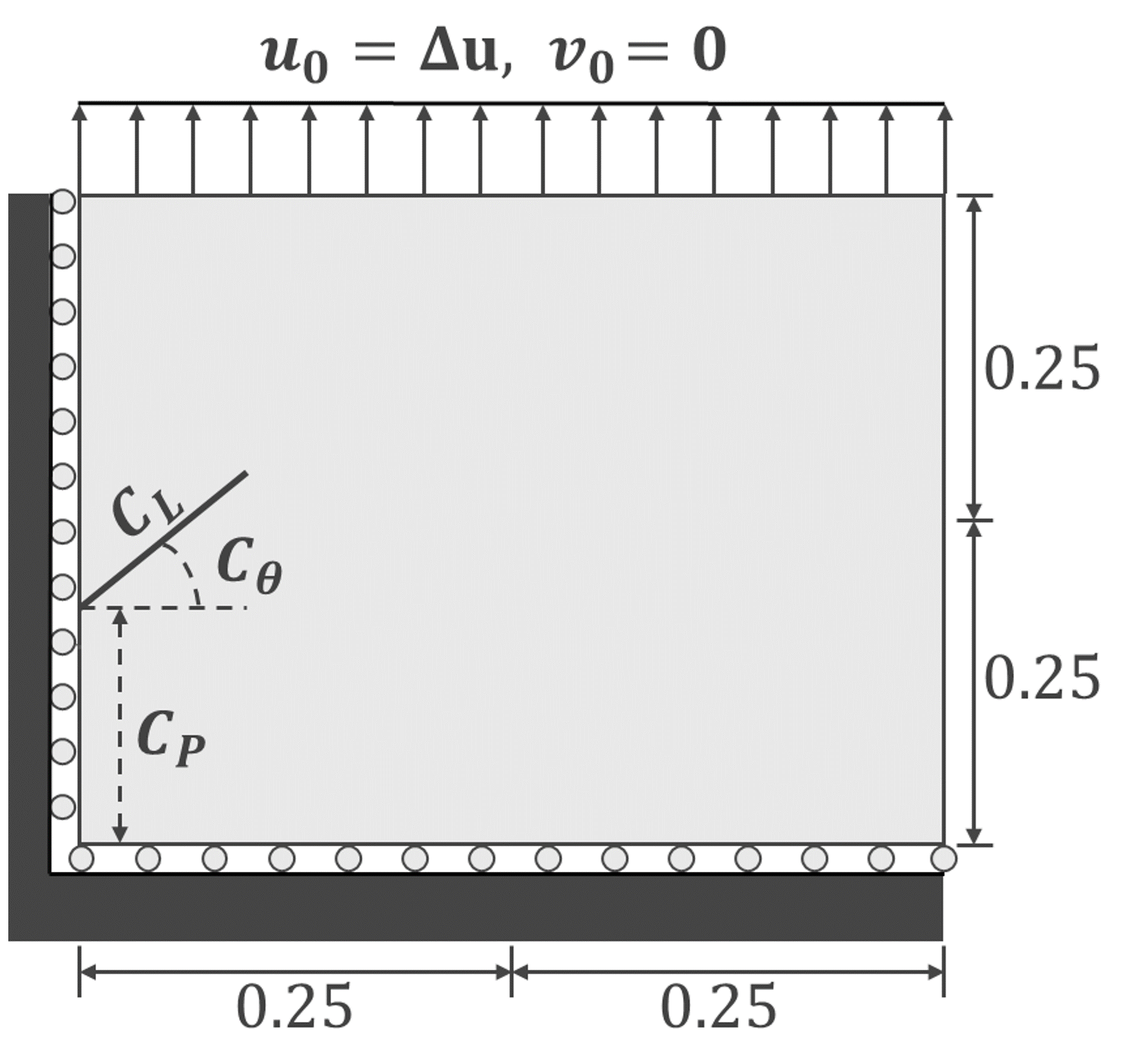}
				\caption{Left-edge crack in tension}
				\label{subfig:leftedge_crack}
			\end{subfigure}
			\centering
			\begin{subfigure}[t]{0.24\textwidth}
				\centering
				\includegraphics[width=1\linewidth]{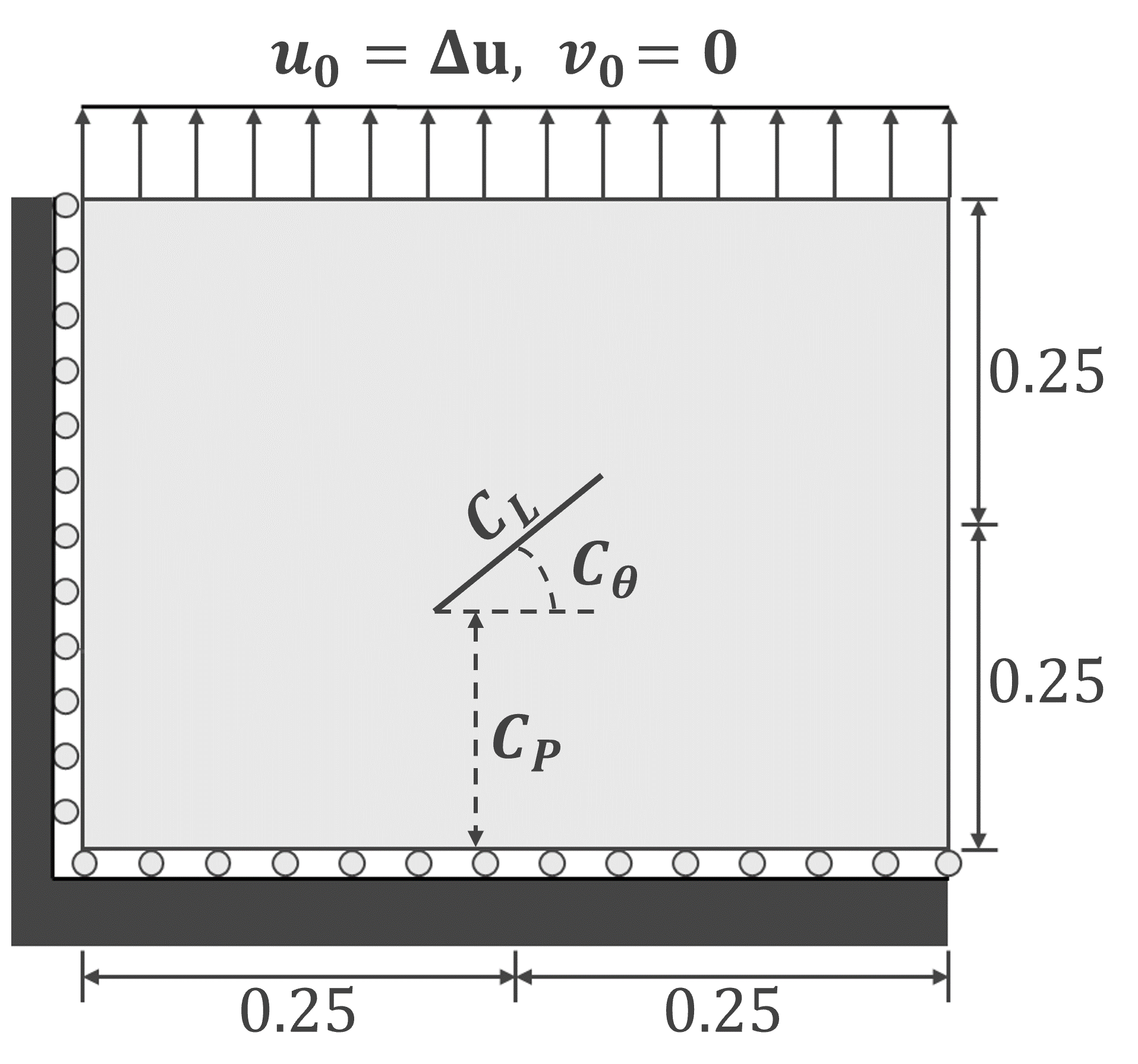}
				\caption{Center crack in tension}
				\label{subfig:center_crack}
			\end{subfigure}
			\centering
			\begin{subfigure}[t]{0.24\textwidth}
				\centering
				\includegraphics[width=1\linewidth]{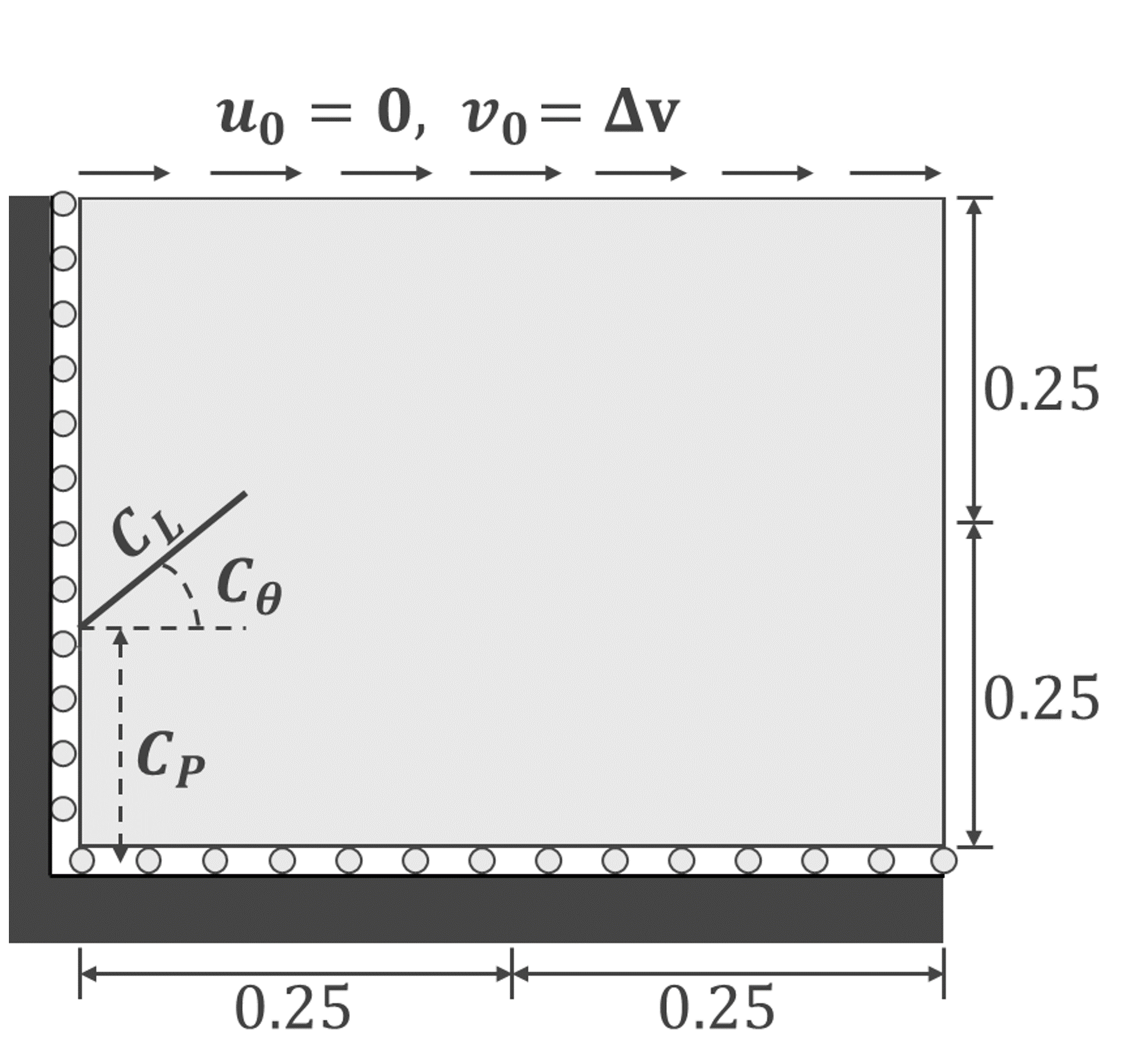}
				\centering
				\caption[r]{Left-edge crack in shear}
				\label{subfig:shearload}
			\end{subfigure}
			\centering
			\begin{subfigure}[t]{0.24\textwidth}
				\centering
				\includegraphics[width=1\linewidth]{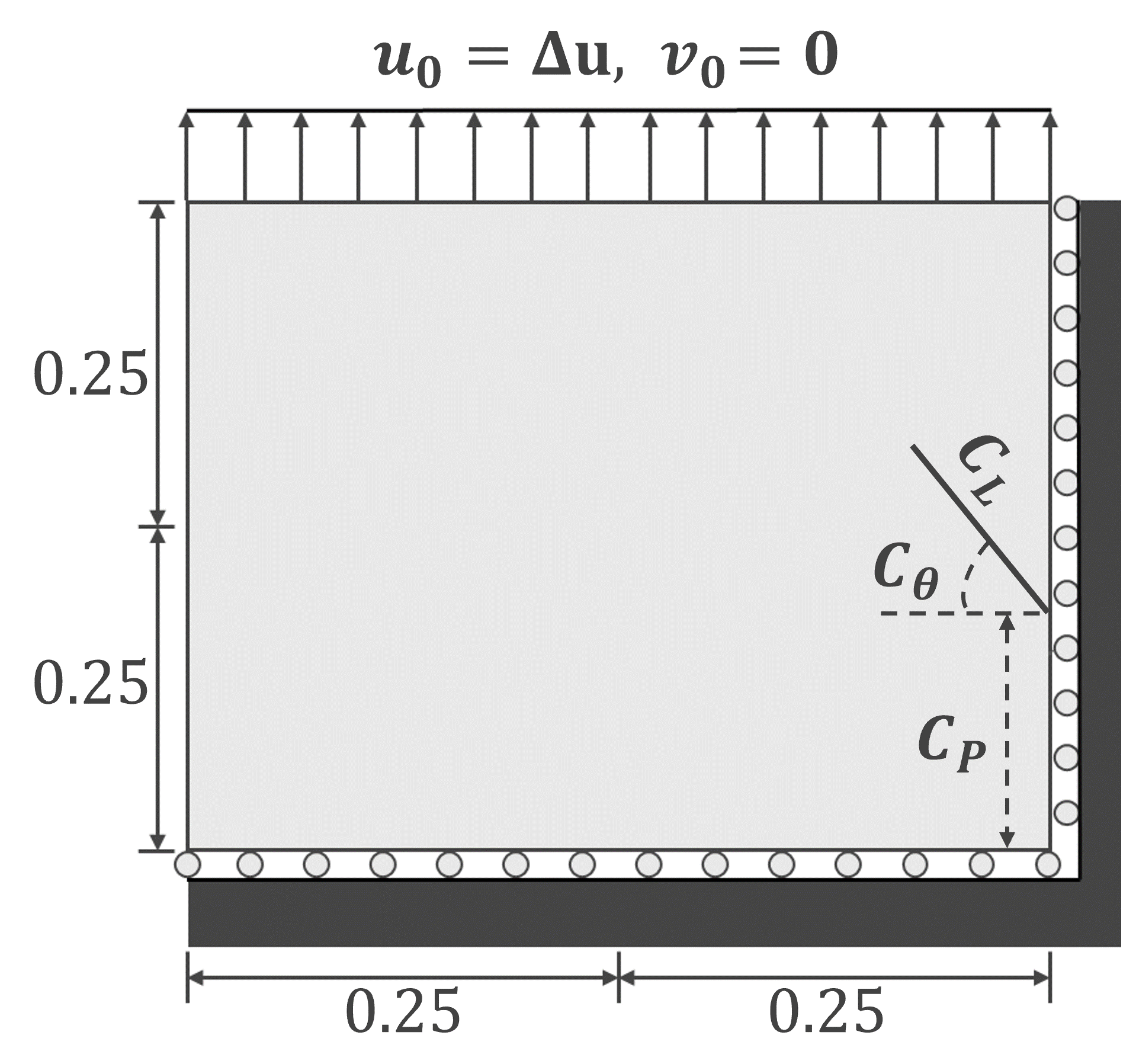}
				\centering
				\caption[r]{Right-edge crack in tension}
				\label{subfig:rightedge_crack}
			\end{subfigure}
			\caption{Problem geometry and input parameters $\{C_{P},C_{\theta},C_{L}\}$ set-up for a) initial case of left-edge cracks subjected to tension, b) center cracks subjected to tension, c) left-edge cracks subjected to shear, and c) right-edge cracks subjected to tension.}
			\label{fig:crack_cases}
		\end{figure}

		First, we trained the multiscale GNN framework for the left-edge notched system under tensile loading, shown in Figure \ref{subfig:leftedge_crack}.
		We obtained a dataset of 1100 PF fracture simulations involving single-edge notched systems under tensile loading from \cite{perera2023dynamic}.  
		Each of these simulations consisted of a unique crack configuration where the initial length $C_{L}$, edge position $C_{P}$, and orientation of the crack $C_{\theta}$ were varied.
		Additional details for the simulation set-up can be found in \cite{perera2023dynamic}.

		To perform TL, we first trained the multiscale GNN framework on the large dataset of 1100 single-edge notched simulations gathered from \cite{perera2023dynamic}.
		From the trained GNN framework, we transferred the weights of the encoder MLP (i.e., $MLP_{\text{in}}$), and the first MP model (i.e., GNN$_{D1}$) to a new GNN framework with similar architecture.
		Using this approach, we performed a series of sequential TL update steps.
		For each TL update step, we transferred the resulting pretrained weights to the next case study.
		We considered the following 4 case studies.
		
		\begin{itemize}
			\item Case 1: Left-edge cracks subjected to Mode I loading as shown in Figure \ref{subfig:leftedge_crack}.
			\item Case 2: Center cracks subjected to Mode I loading as shown in Figure \ref{subfig:center_crack}.
			\item Case 3: Left-edge cracks subjected to Mode II loading as shown in Figure \ref{subfig:shearload}.
			\item Case 4: Right-edge cracks subjected to Mode I loading as shown in Figure \ref{subfig:rightedge_crack}.
		\end{itemize}

		Unlike Case 1 where left-edge cracks are only allowed to propagate towards the right, in Case 2 center cracks can propagate in both the left and right directions, thus, increasing the framework's understanding.
        For Cases 2-3, we leveraged the PF model from \cite{goswami2020adaptive} to generate new datasets for center cracks and shear loading scenarios.
		For each of these cases, we generated a dataset of 30 simulations (i.e., 15 for training, and 15 for testing).
		For Case 4 we leveraged the symmetry of the left-edge crack case study (i.e., Case 1) to mirror 30 randomly chosen simulations from the training dataset used in Case 1.   
		We emphasize the significant decrease in the required size of the training dataset from 1100 (Case 1) to 15 simulations using TL (i.e., approximately 70 times smaller).
		The implementation of TL allows the multiscale GNN framework to be extended to other crack problems with a fast simulation time and decreased computational cost.


\section{Results and discussion}\label{sec:results}

	\subsection{Prediction and error analysis for FSR, TSR and SSR}\label{subsec:results_FSR_TSR_SSR_error}

		\begin{figure}
			\centering
			\begin{subfigure}[t]{1\textwidth}
				\includegraphics[width=1\linewidth]{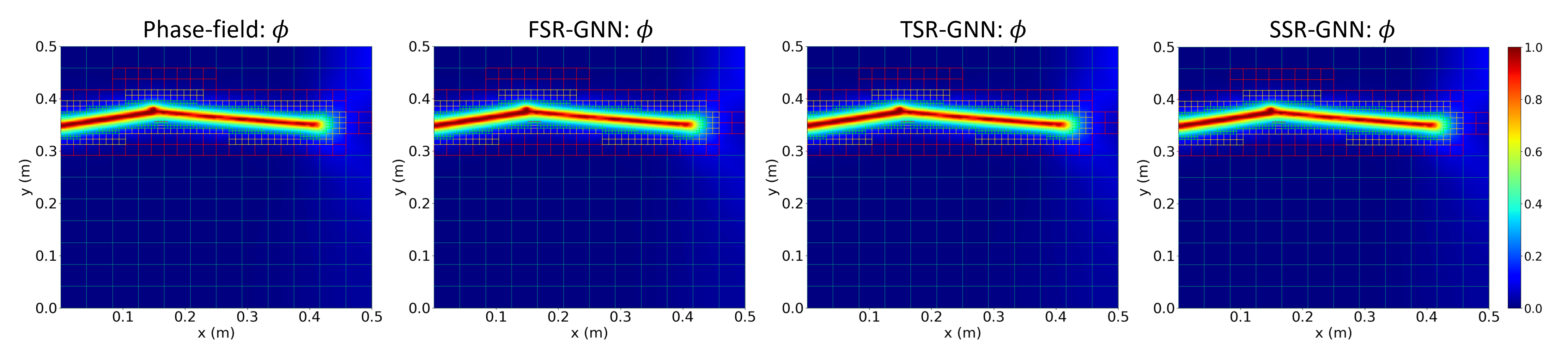}
				\centering
				\caption{Crack field, $\phi$.}
				\label{fig:leftedge_cPhi_FSR_TSR_SSR_sim}
			\end{subfigure}
			\centering
			\begin{subfigure}[c]{1\textwidth}
				\centering
				\includegraphics[width=1\linewidth]{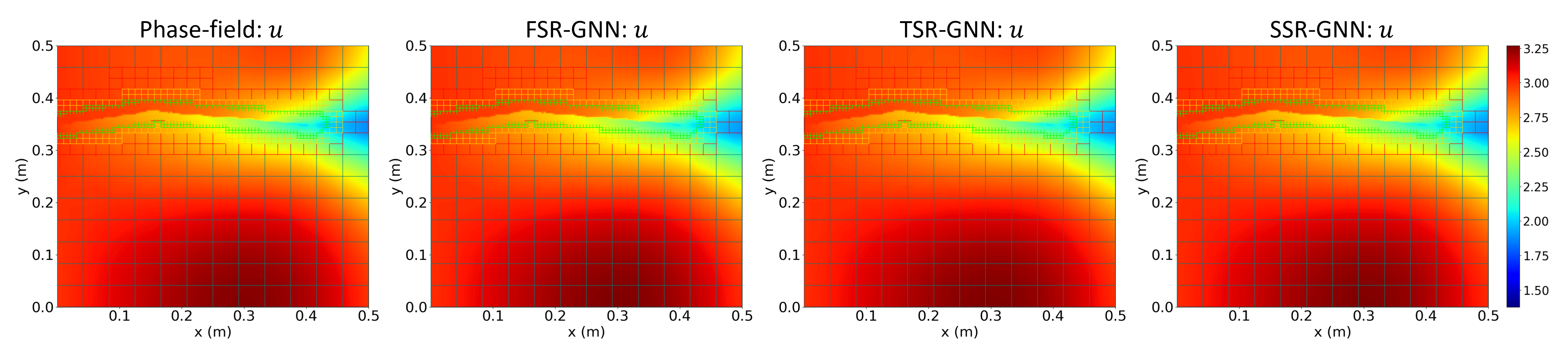}
				\centering
				\caption{X-displacement fields, $u$}
				\label{fig:leftedge_XDisp_FSR_TSR_SSR_sim}
			\end{subfigure}
			\centering
			\begin{subfigure}[b]{1\textwidth}
				\includegraphics[width=1\linewidth]{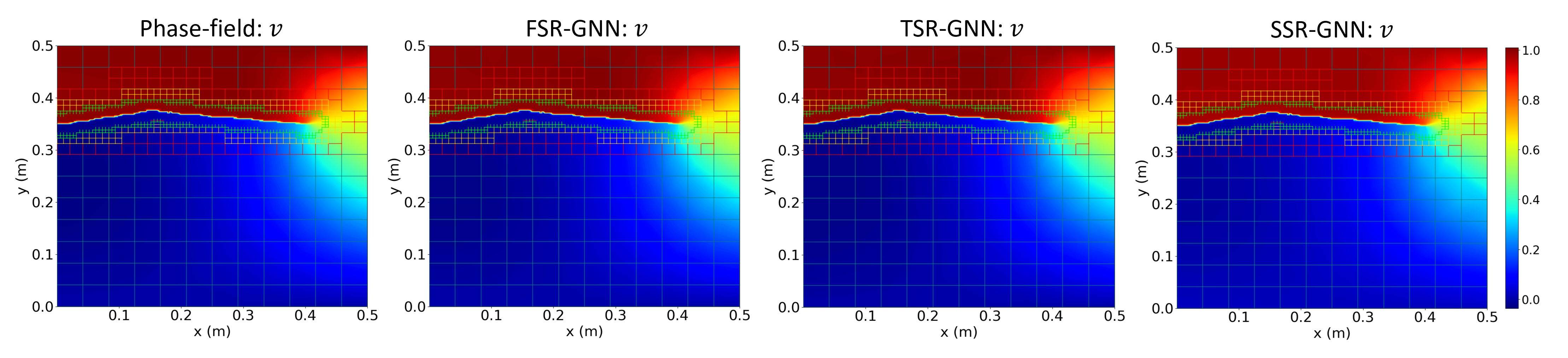}
				\centering
				\caption{Y-displacement fields, $v$}
				\label{fig:leftedge_YDisp_FSR_TSR_SSR_sim}
			\end{subfigure}
			\caption{Comparison of the FSR, TSR, and SSR frameworks versus the high-fidelity PF model for predicting a) crack field, $\phi$, b) x-displacement fields, and c) y-displacement fields in left-edge crack cases subjected to tension.}
		\end{figure}

        Here, we compare the performance of FSR, TSR, and SSR architectures of the multiscale GNN framework.
		We obtained the predictions of the crack variable $\phi$, $x$-displacement, and $y$-displacement for the left-edge crack cases.
		Figures \ref{fig:leftedge_cPhi_FSR_TSR_SSR_sim} - \ref{fig:leftedge_YDisp_FSR_TSR_SSR_sim} depict the predicted $\phi$ values, $x$-displacement, and $y$-displacement for a randomly chosen simulation from the test dataset of left-edge crack cases.
		The simulation chosen shows a crack of positive orientation, positioned towards the top of the left edge of the domain.
		The crack can also be seen approaching the right edge of the domain for complete material failure. 
		From Figure \ref{fig:leftedge_cPhi_FSR_TSR_SSR_sim} all three architectures predicted $\phi$ with nearly identical results to the ground truth (i.e., left-most case).
		We obtained a similar result for x- and y-displacements shown in Figures \ref{fig:leftedge_XDisp_FSR_TSR_SSR_sim} - \ref{fig:leftedge_YDisp_FSR_TSR_SSR_sim}, where qualitatively the FSR, TSR, and SSR architectures predicted displacements with high accuracy.
		This qualitative analysis demonstrates that the TSR and SSR were able to maintain their prediction accuracy despite the reduced downsampling and upsampling steps.

\begin{figure}
			\centering
			\begin{subfigure}[t]{1\textwidth}
				\begin{subfigure}[t]{0.32\textwidth}
					\centering
					\includegraphics[width=1\linewidth]{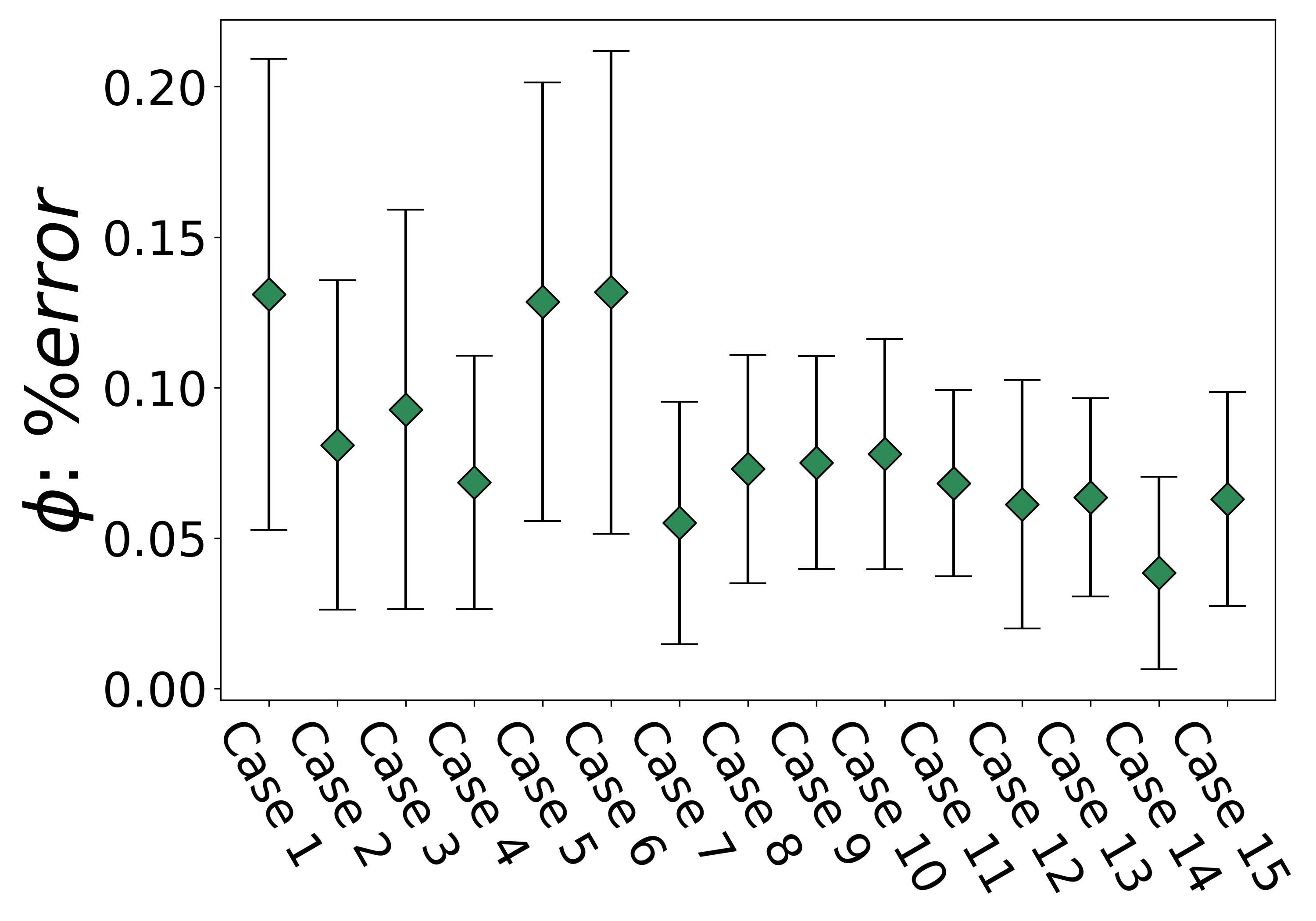}
					\caption{FSR $\%$ error: crack field, $\phi$}
					\label{fig:leftedge_Multiscale_ave_error_phi}
				\end{subfigure}
				\centering
				\begin{subfigure}[t]{0.32\textwidth}
					\centering
					\includegraphics[width=1\linewidth]{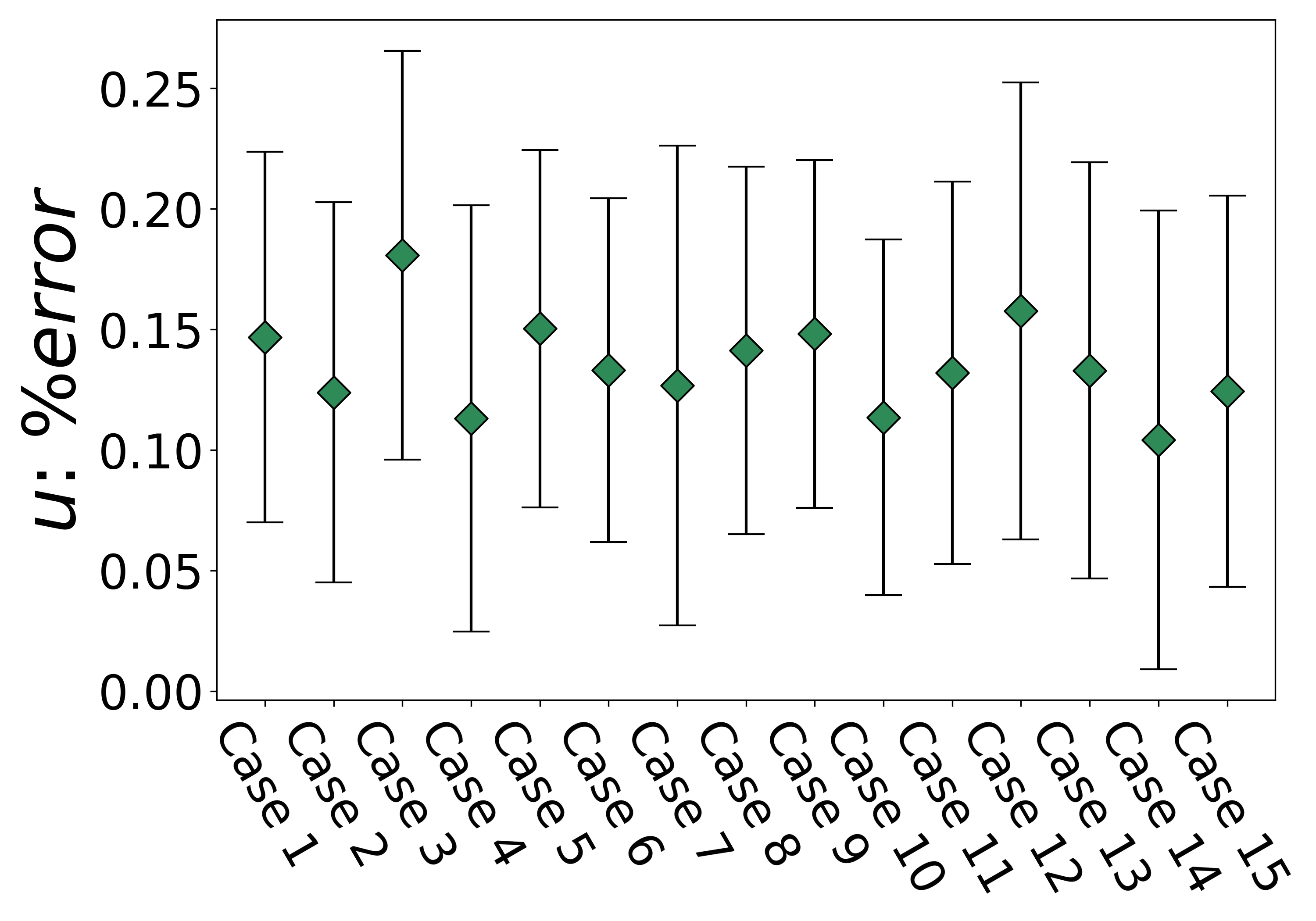}
					\caption{FSR $\%$ error: x-displacement, $u$}
					\label{fig:leftedge_Multiscale_ave_error_u}
				\end{subfigure}
				\centering
				\begin{subfigure}[t]{0.32\textwidth}
					\centering
					\includegraphics[width=1\linewidth]{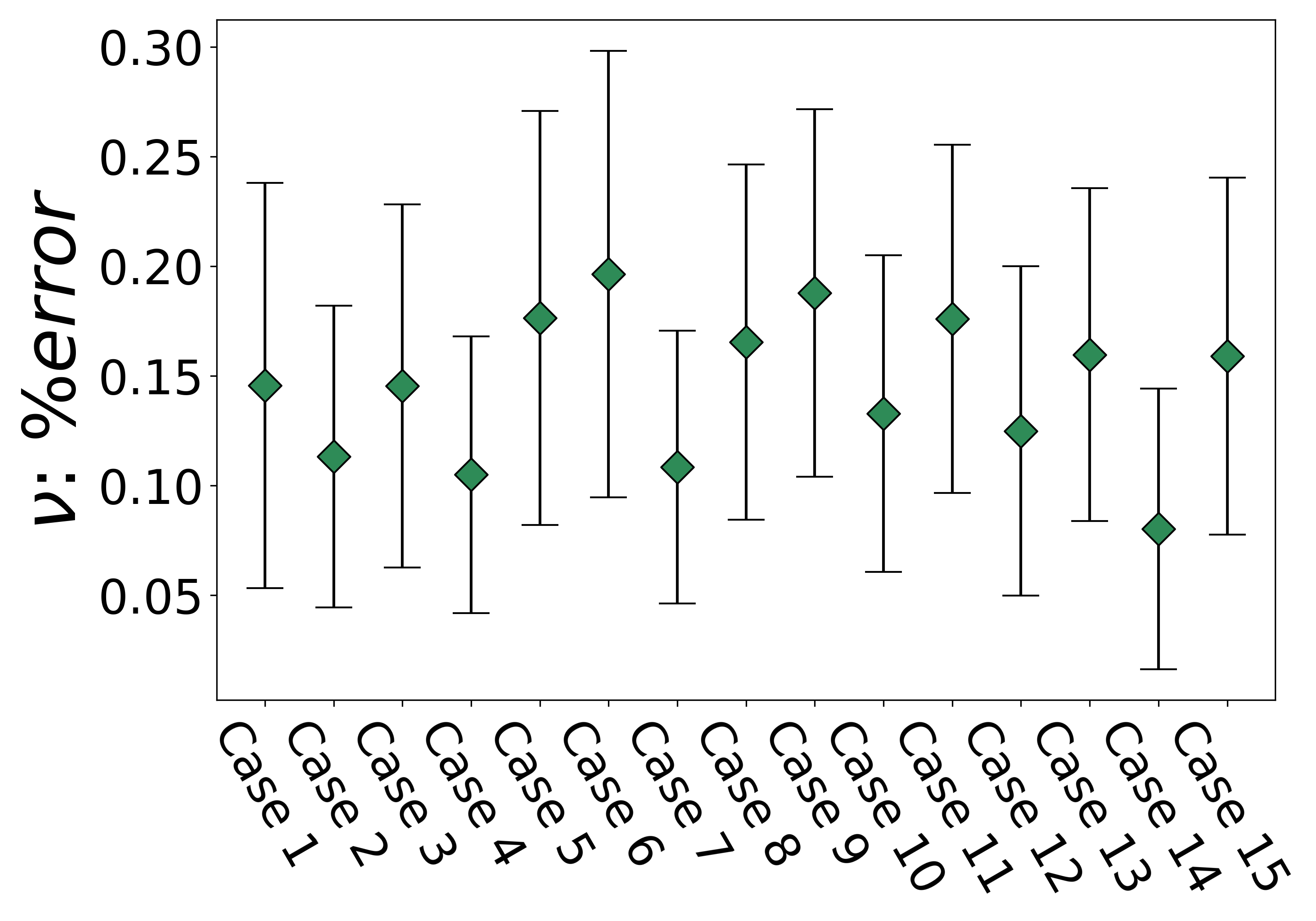}
					\caption{FSR $\%$ error: y-displacement, $v$}
					\label{fig:leftedge_Multiscale_ave_error_v}
				\end{subfigure}
			\end{subfigure}
			\centering
			\begin{subfigure}[c]{1\textwidth}
				\begin{subfigure}[t]{0.32\textwidth}
					\centering
					\includegraphics[width=1\linewidth]{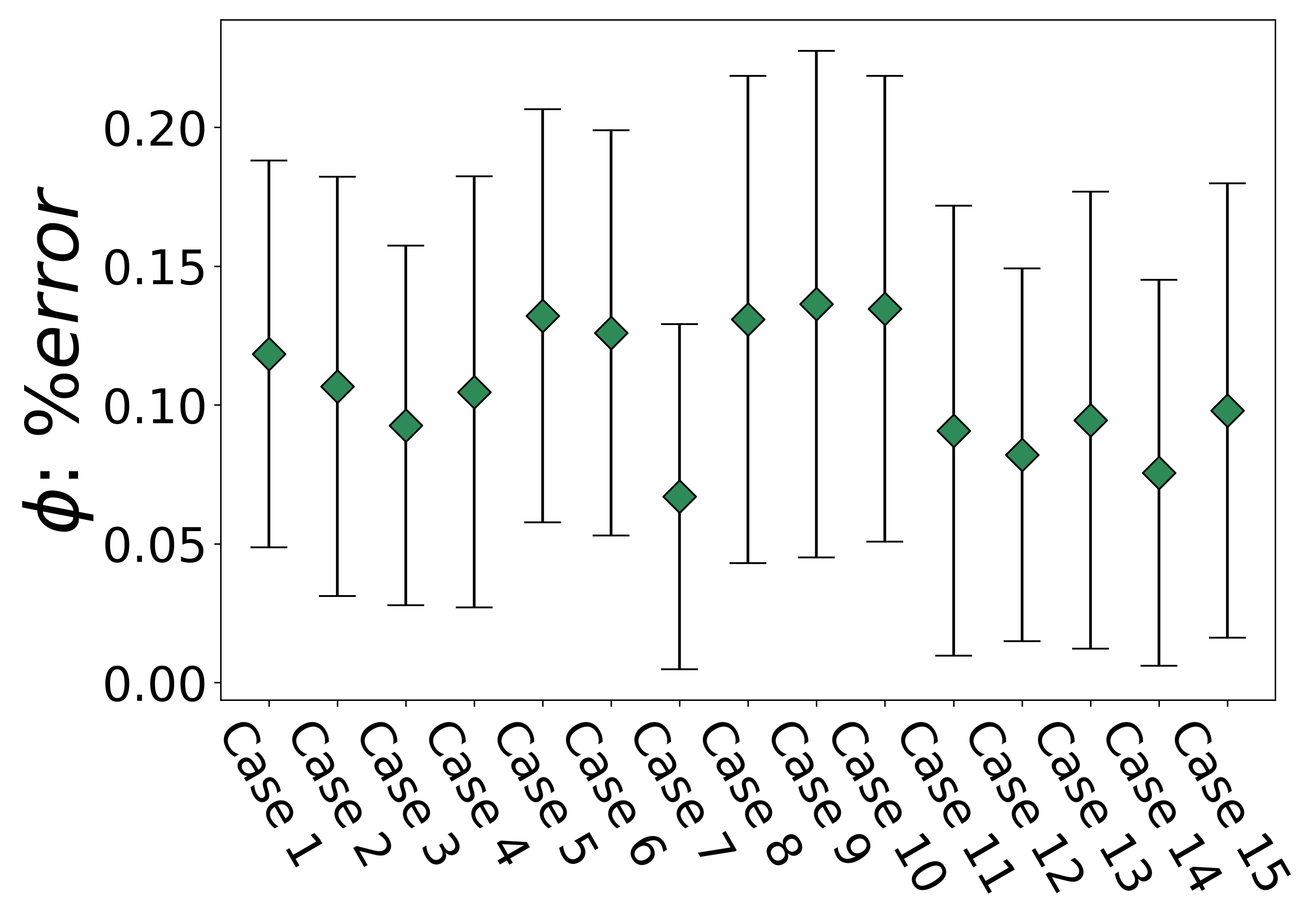}
					\caption{TSR $\%$ error: crack field, $\phi$}
					\label{fig:leftedge_MultiscaleTwoStage_ave_error_phi}
				\end{subfigure}
				\centering
				\begin{subfigure}[t]{0.32\textwidth}
					\centering
					\includegraphics[width=1\linewidth]{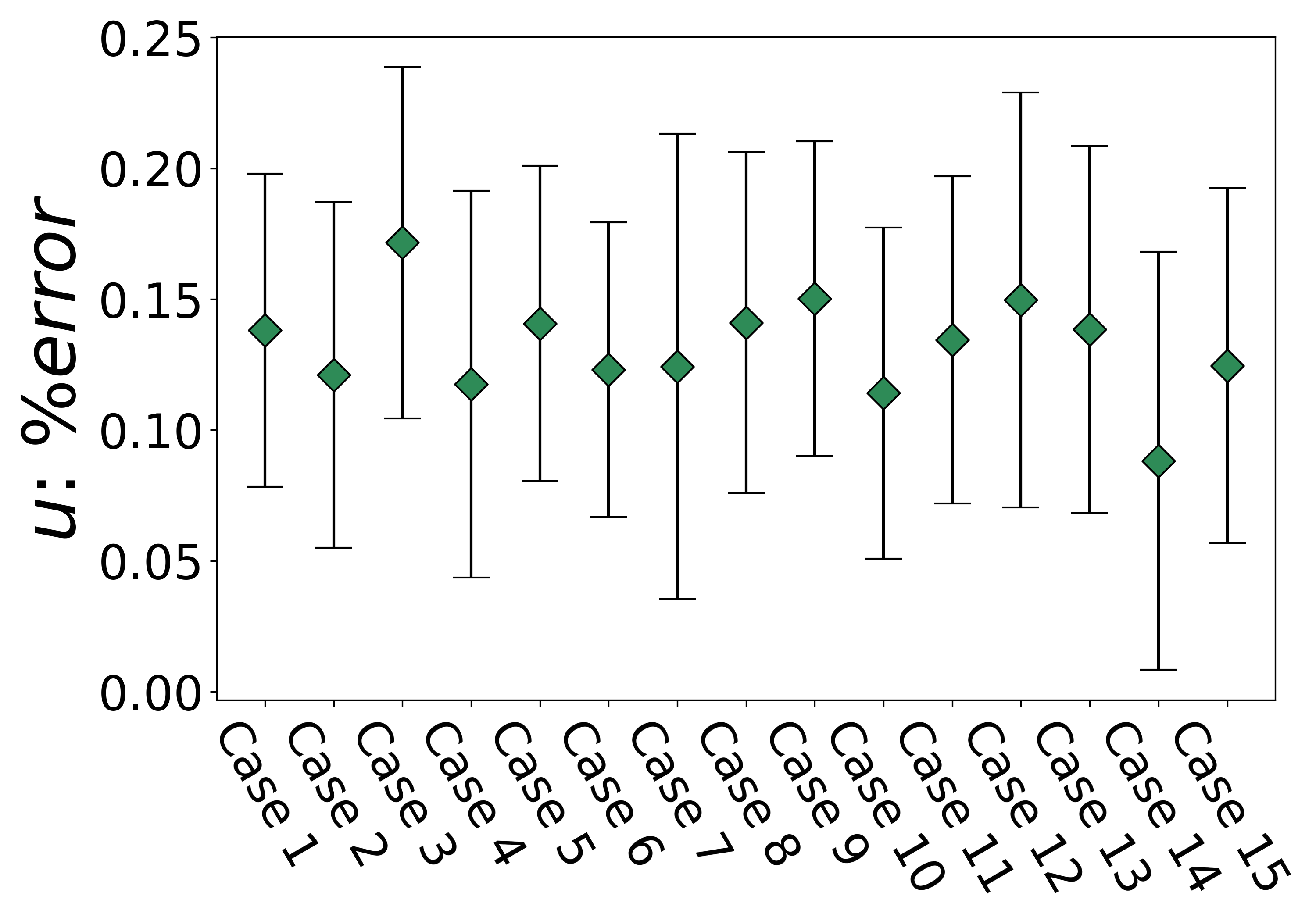}
					\caption{TSR $\%$ error: x-displacement, $u$}
					\label{fig:leftedge_MultiscaleTwoStage_ave_error_u}
				\end{subfigure}
				\centering
				\begin{subfigure}[t]{0.32\textwidth}
					\centering
					\includegraphics[width=1\linewidth]{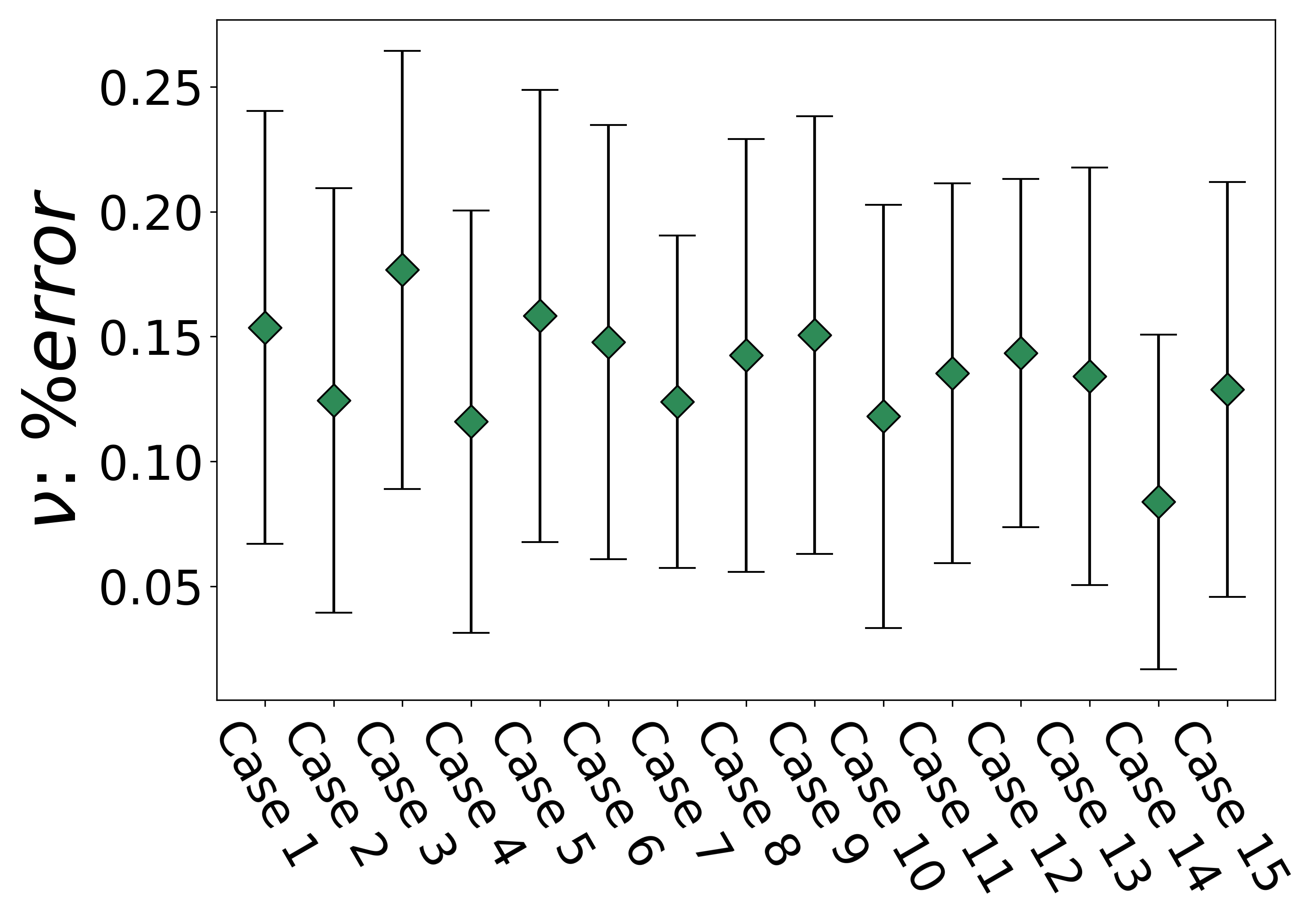}
					\caption{TSR $\%$ error: y-displacement, $v$}
					\label{fig:leftedge_MultiscaleTwoStage_ave_error_v}
				\end{subfigure}
			\end{subfigure}
			\centering
			\begin{subfigure}[b]{1\textwidth}
				\begin{subfigure}[t]{0.32\textwidth}
					\centering
					\includegraphics[width=1\linewidth]{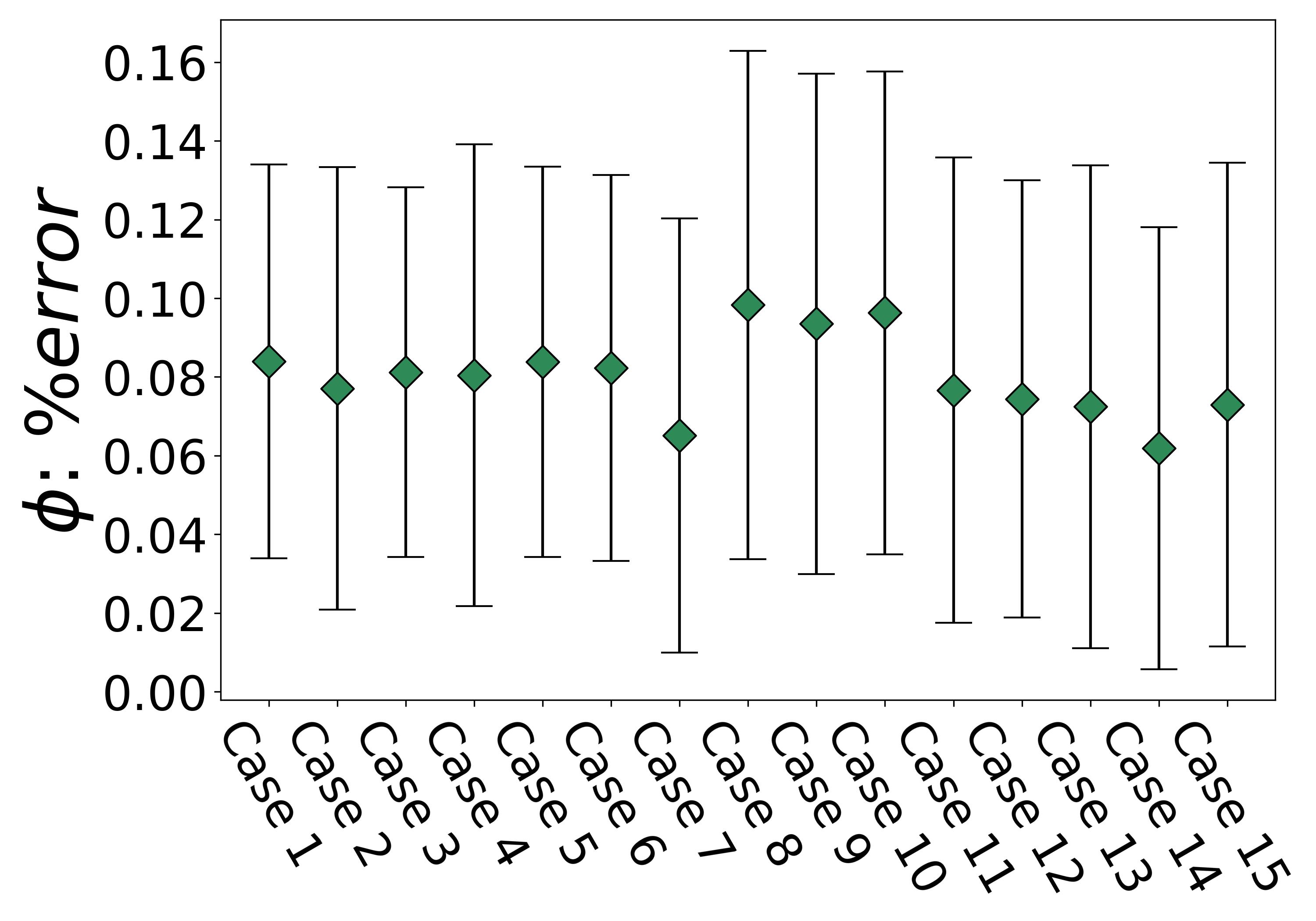}
					\caption{SSR $\%$ error: crack field, $\phi$}
					\label{fig:leftedge_MultiscaleSingleStage_ave_error_phi}
				\end{subfigure}
				\centering
				\begin{subfigure}[t]{0.32\textwidth}
					\centering
					\includegraphics[width=1\linewidth]{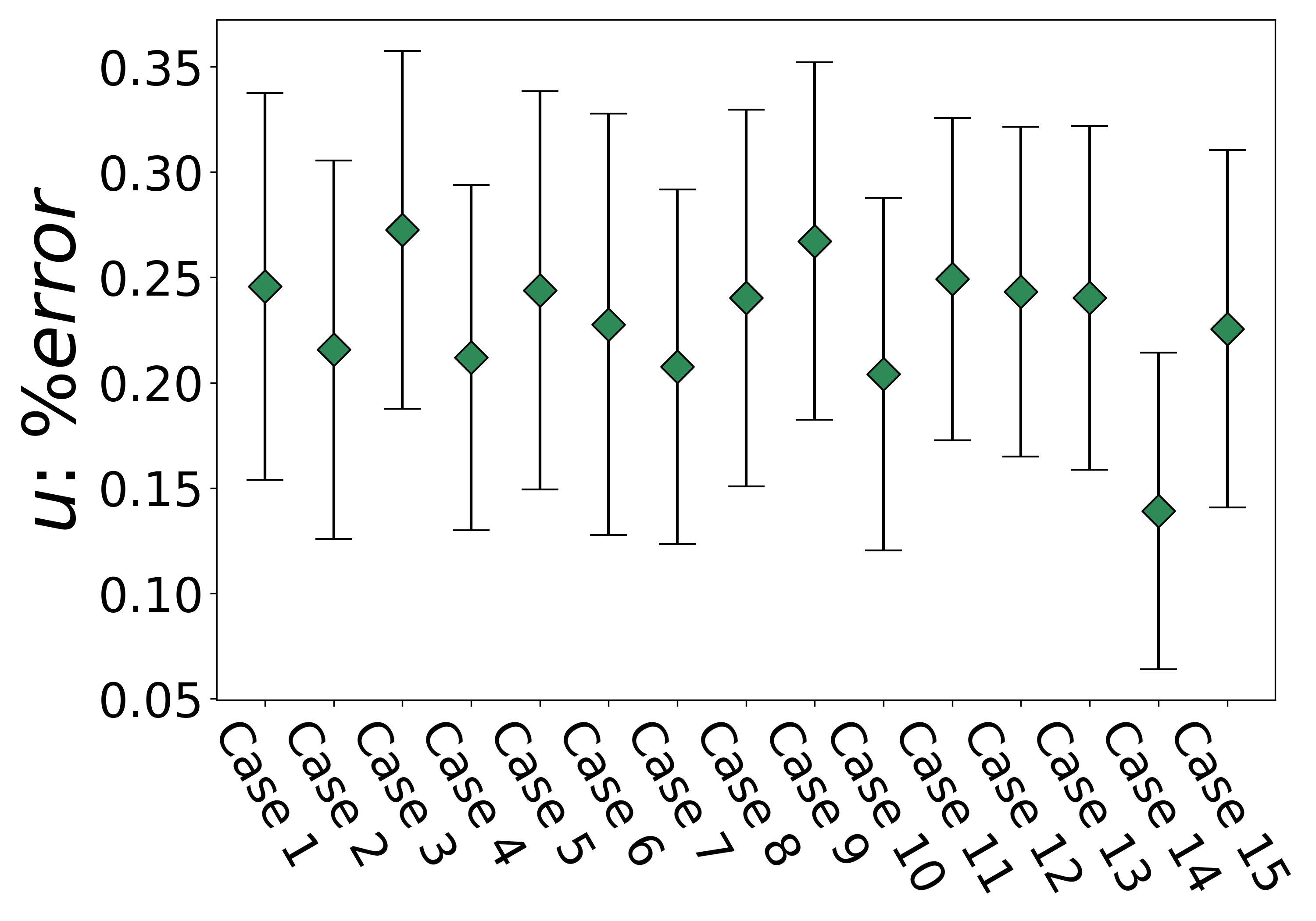}
					\caption{SSR $\%$ error: x-displacement, $u$}
					\label{fig:leftedge_MultiscaleSingleStage_ave_error_u}
				\end{subfigure}
				\centering
				\begin{subfigure}[t]{0.32\textwidth}
					\centering
					\includegraphics[width=1\linewidth]{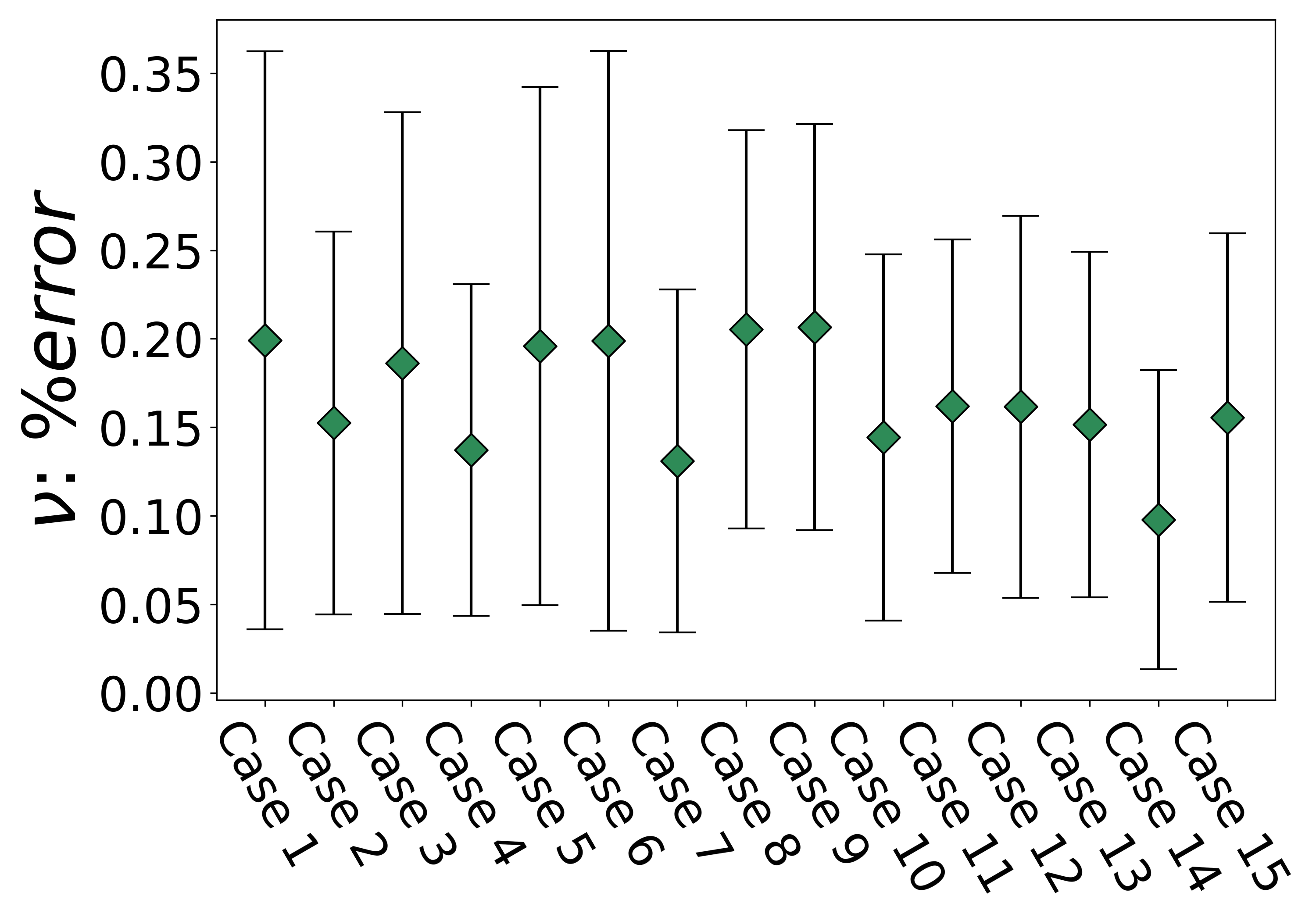}
					\caption{SSR $\%$ error: y-displacement, $v$}
					\label{fig:leftedge_MultiscaleSingleStage_ave_error_v}
				\end{subfigure}
			\end{subfigure}
			\caption{Comparison of average $\%$ errors on left-edge crack cases for a) FSR GNN crack-field predictions, b) FSR GNN x-displacement predictions, c) FSR GNN y-displacement predictions, d) TSR GNN crack-field predictions, e) TSR GNN x-displacement predictions, f) TSR GNN y-displacement predictions, g) SSR GNN crack-field predictions, h) SSR GNN x-displacement predictions, and i) SSR GNN y-displacement predictions.}
		\end{figure}

		Next, we computed the errors corresponding to each simulation from the test dataset. 
		For each test simulation, we first computed the average percent error across all mesh points in $\mathcal{M}^{ref}$ for each time step.
		We then computed the average of the resulting percent error across all time steps of the simulations.
		For instance, we computed the error for the crack field as
		{\begin{flalign}
			&& \phi_{error} = \frac{1}{T_{f}} \sum_{t}^{T_{f}} \left[ \frac{1}{M} \sum_{s}^{M} \frac{| \phi_{s}^{pred} - \phi_{s}^{true} |}{\phi_{s}^{true}} \times 100 \right]_{t} && . 
			\label{eq:average_error}
		\end{flalign}}
		were $\phi_{s}^{pred}$ and $\phi_{s}^{true}$ denote the predicted and ground truth crack field value at each mesh point $s\in\mathcal{M}^{ref}$, respectively, $M$ is the total number of mesh points in $\mathcal{M}^{ref}$, $t$ is the first predicted time-step, and $T_{f}$ the final time-step. 
		Using equation (\ref{eq:average_error}), we gathered the $\phi$ and displacement errors corresponding to each test simulation for the FSR, TSR, and SSR architectures.
		Figures \ref{fig:leftedge_Multiscale_ave_error_phi} - \ref{fig:leftedge_Multiscale_ave_error_v} show the resulting crack field and displacement errors for the FSR architecture.
		Figures \ref{fig:leftedge_MultiscaleTwoStage_ave_error_phi} - \ref{fig:leftedge_MultiscaleTwoStage_ave_error_v} depict the errors for the TSR architecture. 
		Similarly, Figure \ref{fig:leftedge_MultiscaleSingleStage_ave_error_phi} - \ref{fig:leftedge_MultiscaleSingleStage_ave_error_v} shows the resulting average errors for the SSR architecture.

		Comparing $\phi$ predictions (i.e., left-most), we note that the FSR, TSR, and SSR architectures resulted in average percent errors below 0.3$\%$.
		The errors in $x$-displacement and $y$-displacement also remained below 0.35$\%$ for all architectures. 
		These results confirm our previous observations from the qualitative analysis which indicated that TSR and SSR were capable of maintaining high prediction accuracy despite their reduced downsampling and upsampling steps.
  
		Next, we computed the average errors across all testing simulations for each architecture \ref{fig:all_errors}.
		From Figure \ref{fig:all_errors}, we note that FSR architecture demonstrated the highest accuracy when predicting the crack field $\phi$.
		However, the SSR architecture showed lower errors in crack field predictions than the TSR architecture.
		The TSR architecture resulted in the lowest error for $x$-displacement prediction, while the FSR and SSR architectures showed similar errors at approximately 0.08$\%$.
		For $y$-displacement predictions, the FSR and TSR architectures showed similar low errors at approximately 0.08$\%$, while the SSR architecture showed the highest error close to 0.12$\%$.
		Finally, Figure \ref{fig:all_errors} shows that reducing the number of refinement operations for TSR and SSR did not significantly increase prediction error.
		The highest prediction error of 0.12$\%$ for SSR on $y$-displacement predictions is still considerably low compared to previous work \cite{perera2023dynamic}.  

	\subsection{Simulation time analysis for FSR, TSR, and SSR}\label{subsec:results_simulation_time}

\begin{figure}
			\centering
			\begin{subfigure}[t]{0.49\textwidth}
				\centering
				\includegraphics[width=1\linewidth]{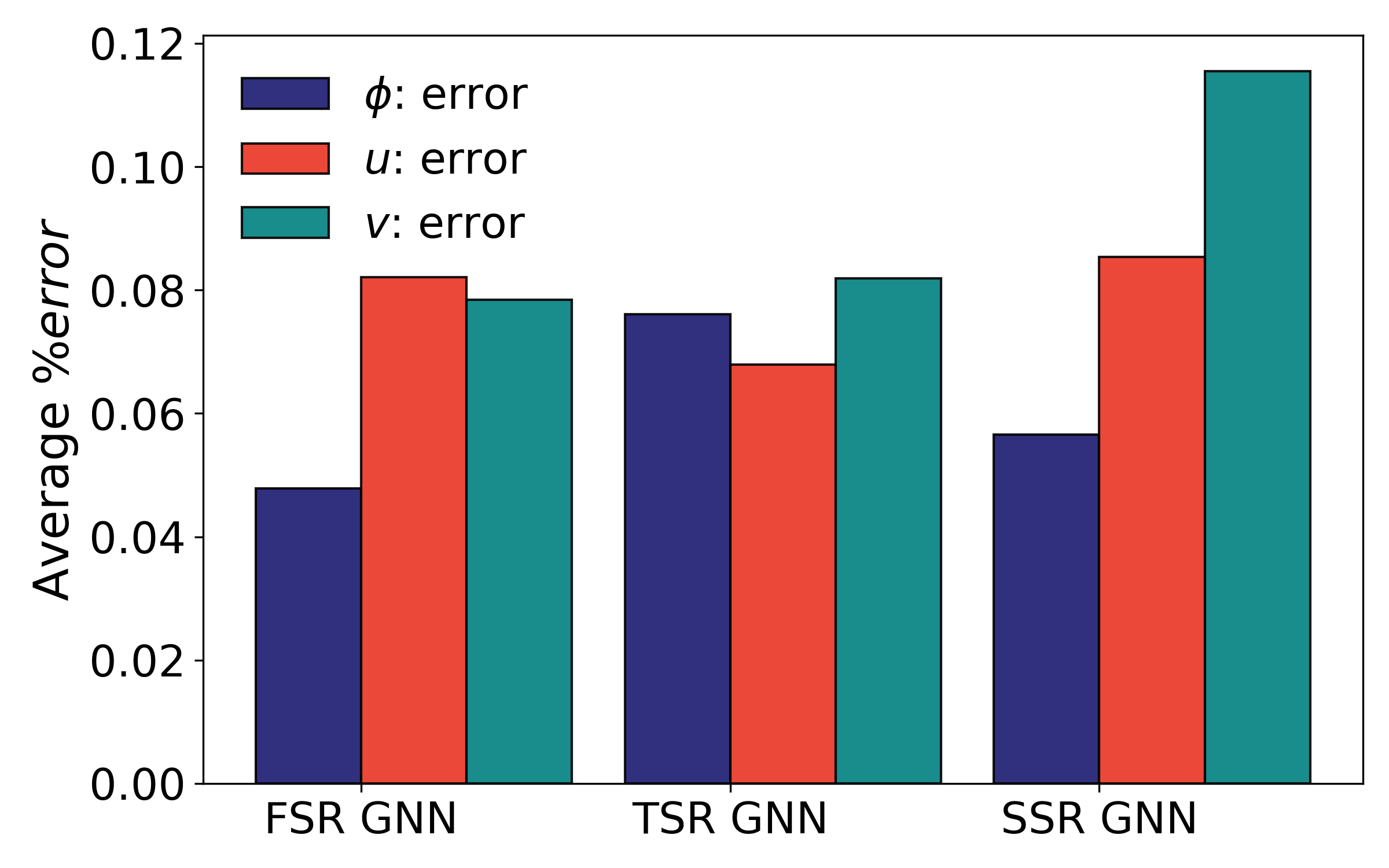}
				\caption{Average $\%$ errors for FSR, TSR, and SSR.}
				\label{fig:all_errors}
			\end{subfigure}
			\centering
			\begin{subfigure}[t]{0.49\textwidth}
				\centering
				\includegraphics[width=1\linewidth]{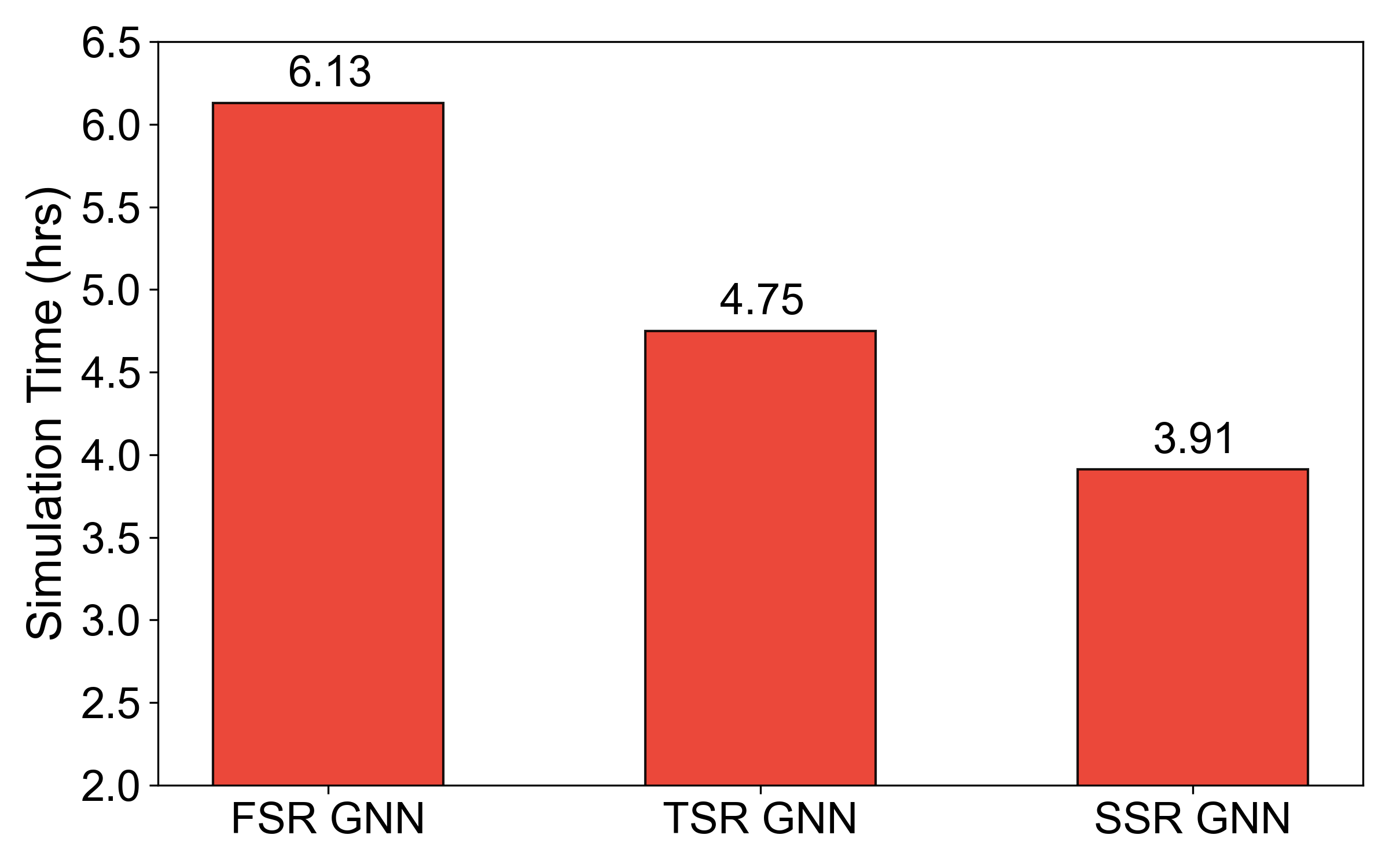}
				\caption{Simulation time (hrs) for FSR, TSR, and SSR.}
				\label{fig:simulation_time}
			\end{subfigure}
			\caption{a) Error analysis of the FSR, TSR, and SSR frameworks from the resulting average percent error across all testing simulations. b) Simulation time analysis (hrs) required for the FSR, TSR, and SSR frameworks to generate 30 simulations.}
			\label{superfig:ave_errors_and_simulation_time}
		\end{figure}

		We compared the computational cost for FSR, TSR, and SSR architectures by calculating the time required to generate 30 simulations.
		In Section \ref{subsec:results_FSR_TSR_SSR_error}, we showed that the prediction errors did not significantly increase for the TSR and SSR architectures despite their lower number of downscaling and upscaling operations.
		Each operation increases computational costs because it requires storing the resulting mesh configurations and their node embeddings.
        Each operation also comes with added MP GNNs.
		Figure \ref{fig:simulation_time} depicts the total time in hours required for each framework to generate 30 randomly chosen simulations from the training dataset.
		As shown in Figure \ref{fig:simulation_time}, the FSR GNN architecture required the longest simulation time of 6.13 hours.
		The TSR architecture was the second most computationally demanding, requiring 4.75 hours to simulate 30 cases.
		As expected, the fastest architecture with the lowest computational costs was the SSR, which required only 3.91 hours.
		In contrast, we note from \cite{perera2023dynamic} that the high-fidelity PF model required approximately 43.5 hours to generate 30 cases.
        Due to its lower simulation time and high prediction accuracy, we chose the SSR architecture for subsequent TL steps.

	\subsection{Center crack cases}\label{subsec:results_center_cracks}

		\begin{figure}
			\centering
			\begin{subfigure}[t]{\textwidth}
				\centering
				\begin{subfigure}[l]{0.49\textwidth}
					\centering
					\includegraphics[width=1\linewidth]{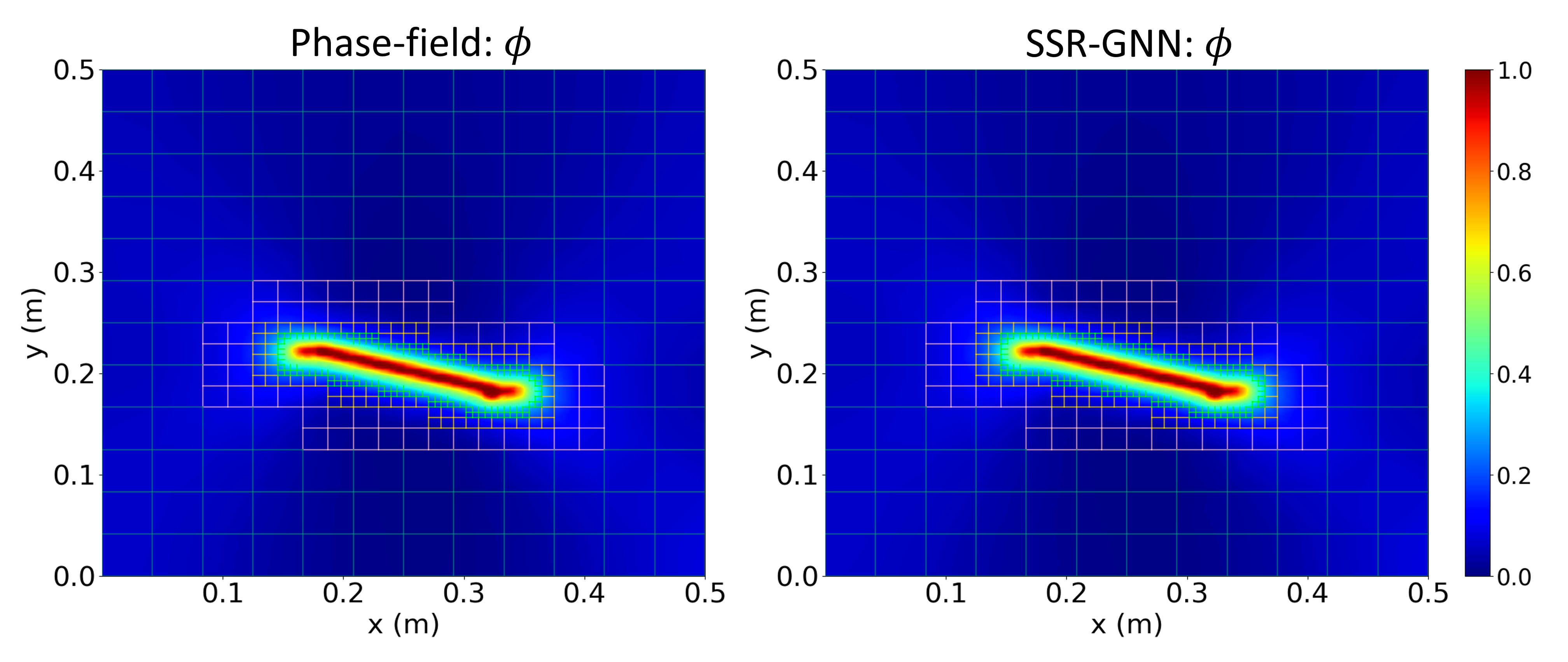}
					\caption{$\phi$ at time $t_{0}$}
					\label{subfig:CenterCrack_cPhi_sim_t0}
				\end{subfigure}
				\centering
				\begin{subfigure}[r]{0.49\textwidth}
					\centering
					\includegraphics[width=1\linewidth]{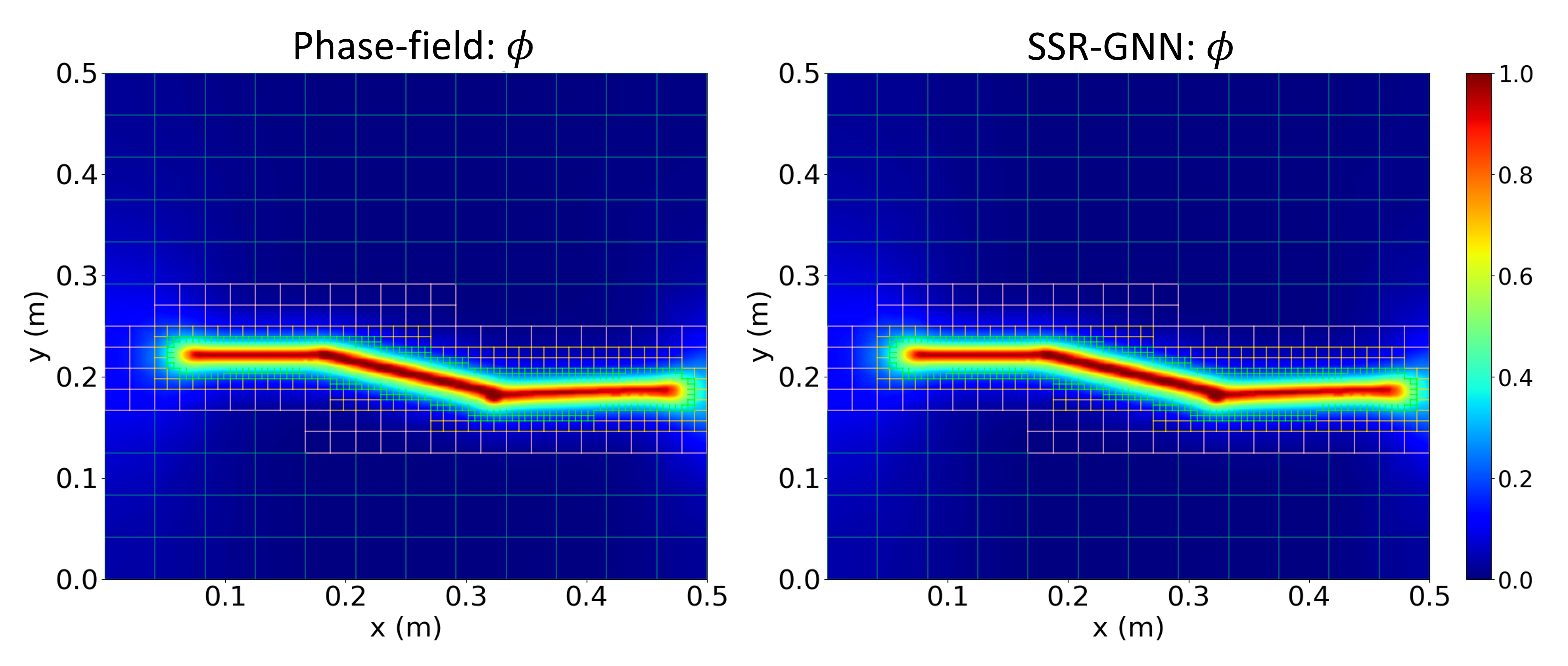}
					\caption{$\phi$ at time $t_{f}$}
					\label{subfig:CenterCrack_cPhi_sim_tf}
				\end{subfigure}
			\end{subfigure}
			\centering
			\begin{subfigure}[c]{\textwidth}
				\centering
				\begin{subfigure}[l]{0.49\textwidth}
					\centering
					\includegraphics[width=1\linewidth]{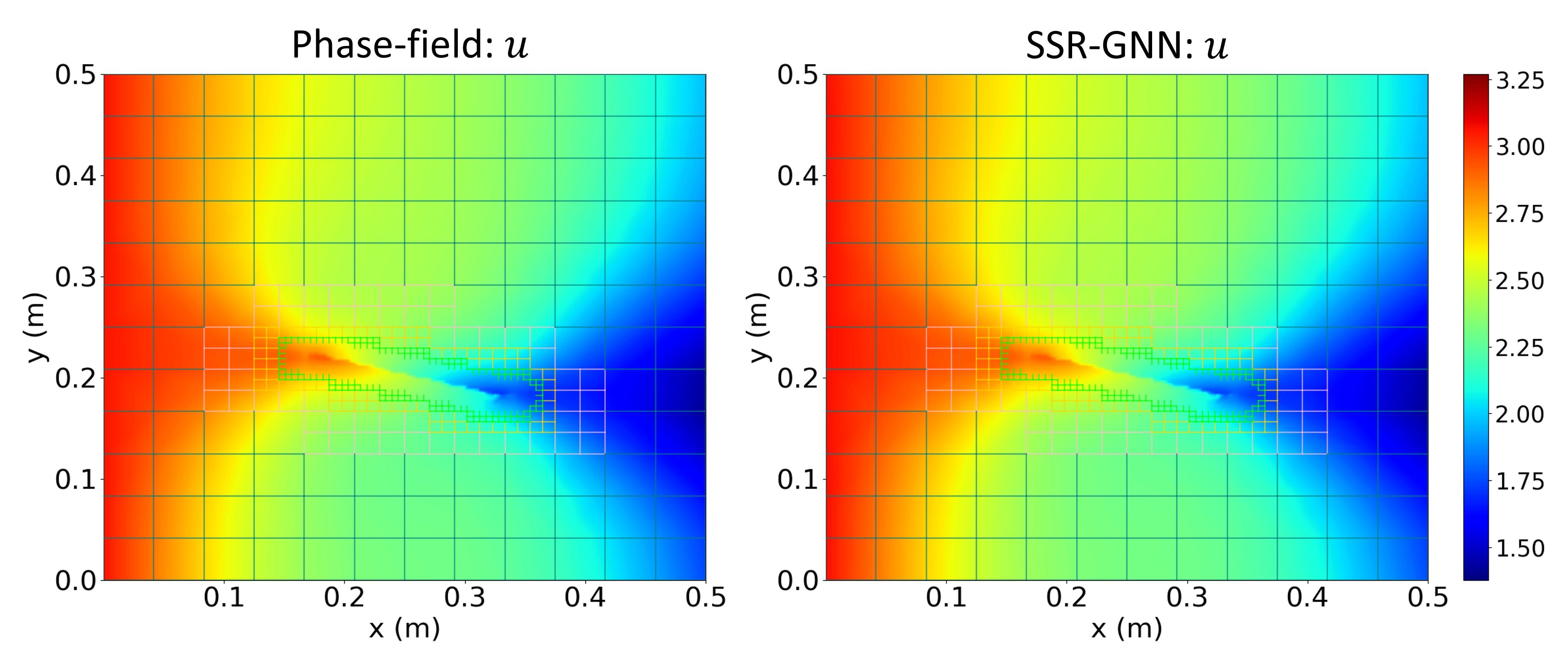}
					\caption{$u$: time $t_{0}$}
					\label{subfig:CenterCrack_XDisp_sim_t0}
				\end{subfigure}
				\centering
				\begin{subfigure}[r]{0.49\textwidth}
					\centering
					\includegraphics[width=1\linewidth]{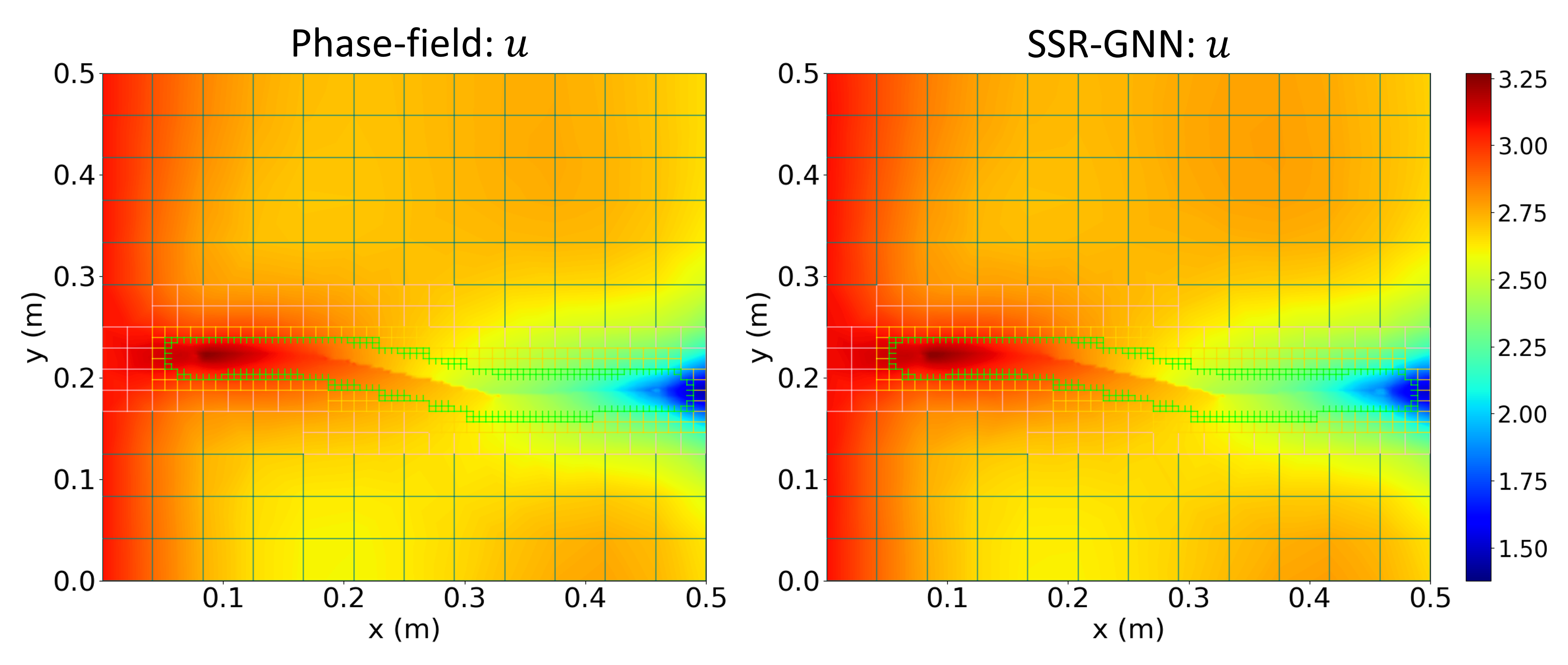}
					\caption{$u$: time $t_{f}$}
					\label{subfig:CenterCrack_XDisp_sim_tf}
				\end{subfigure}
			\end{subfigure}
			\centering
			\begin{subfigure}[b]{\textwidth}
				\centering
				\begin{subfigure}[l]{0.49\textwidth}
					\centering
					\includegraphics[width=1\linewidth]{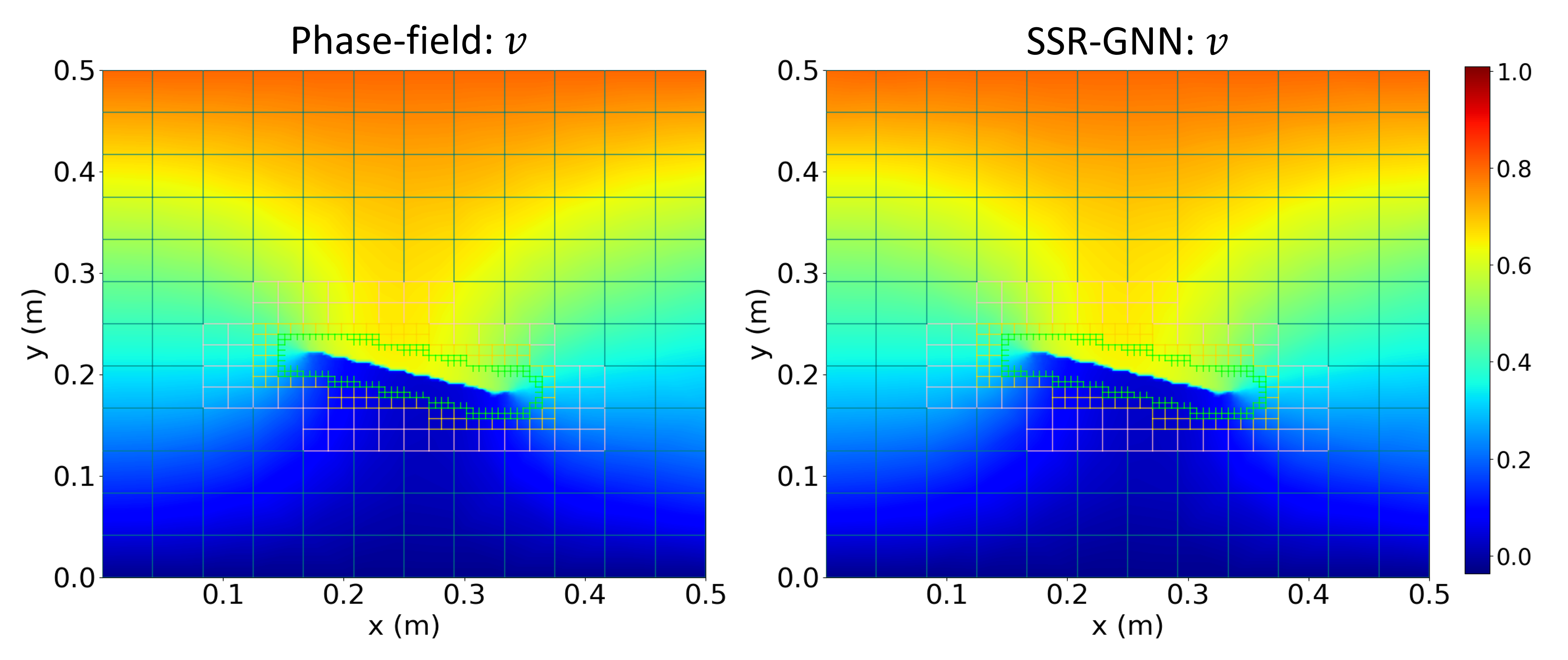}
					\caption{$v$: time $t_{0}$}
					\label{subfig:CenterCrack_YDisp_sim_t0}
				\end{subfigure}
				\centering
				\begin{subfigure}[r]{0.49\textwidth}
					\centering
					\includegraphics[width=1\linewidth]{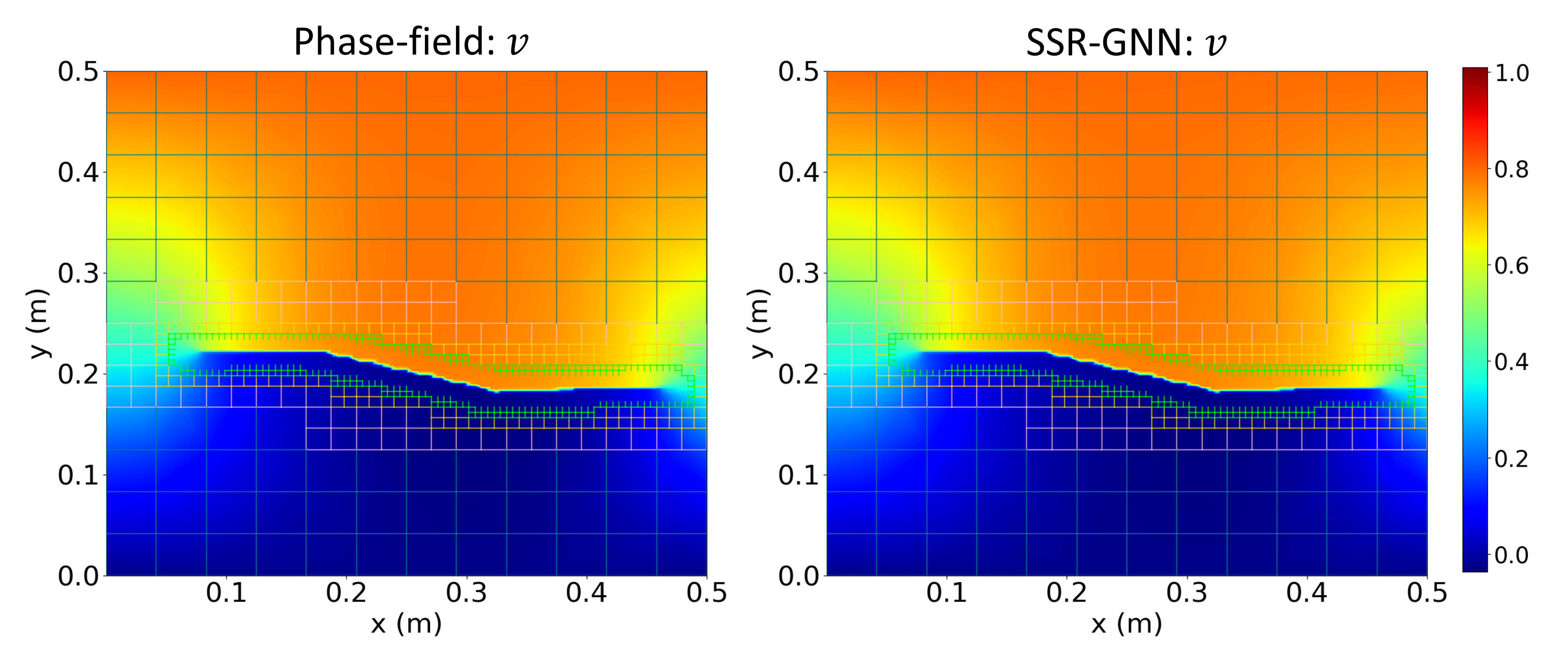}
					\caption{$v$: time $t_{f}$}
					\label{subfig:CenterCrack_YDisp_sim_tf}
				\end{subfigure}
			\end{subfigure}
			\caption{PF model versus the SSR framework on the predicted evolution of the crack field, $\phi$ (a-b), x-displacement field, $u$ (c-d), and y-displacement field, $v$ (e-f), for a center crack test simulation at initial time-step, $t_{0}$, and final time-step, $t_{f}$.}
			\label{fig:CenterCrack_sim}
		\end{figure}

		In Sections \ref{subsec:results_FSR_TSR_SSR_error} - \ref{subsec:results_simulation_time}, we determined that the SSR architecture provided high prediction accuracy at significantly lower computational costs. 
        We implemented TL to the multiscale GNN framework with SSR architecture for cases with center cracks as shown in Figure \ref{subfig:center_crack}.
        We tested the extended SSR-based framework to predict the crack and displacement fields for a randomly chosen case from the testing data.
		For this analysis, we first obtained the initial time-step prediction ($t_{0}$) as shown in Figure \ref{subfig:CenterCrack_cPhi_sim_t0} for the ground truth versus predicted crack field, Figure \ref{subfig:CenterCrack_XDisp_sim_t0} for the ground truth versus predicted $x$-displacement, and Figure \ref{subfig:CenterCrack_YDisp_sim_t0} for the ground truth versus predicted $y$-displacement. 
		Next, we propagated the simulation in time and obtained the corresponding predictions of crack and displacement fields for a time-step approaching complete material failure ($t_{f}$).
  
		Figures \ref{subfig:CenterCrack_cPhi_sim_tf}, \ref{subfig:CenterCrack_XDisp_sim_tf}, and \ref{subfig:CenterCrack_YDisp_sim_tf} depict the ground truth versus predicted crack field, $x$-displacement, and $y$-displacement, respectively, at $t_{f}$.
		These results depict similar accuracies compared to the left-edge crack cases.
		During both time steps (i.e., $t_{0}$ and $t_{f}$) the SSR-based framework's predictions are virtually indistinguishable from the high-fidelity PF model. 
		These qualitative results demonstrate that the SSR-based framework was capable of predicting crack propagation for center crack cases through TL. 

\begin{figure}
			\centering
			\begin{subfigure}[t]{0.32\textwidth}
				\centering
				\includegraphics[width=1\linewidth]{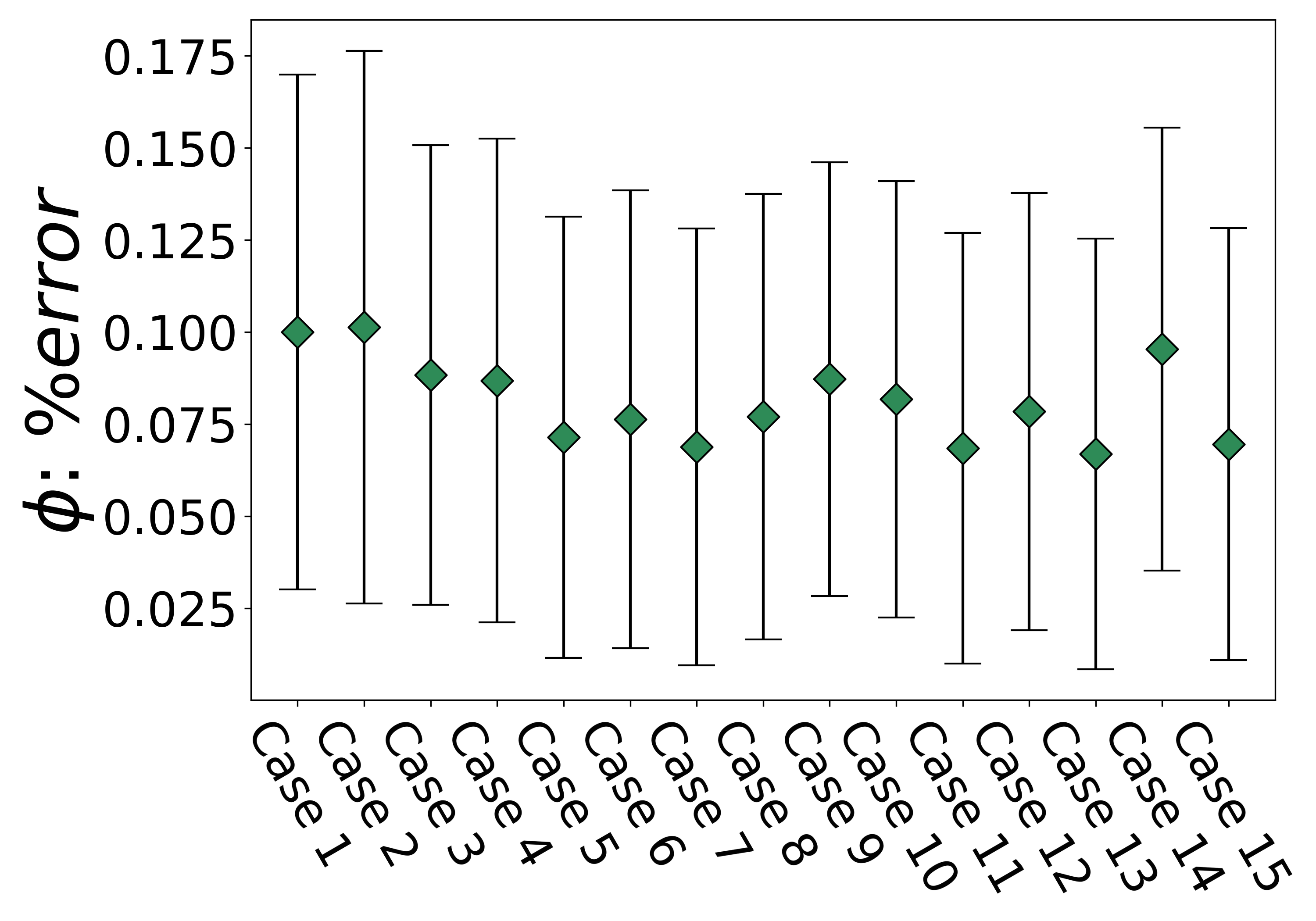}
				\caption{crack field, $\phi$}
			\end{subfigure}
			\centering
			\begin{subfigure}[t]{0.32\textwidth}
				\centering
				\includegraphics[width=1\linewidth]{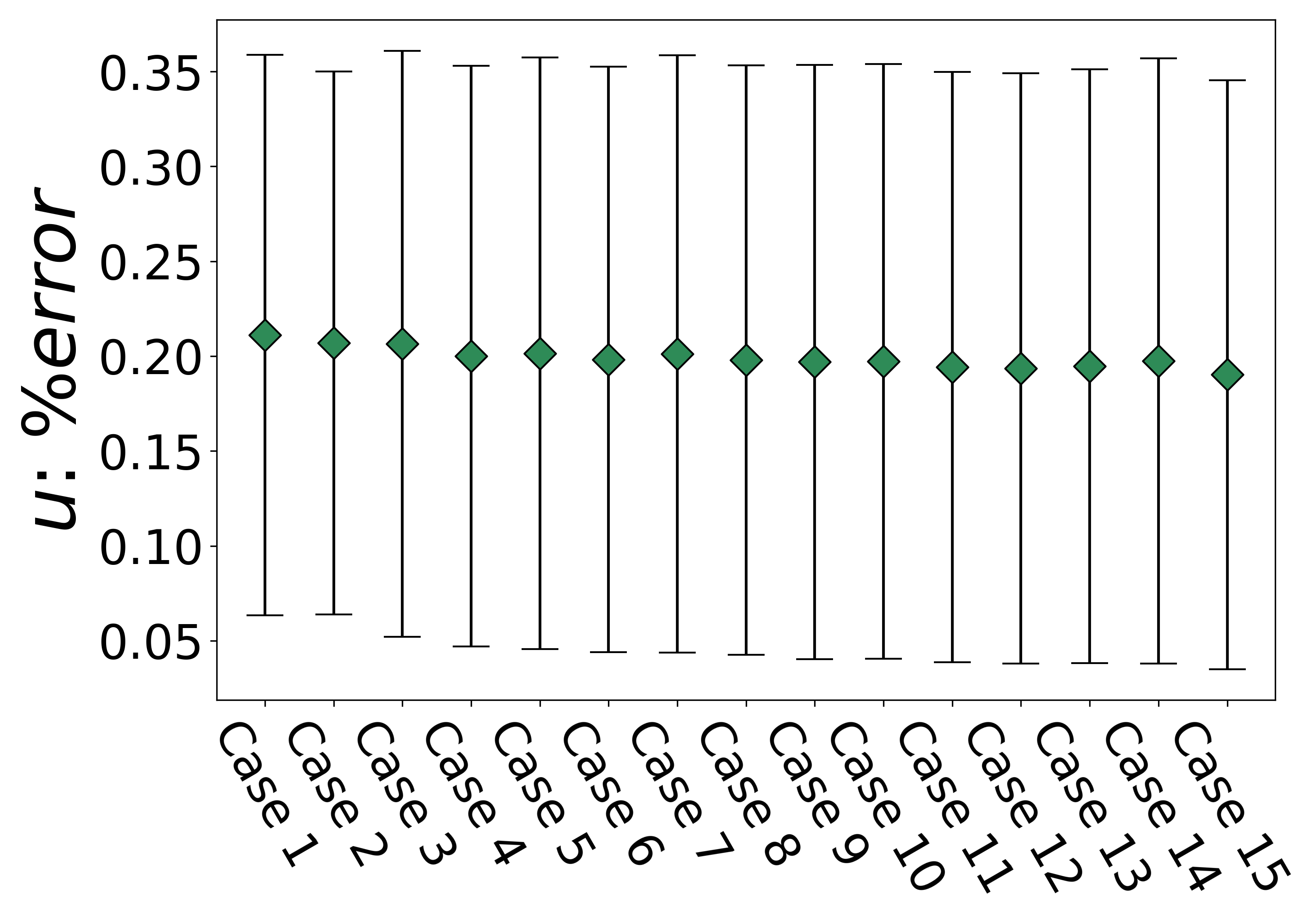}
				\caption{X-displacement, $u$}
			\end{subfigure}
			\centering
			\begin{subfigure}[t]{0.32\textwidth}
				\centering
				\includegraphics[width=1\linewidth]{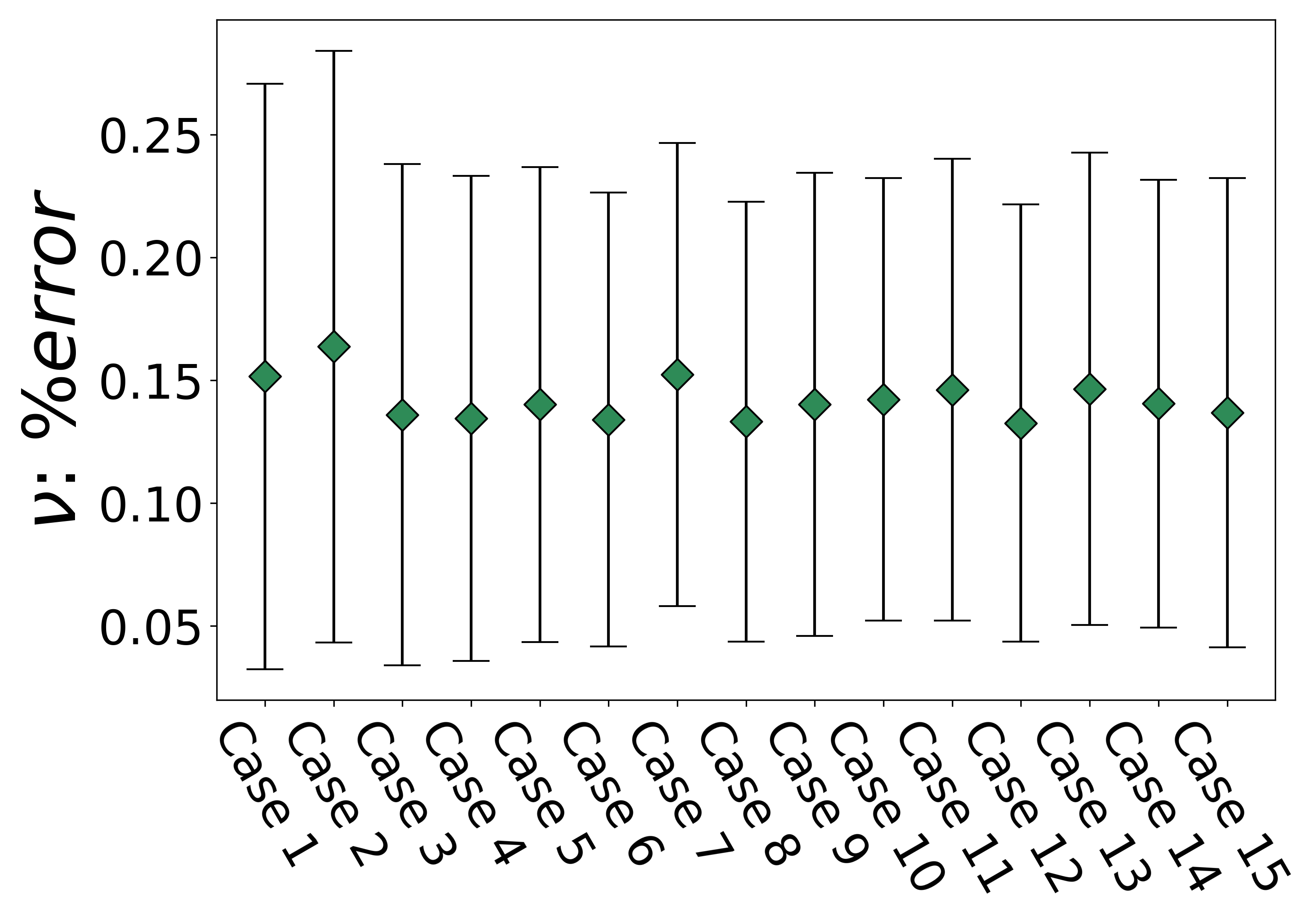}
				\caption{Y-displacement, $v$}
			\end{subfigure}
			\caption{Average percent errors for each simulation in the test dataset of center crack cases for a) crack-field predictions, $\phi$, b) x-displacement predictions, $u$, and c) y-displacement predictions, $v$.}
			\label{fig:CenterCrack_ave_error}
		\end{figure}

		We also performed a quantitative analysis to verify these qualitative observations by generating the errors of each test simulation. 
		Figure \ref{fig:CenterCrack_ave_error} depicts the computed average percent errors for center crack test cases obtained using equation (\ref{eq:average_error}).
		For crack field errors, the extended SSR framework maintained high prediction accuracy (less than 0.125$\%$ error) across all test cases.
		Similarly, the $x$-displacement errors remained below 0.25$\%$ error, and y-displacement errors below 0.20$\%$.
		These results emphasize the high prediction accuracy of the SSR framework by implementing TL using significantly smaller training datasets.  
		Additionally, extending the SSR framework from left-edge cracks to center cracks increases the framework's capabilities in predicting cases where cracks propagate in both the left and right directions.

	\subsection{Shear load cases}

		\begin{figure}
			\centering
			\begin{subfigure}[t]{\textwidth}
				\centering
				\begin{subfigure}[l]{0.49\textwidth}
					\centering
					\includegraphics[width=1\linewidth]{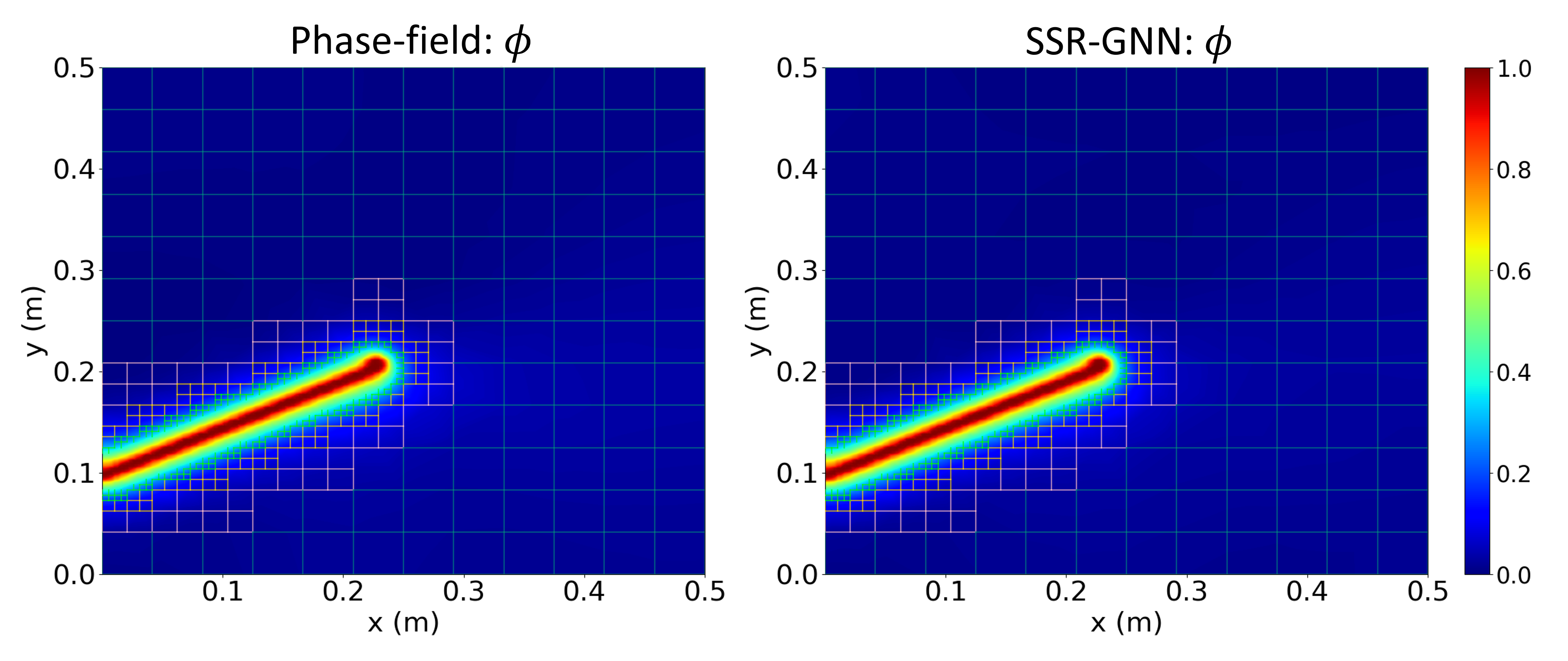}
					\caption{$\phi$ at time $t_{0}$}
					\label{subfig:ShearLoad_cPhi_sim_t0}
				\end{subfigure}
				\centering
				\begin{subfigure}[r]{0.49\textwidth}
					\centering
					\includegraphics[width=1\linewidth]{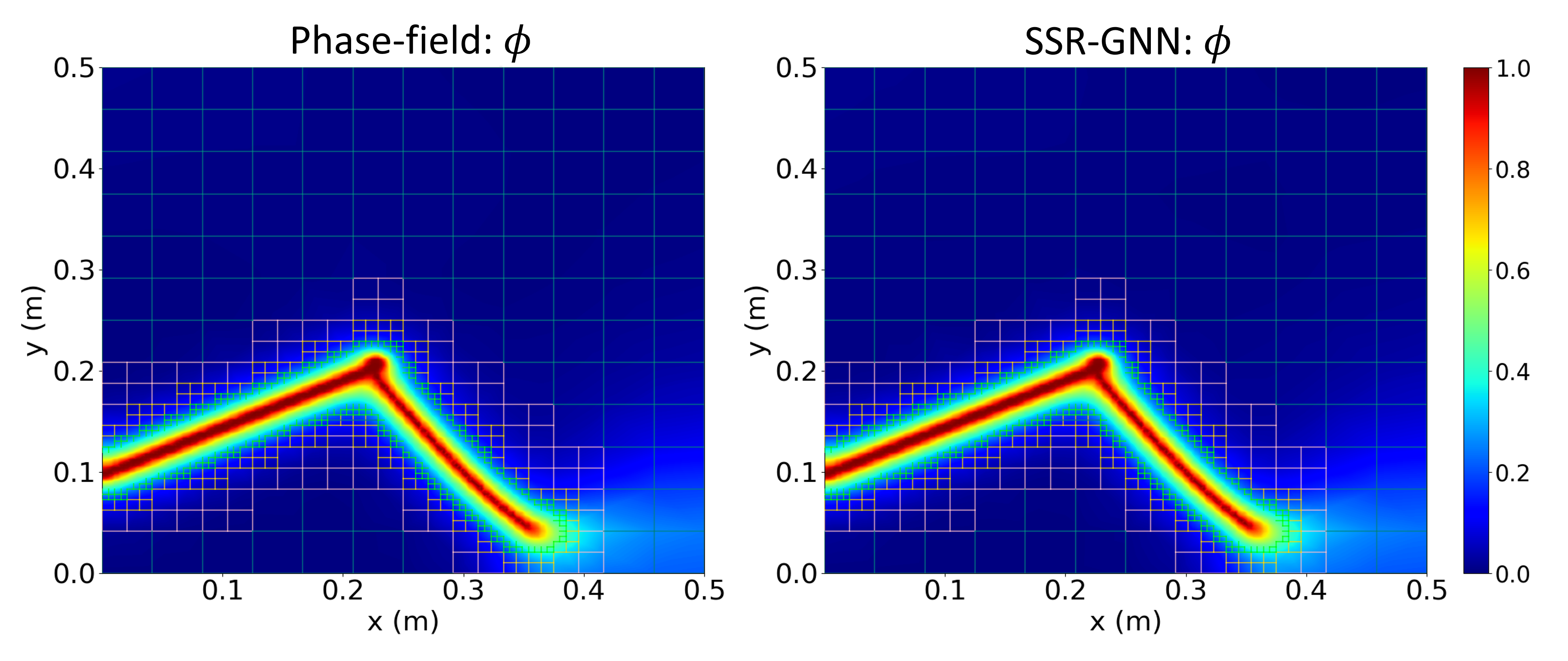}
					\caption{$\phi$ at time $t_{f}$}
					\label{subfig:ShearLoad_cPhi_sim_tf}
				\end{subfigure}
			\end{subfigure}
			\centering
			\begin{subfigure}[c]{\textwidth}
				\centering
				\begin{subfigure}[l]{0.49\textwidth}
					\centering
					\includegraphics[width=1\linewidth]{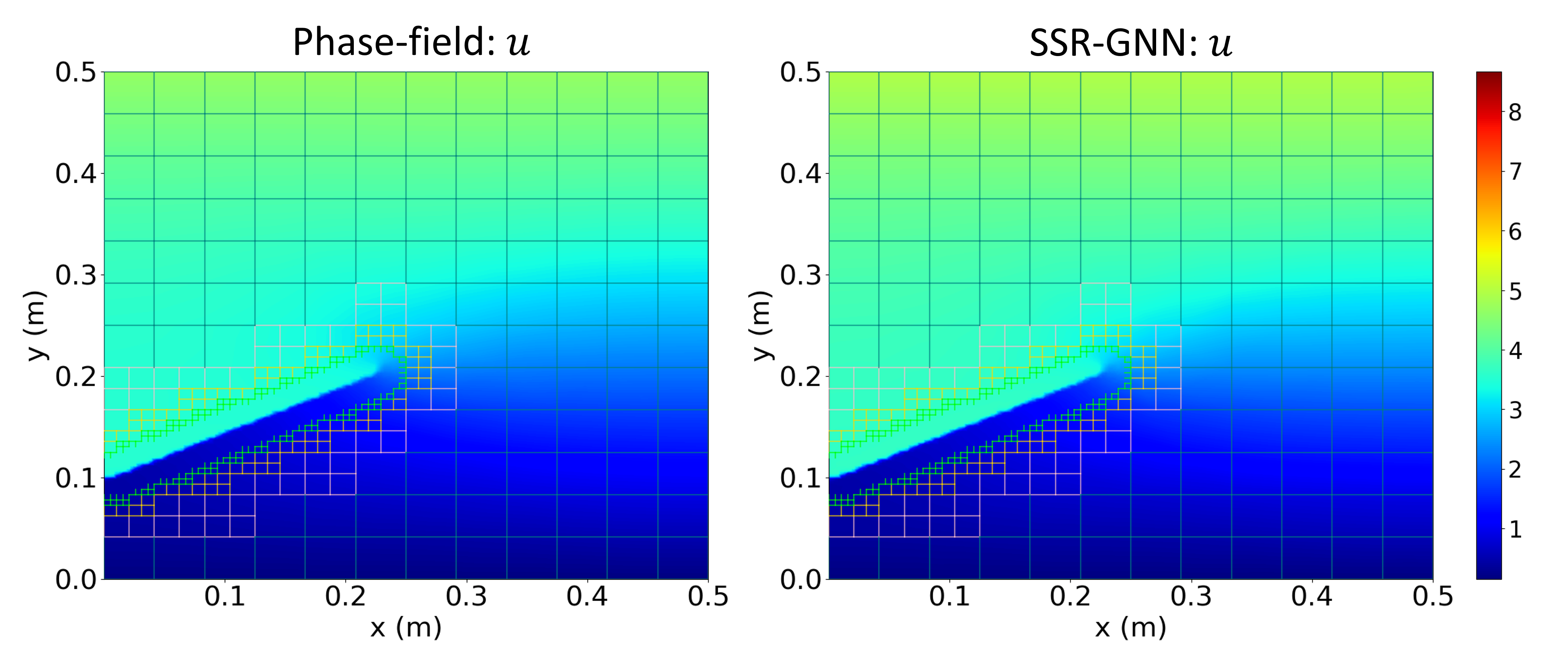}
					\caption{$u$: time $t_{0}$}
					\label{subfig:ShearLoad_XDisp_sim_t0}
				\end{subfigure}
				\centering
				\begin{subfigure}[r]{0.49\textwidth}
					\centering
					\includegraphics[width=1\linewidth]{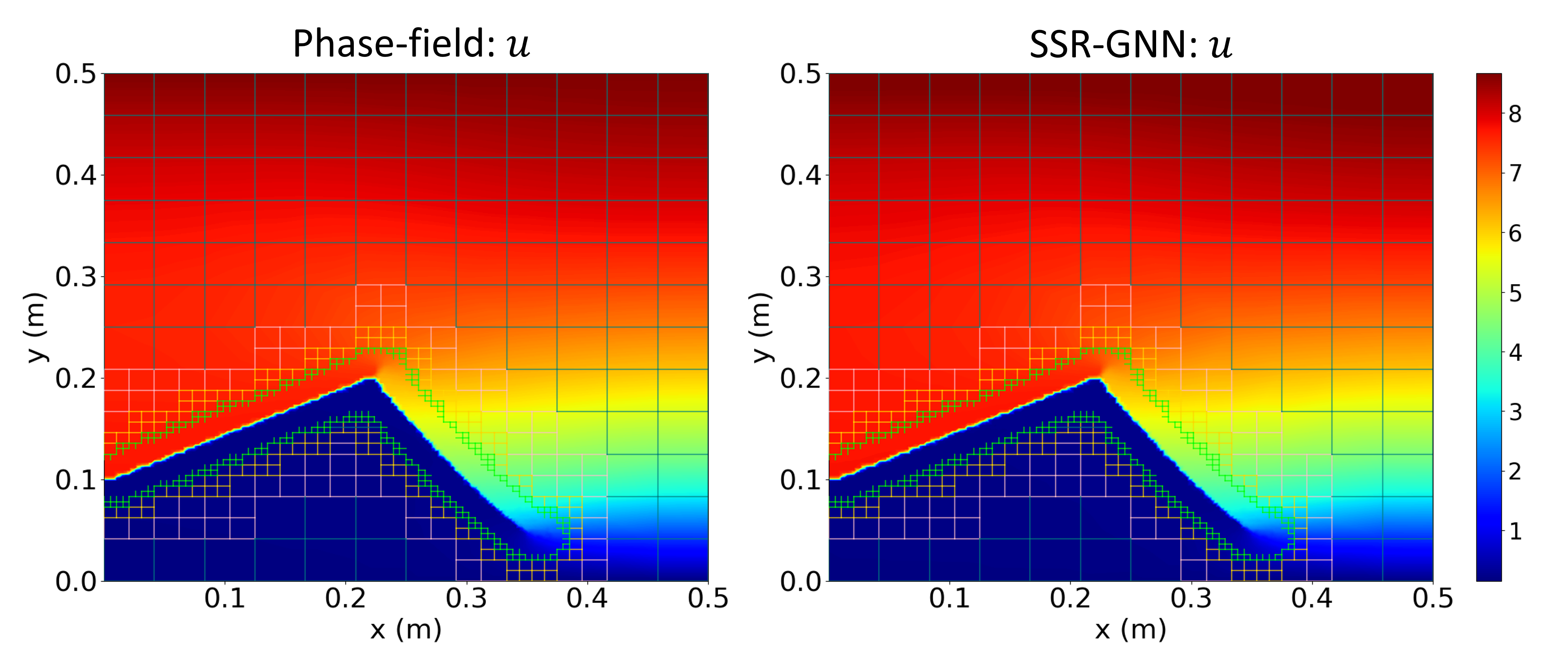}
					\caption{$u$: time $t_{f}$}
					\label{subfig:ShearLoad_XDisp_sim_tf}
				\end{subfigure}
			\end{subfigure}
			\centering
			\begin{subfigure}[b]{\textwidth}
				\centering
				\begin{subfigure}[l]{0.49\textwidth}
					\centering
					\includegraphics[width=1\linewidth]{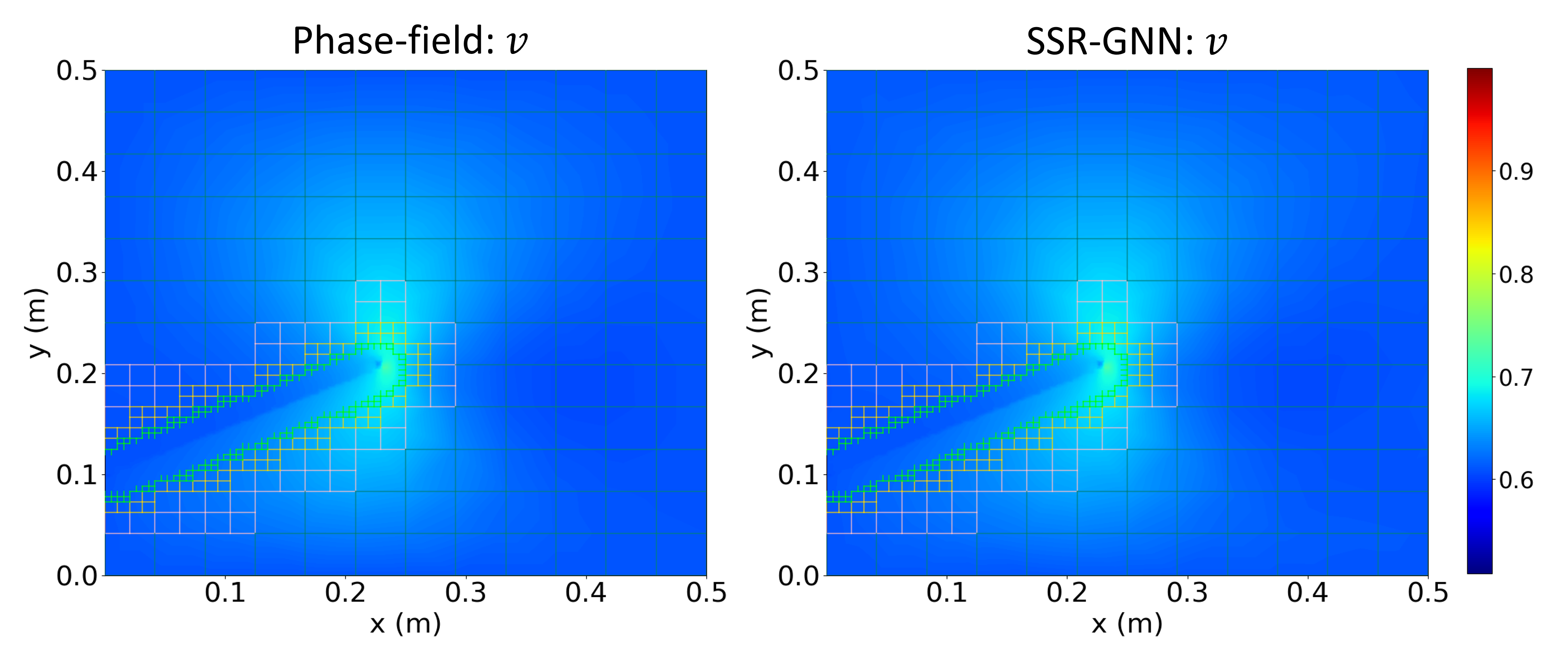}
					\caption{$v$: time $t_{0}$}
					\label{subfig:ShearLoad_YDisp_sim_t0}
				\end{subfigure}
				\centering
				\begin{subfigure}[r]{0.49\textwidth}
					\centering
					\includegraphics[width=1\linewidth]{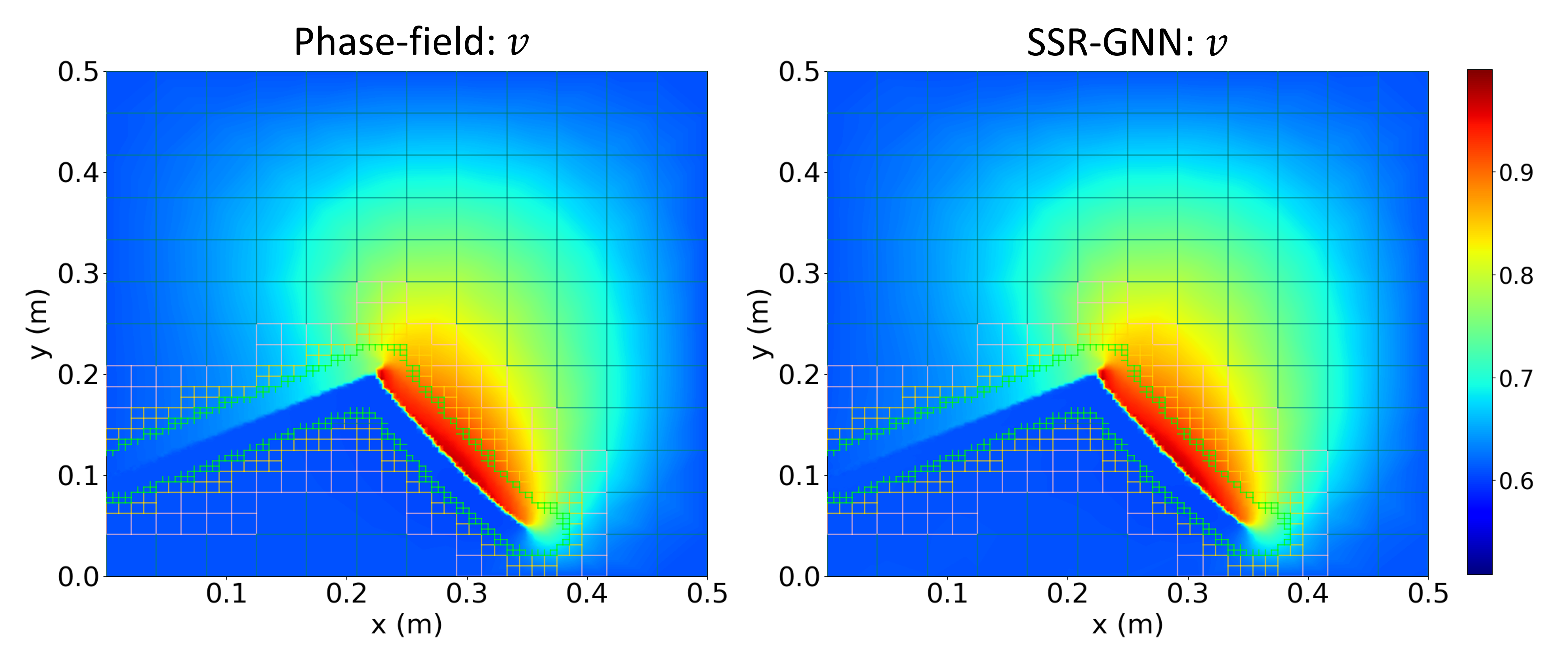}
					\caption{$v$: time $t_{f}$}
					\label{subfig:ShearLoad_YDisp_sim_tf}
				\end{subfigure}
			\end{subfigure}
			\caption{PF model versus the SSR framework on the predicted evolution of the crack field, $\phi$ (a-b), x-displacement field, $u$ (c-d), and y-displacement field, $v$ (e-f), for a shear load test simulation at initial time-step, $t_{0}$, and final time-step, $t_{f}$.}
			\label{fig:ShearLoad_sim}
		\end{figure}

		In this step, we apply TL update to the extended SSR-based framework obtained in Section \ref{subsec:results_center_cracks}.
        As mentioned in Section \ref{subsec:methods_graph_network}, the node features $\{u_{0},v_{0}\}$ allow the framework to distinguish between tensile loading (e.g., $u_{0}=\Delta u$ and $v_{0} = 0$) and shear loading (e.g., $u_{0}=0$ and $v_{0} = \Delta v$) cases.
		Following a similar approach as for center cracks, we evaluated the resulting framework for shear load cases to predict the crack field, and displacement fields at an initial time-step, $t_{0}$, and at a later time-step approaching material failure, $t_{f}$.
		
\begin{figure}
			\centering
			\begin{subfigure}[t]{0.32\textwidth}
				\centering
				\includegraphics[width=1\linewidth]{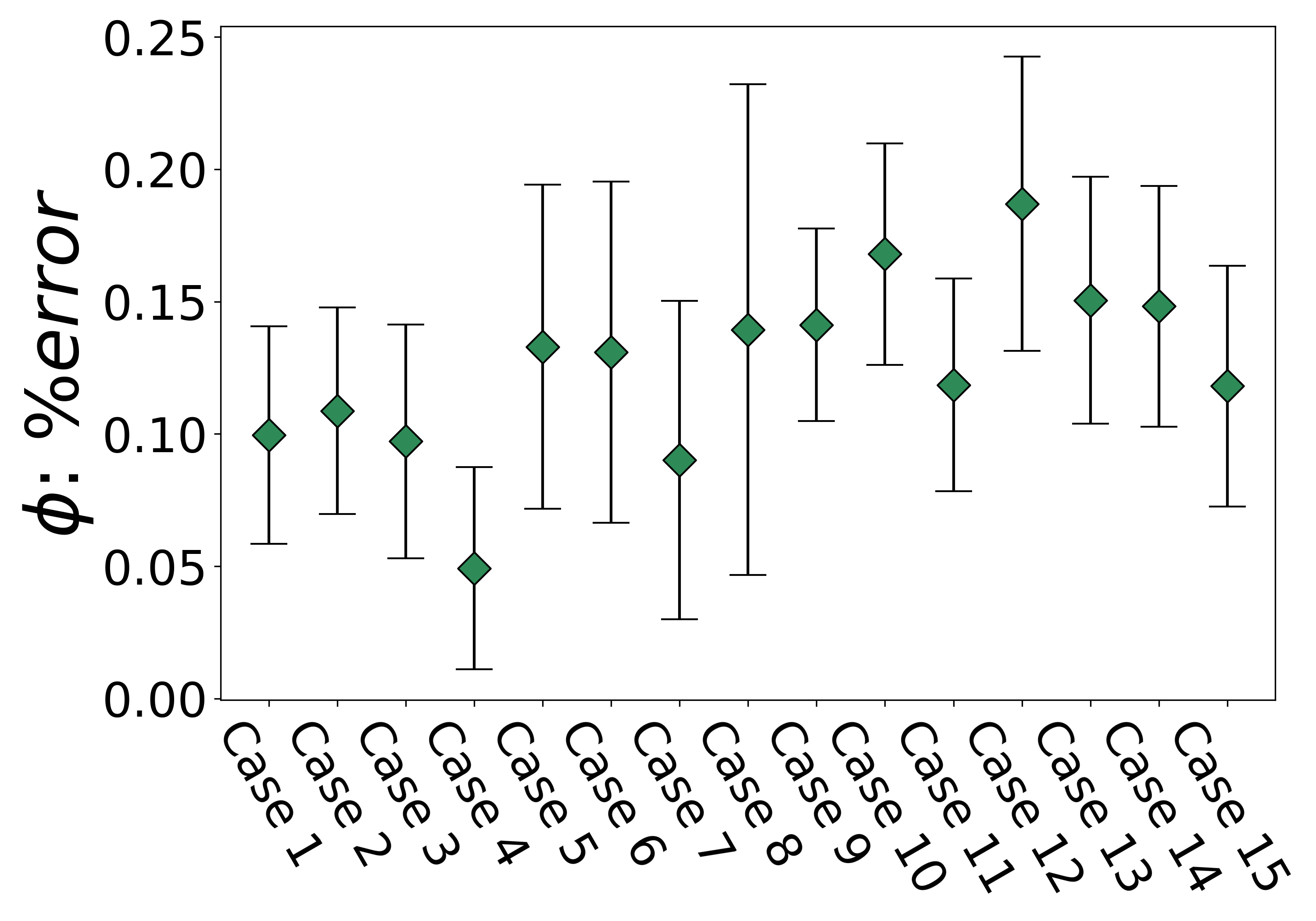}
				\caption{crack field, $\phi$}
			\end{subfigure}
			\centering
			\begin{subfigure}[t]{0.32\textwidth}
				\centering
				\includegraphics[width=1\linewidth]{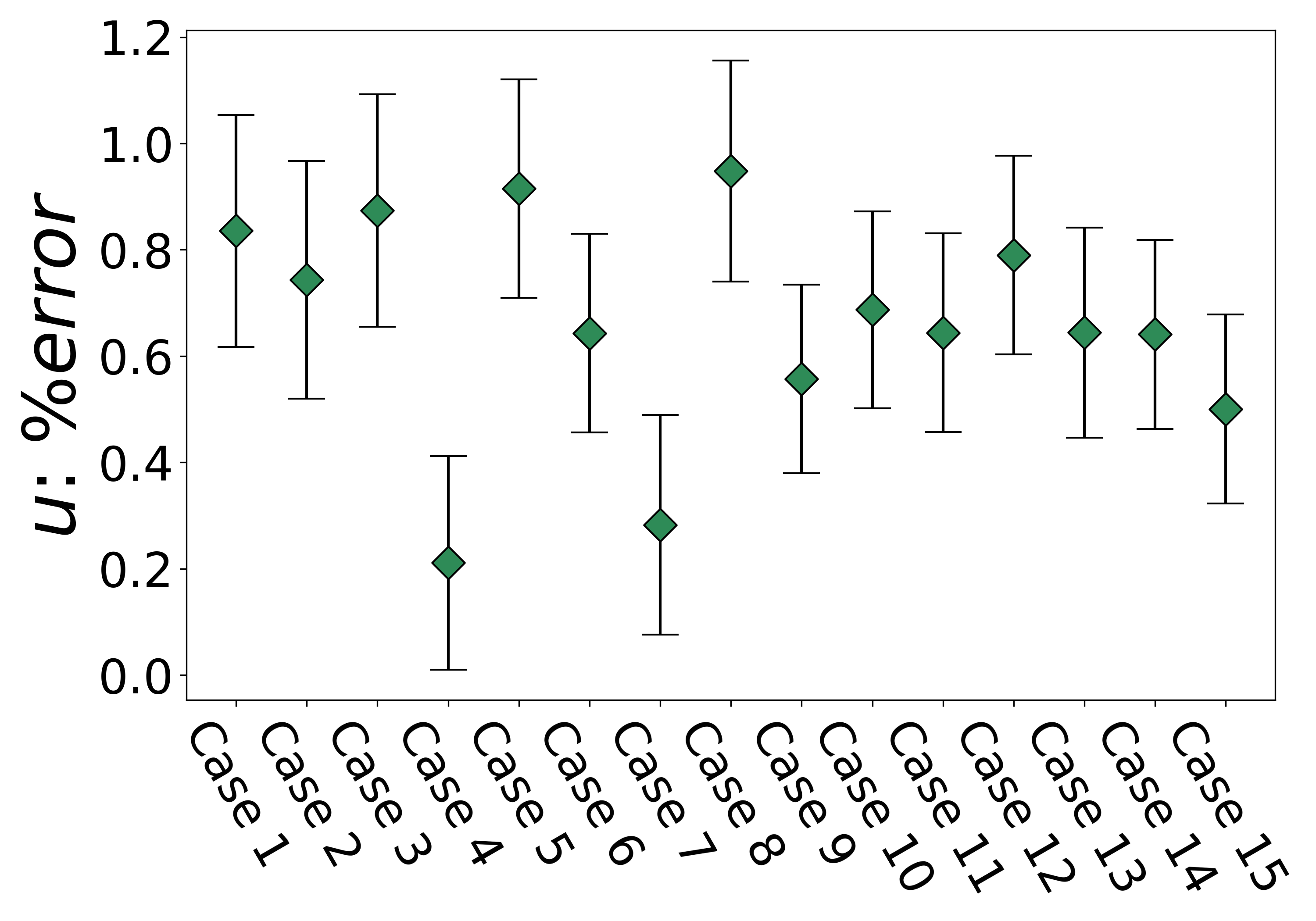}
				\caption{X-displacement, $u$}
			\end{subfigure}
			\centering
			\begin{subfigure}[t]{0.32\textwidth}
				\centering
				\includegraphics[width=1\linewidth]{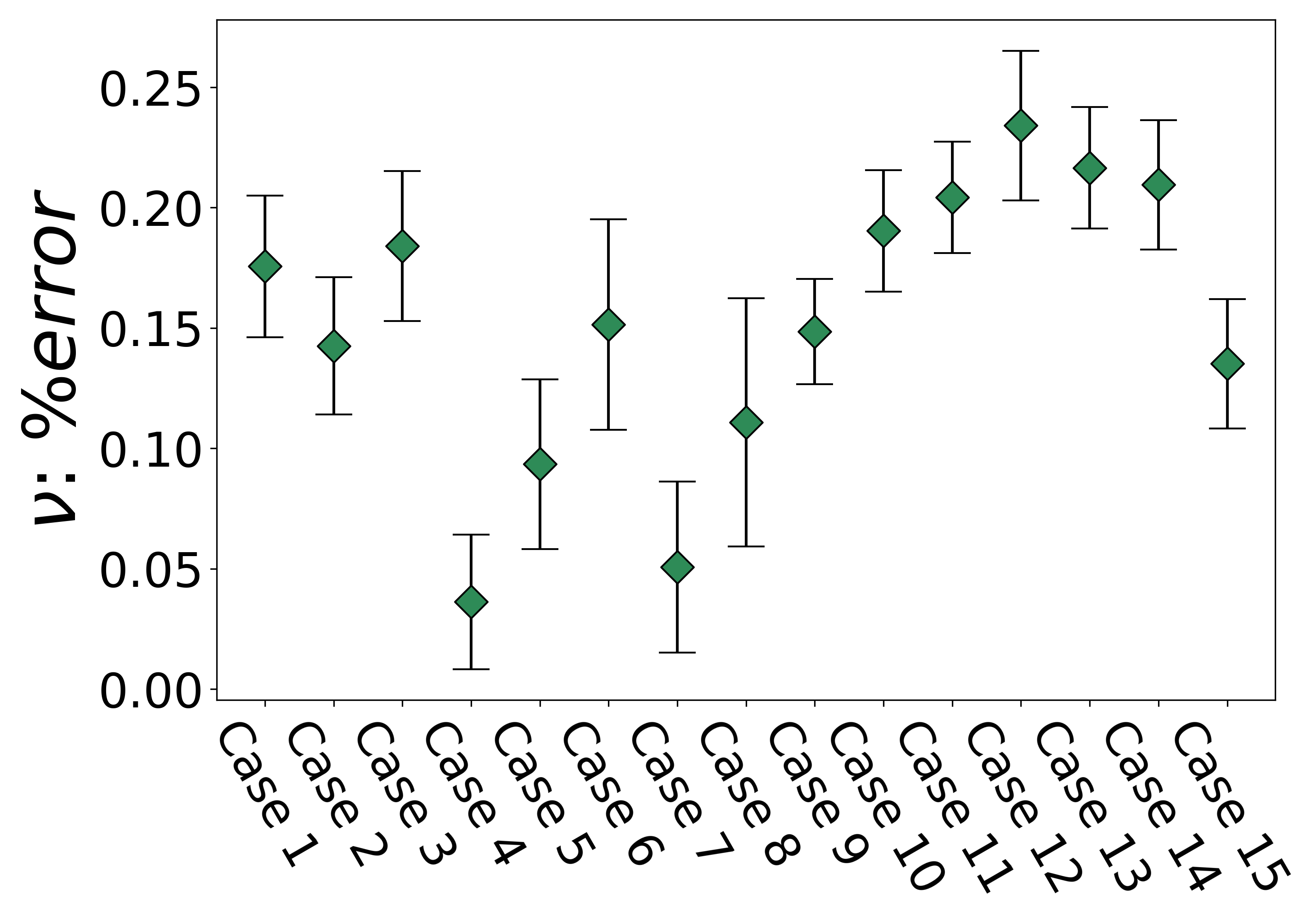}
				\caption{Y-displacement, $v$}
			\end{subfigure}
			\caption{Average percent errors for each simulation in the test dataset of shear load cases for a) crack-field predictions, $\phi$, b) x-displacement predictions, $u$, and c) y-displacement predictions, $v$.}
			\label{fig:Shear_ave_error}
		\end{figure}

        Figures \ref{subfig:ShearLoad_cPhi_sim_t0}, \ref{subfig:ShearLoad_XDisp_sim_t0}, and \ref{subfig:ShearLoad_YDisp_sim_t0} compare GNN predictions and PF predictions for the crack field, $x$-displacement, and $y$-displacement fields, respectively for initial time steps.
		Figures \ref{subfig:ShearLoad_cPhi_sim_tf}, \ref{subfig:ShearLoad_XDisp_sim_tf}, and \ref{subfig:ShearLoad_YDisp_sim_tf} compare GNN and PF predictions for the crack field, $x$-displacement, and $y$-displacement fields, respectively for the final time step.
		These figures show that the new extended SSR framework was able to capture cases involving shear loads and generate accurate predictions at both time steps with virtually identical results compared to the ground truth.  
		For quantitative analysis, we computed the average errors for each test simulation following the approach described in Section \ref{subsec:results_FSR_TSR_SSR_error}.
		Figure \ref{fig:Shear_ave_error} shows the resulting average percent errors for all test cases under shear loads.
		As shown, the SSR also predicted shear cases with high accuracy.
		Average percent errors for the crack field, $x$-displacement, and $y$-displacement resulted below 0.25$\%$, 1.20$\%$, and 0.25$\%$, respectively.
		
	\subsection{Right-edge crack cases}

\begin{figure}
			\centering
			\begin{subfigure}[t]{\textwidth}
				\centering
				\begin{subfigure}[l]{0.49\textwidth}
					\centering
					\includegraphics[width=1\linewidth]{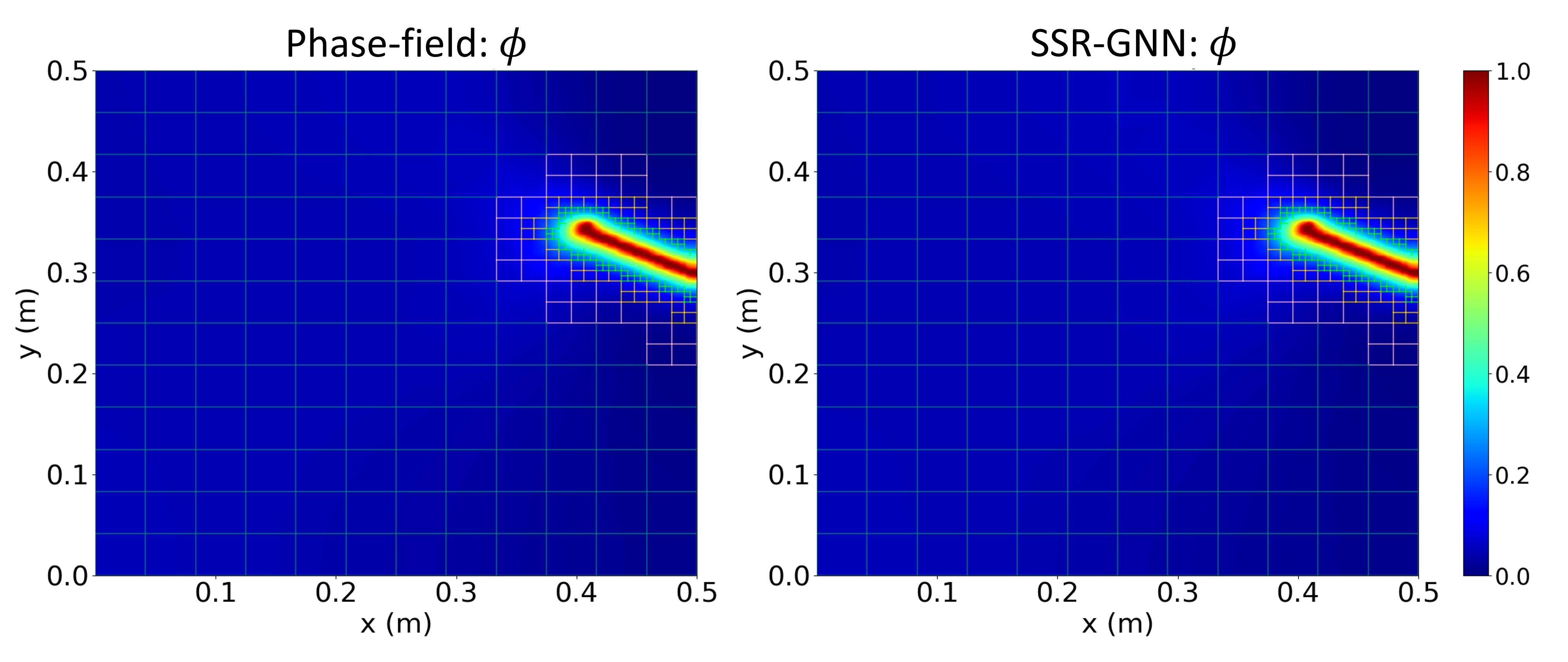}
					\caption{$\phi$ at time $t_{0}$}
					\label{subfig:Flipped_cPhi_sim_t0}
				\end{subfigure}
				\centering
				\begin{subfigure}[r]{0.49\textwidth}
					\centering
					\includegraphics[width=1\linewidth]{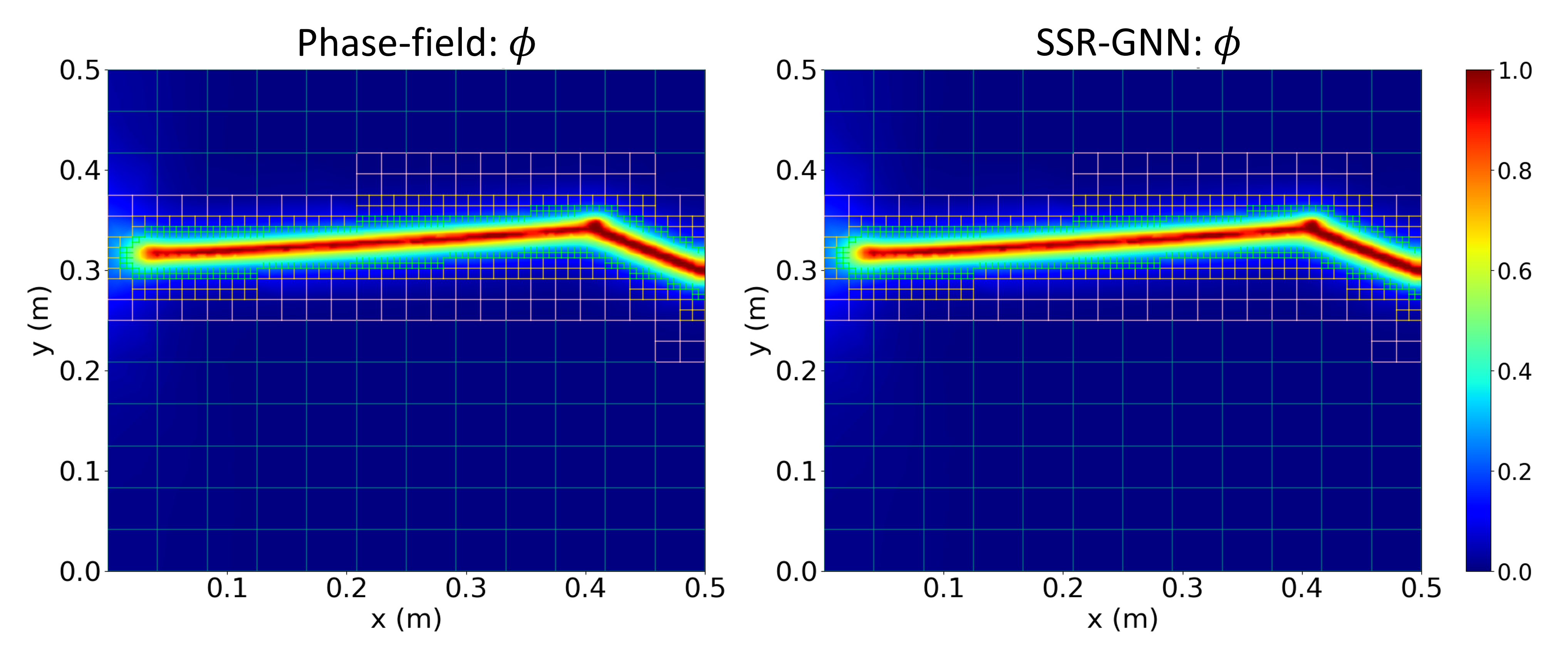}
					\caption{$\phi$ at time $t_{f}$}
					\label{subfig:Flipped_cPhi_sim_tf}
				\end{subfigure}
			\end{subfigure}
			\centering
			\begin{subfigure}[c]{\textwidth}
				\centering
				\begin{subfigure}[l]{0.49\textwidth}
					\centering
					\includegraphics[width=1\linewidth]{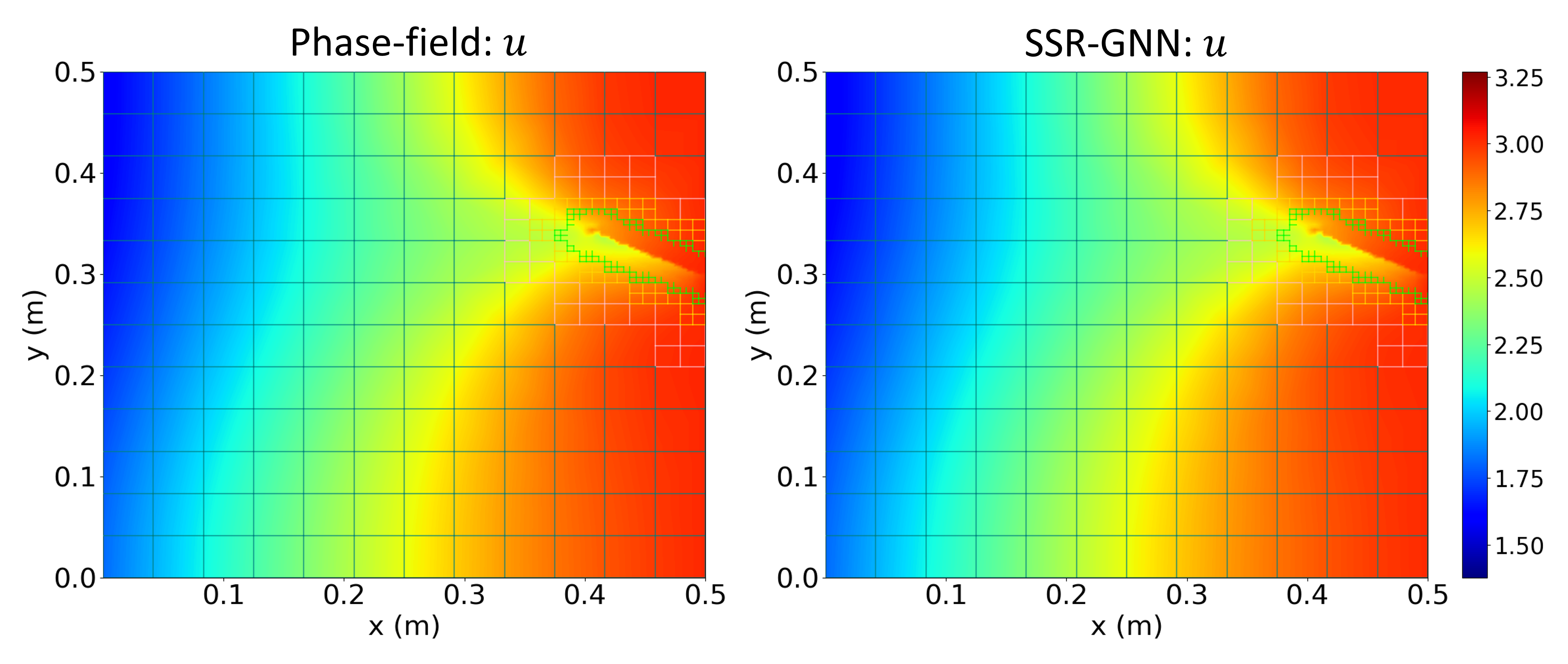}
					\caption{$u$: time $t_{0}$}
					\label{subfig:Flipped_XDisp_sim_t0}
				\end{subfigure}
				\centering
				\begin{subfigure}[r]{0.49\textwidth}
					\centering
					\includegraphics[width=1\linewidth]{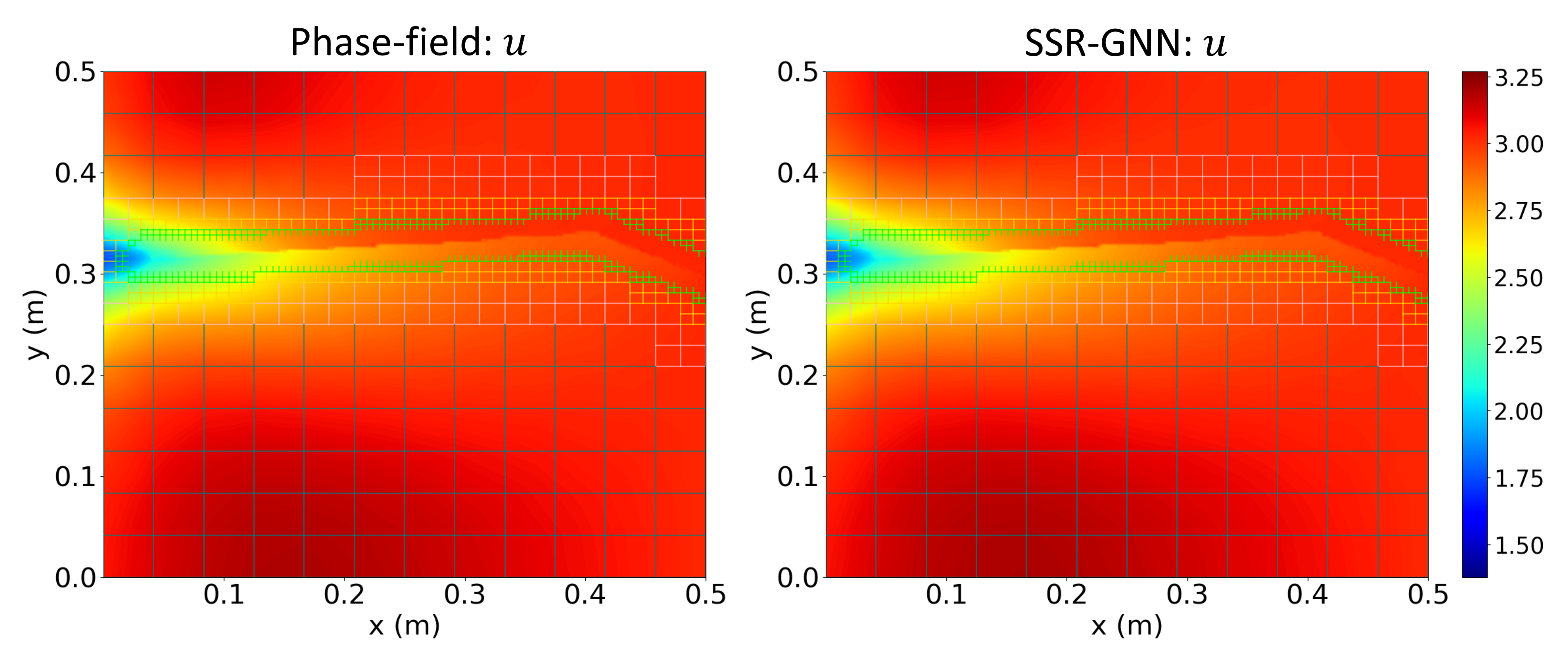}
					\caption{$u$: time $t_{f}$}
					\label{subfig:Flipped_XDisp_sim_tf}
				\end{subfigure}
			\end{subfigure}
			\centering
			\begin{subfigure}[b]{\textwidth}
				\centering
				\begin{subfigure}[l]{0.49\textwidth}
					\centering
					\includegraphics[width=1\linewidth]{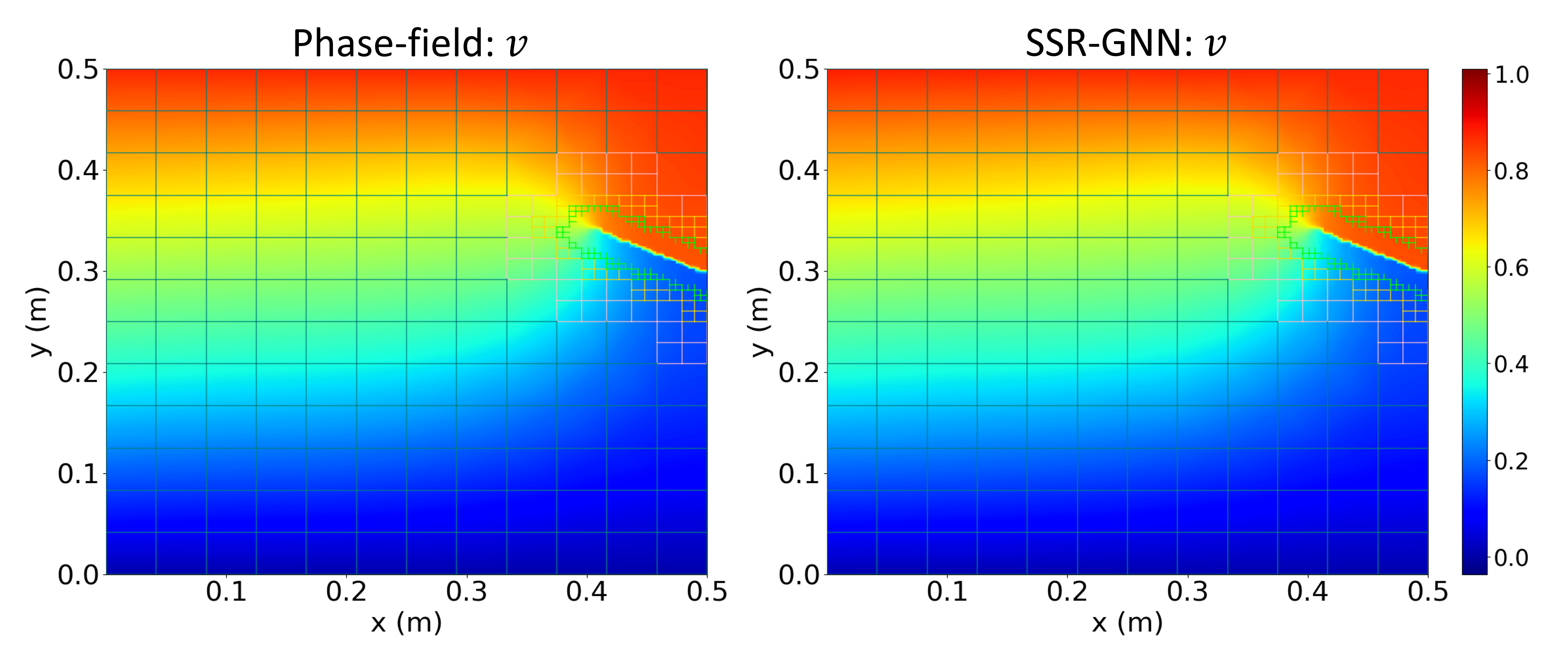}
					\caption{$v$: time $t_{0}$}
					\label{subfig:Flipped_YDisp_sim_t0}
				\end{subfigure}
				\centering
				\begin{subfigure}[r]{0.49\textwidth}
					\centering
					\includegraphics[width=1\linewidth]{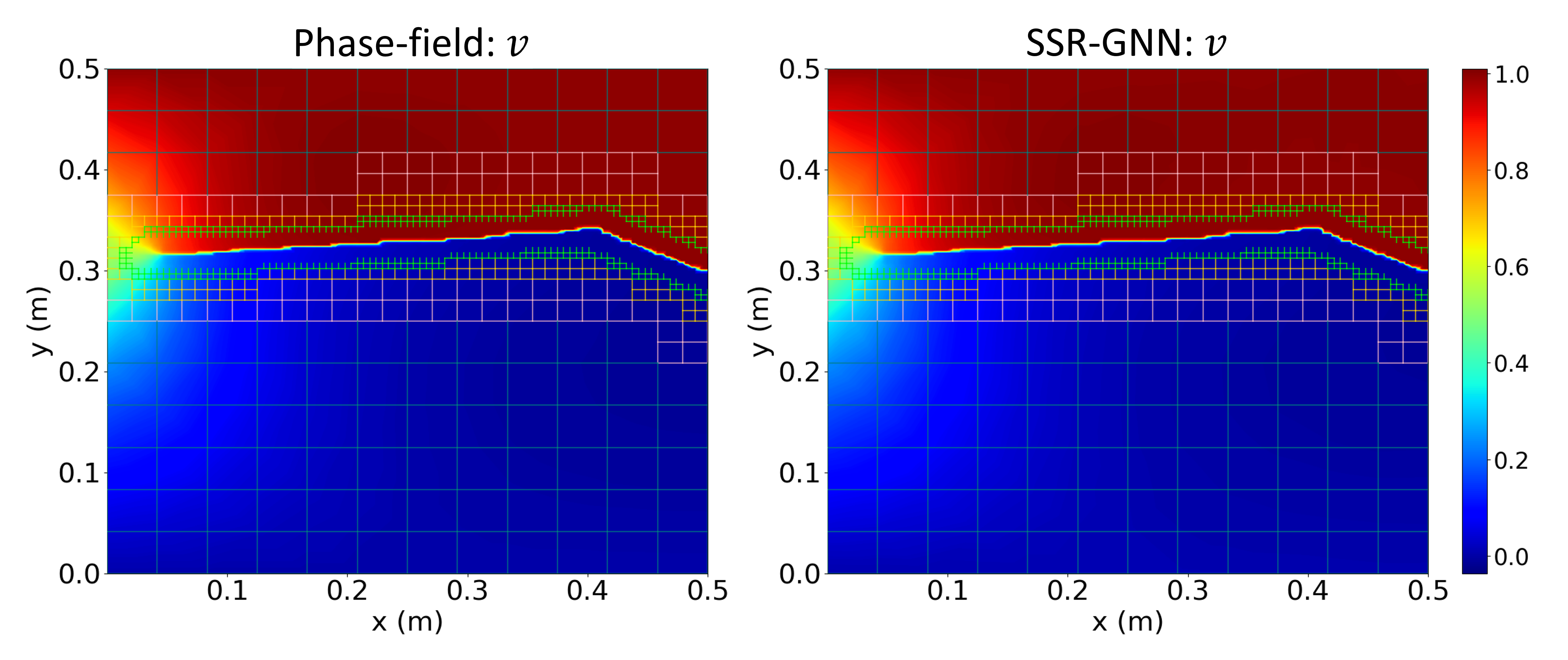}
					\caption{$v$: time $t_{f}$}
					\label{subfig:Flipped_YDisp_sim_tf}
				\end{subfigure}
			\end{subfigure}
			\caption{PF model versus the SSR framework on the predicted evolution of the crack field, $\phi$ (a-b), x-displacement field, $u$ (c-d), and y-displacement field, $v$ (e-f), for a right-edge crack test simulation at initial time-step, $t_{0}$, and final time-step, $t_{f}$.}
			\label{fig:Flipped_sim}
		\end{figure}

\begin{figure}
			\centering
			\begin{subfigure}[t]{0.32\textwidth}
				\centering
				\includegraphics[width=1\linewidth]{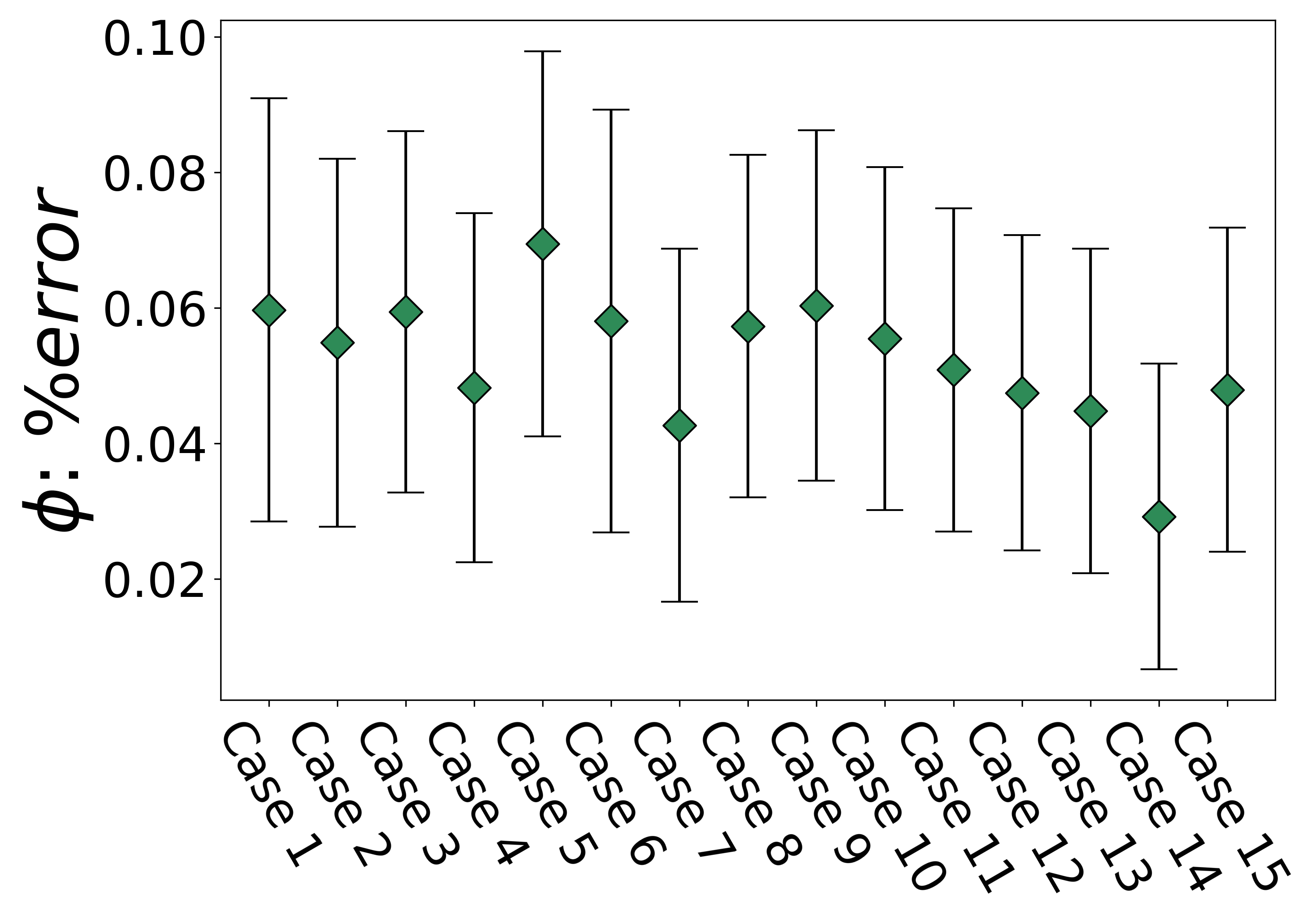}
				\caption{crack field, $\phi$}
				\label{subfig:cPhi_Flipped_ave_error}
			\end{subfigure}
			\centering
			\begin{subfigure}[t]{0.32\textwidth}
				\centering
				\includegraphics[width=1\linewidth]{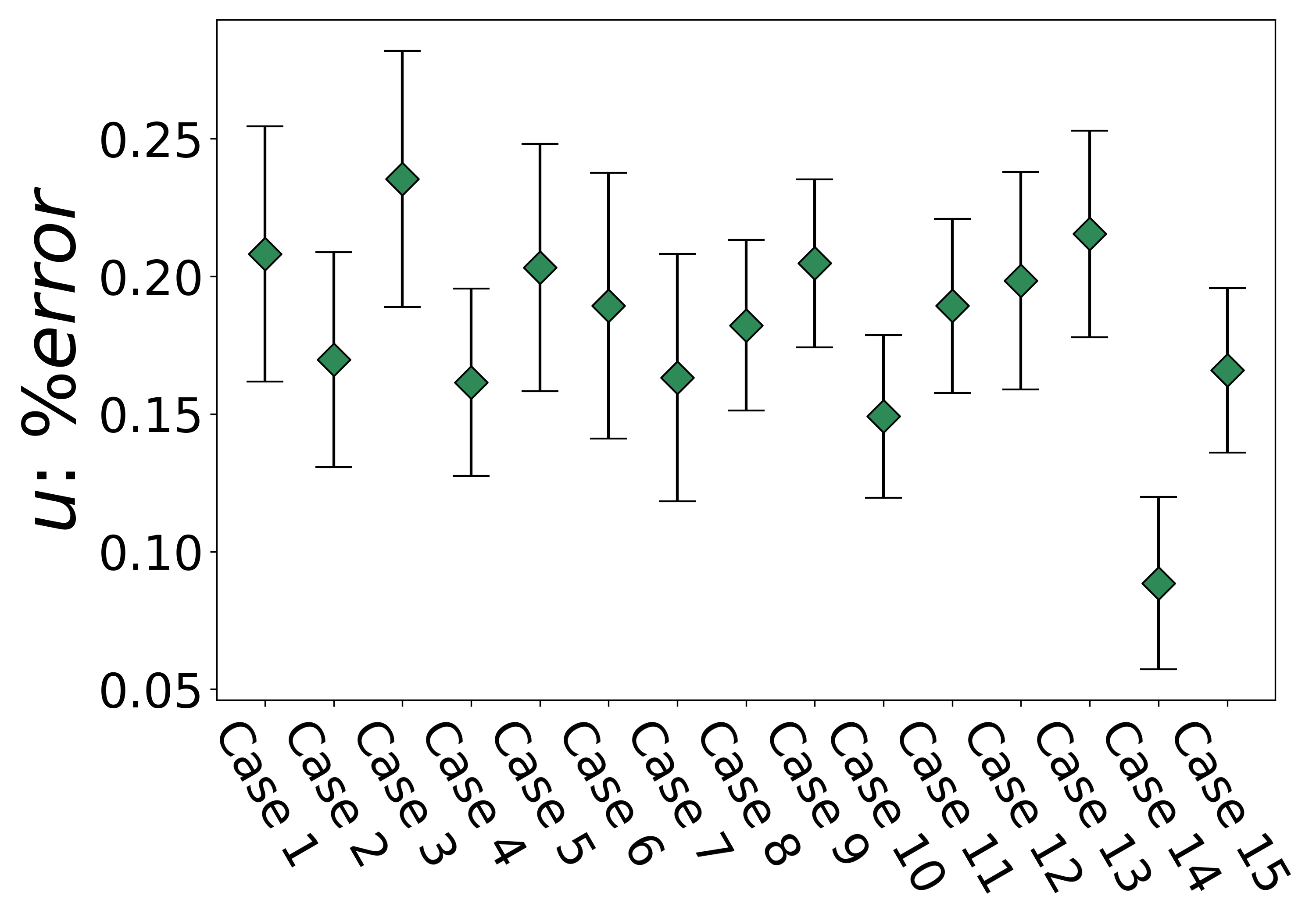}
				\caption{X-displacement, $u$}
				\label{subfig:XDisp_Flipped_ave_error}
			\end{subfigure}
			\centering
			\begin{subfigure}[t]{0.32\textwidth}
				\centering
				\includegraphics[width=1\linewidth]{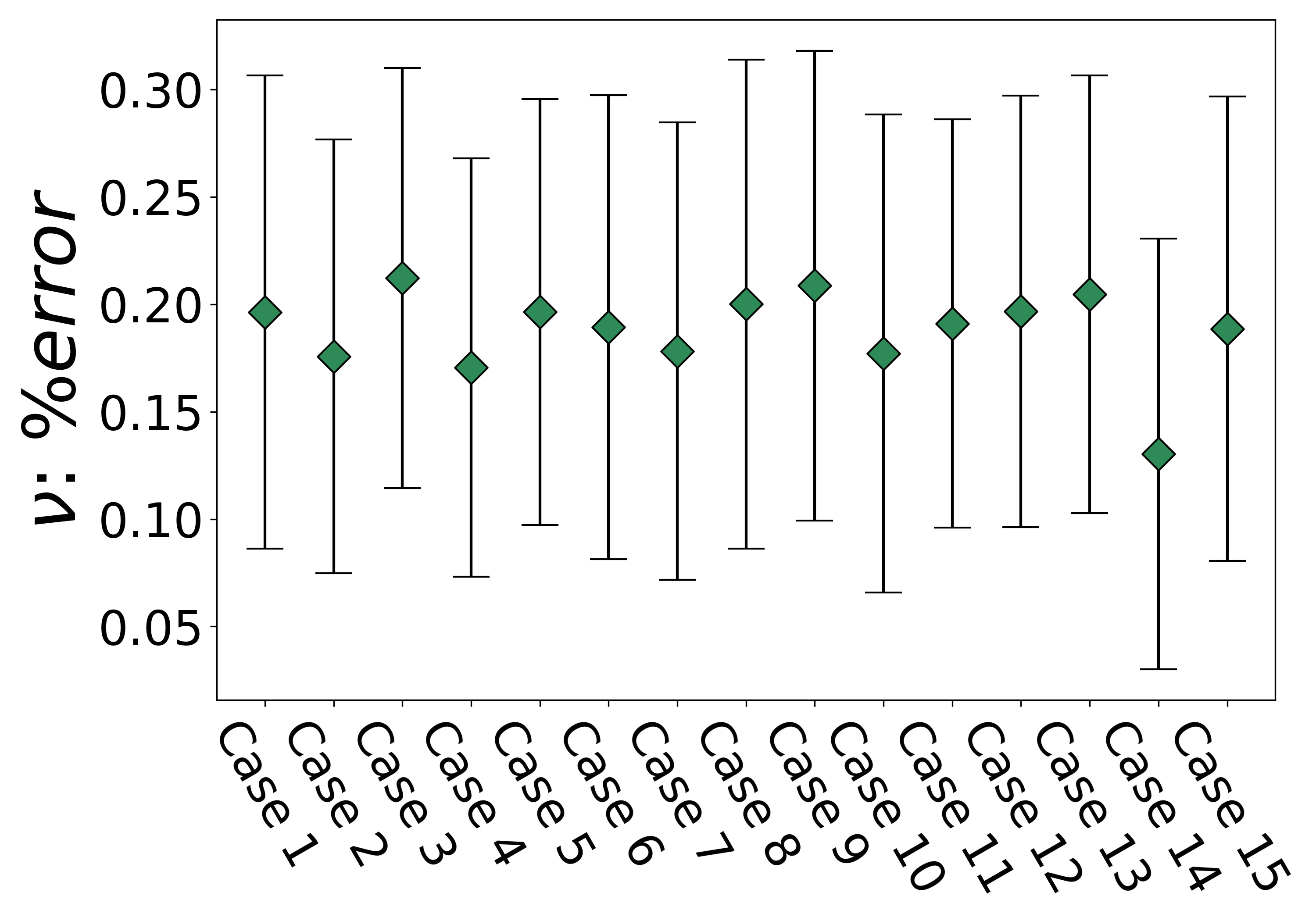}
				\caption{Y-displacement, $v$}
				\label{subfig:YDisp_Flipped_ave_error}
			\end{subfigure}
			\caption{Average percent errors for each simulation in the test dataset of right-edge crack cases for a) crack-field predictions, $\phi$, b) x-displacement predictions, $u$, and c) y-displacement predictions, $v$.}
			\label{fig:Flipped_ave_error}
		\end{figure}

		Lastly, we extended the framework for cases involving right-edge cracks.
		For this TL update step, we mirrored the dataset involving left-edge cracks subjected to tension as described in Section \ref{subsec:methods_transferlearning}.
		Following the same approach as in previous sections, we first tested the resulting trained SSR framework to simulate a randomly chosen simulation from the test dataset.
		Figures \ref{subfig:Flipped_cPhi_sim_t0} - \ref{subfig:Flipped_cPhi_sim_tf} depict the prediction history of the crack field parameter from $t_{0}$ to $t_{f}$ for the chosen test simulation.
		The prediction history depicts qualitatively identical results compared to the PF predictions.
		This high prediction accuracy is verified in Figure \ref{subfig:cPhi_Flipped_ave_error}, where the average error across all simulations remains below 0.10$\%$.
		For the x-displacement and y-displacement predictions, we show the resulting SSR simulation histories in Figures \ref{subfig:Flipped_XDisp_sim_t0} - \ref{subfig:Flipped_YDisp_sim_tf}, respectively.
		The SSR framework also shows high prediction accuracy compared to the PF simulations.
		The obtained average percent errors for x-displacement are shown in Figures \ref{subfig:XDisp_Flipped_ave_error}.
		The errors across all testing samples remained below 0.25$\%$.
		Also, for y-displacement errors shown in Figure \ref{subfig:YDisp_Flipped_ave_error}, it may be noted that the average percent errors remained below 0.30$\%$.
		Ultimately, these results show that by implementing a series of sequential TL update steps, the SSR-based framework was successful in simulating multiple problem configurations with high accuracy.    

\section{Conclusion}
	Complex multiphysics phenomena are often modeled using computationally expensive approaches involving solving coupled multiphysics equations on a mesh.
	Recent ML models such as mesh-based GNNs have emerged as promising tools to simulate multiphysics problems at a reduced cost.
	However, conventional mesh-based GNNs suffer from over-smoothing when working with fine meshes due to a high number of required MP steps.
	This work introduces a mesh-based multiscale GNN framework with AMR for simulating multiphysics problems with a reduced number of MP steps, high prediction accuracy, and accelerated performance times.
	The developed formulation implements sequential coarsening and upscaling operations by removing/adding the highest mesh refinement resolution level at each step.
	This approach results in new graphs with fewer mesh resolution levels providing new distant connections and larger local neighborhoods.
	The framework employs a state-of-the-art Graph Transformer for the MP networks at each coarsening/upscaling operation, and skip-connectors from the coarsening operations to the upscaling operations to avoid information loss.  
	
	Due to the high complexity and computational requirements of multiphysics PF models, this work tested the multiscale GNN on PF crack problems with a near singular operator and coupled equations.      
	First, we studied single-edge notched systems (i.e., left-edge cracks) subjected to tension. 
	We obtained a large dataset for these cases using an open-source PF fracture code.
	We developed and compared three architectures (FSR, TSR, and SSR) for simulating single-edge notched systems using different numbers of MP GNN blocks, and coarsening/upscaling operations.
	The SSR architecture, with the smallest number of operations, demonstrated the fastest prediction times while maintaining high prediction accuracy.
	 
	We then used TL to extend the SSR-based framework to simulate different PF crack propagation problems. 
	We implemented TL to the SSR framework for simulating problems involving (i) center cracks, (ii) left-edge cracks subjected to shear, and (iii) right-edge cracks using only 15 training samples. 
	For all cases (i-iii) the proposed accelerated framework predicted the crack field, and displacement field evolution with high accuracy above 98$\%$.
	These results demonstrated that by using TL, the multiscale SSR-based GNN framework can be extended for different problem configurations with two orders of magnitude smaller training datasets.
	
    Ultimately, this work introduced a new mesh-based multiscale formulation that benefits from the computational efficiencies of the AMR method, mimicking the conventional iterative multigrid scheme, and the TL approach. 
	The resulting AMR mesh-based multiscale GNN provides a framework for simulating additional complex AMR mesh-based engineering and multiphysics problems with high accuracy and accelerated performance. 

\section{Supplementary information}
    Additional information for (i) maximum $\%$ error analysis for the entire test datasets of left-edge crack, center crack, shear load, and right-edge crack cases, and (ii) generated sample simulations for each case can be found in \url{https://github.com/rperera12/Adaptive-mesh-based-Multiscale-Graph-Neural-Network}.

\section{Acknowledgements}
	The authors are grateful for the financial support provided by the U.S. Department of Defense in conjunction with the Naval Air Warfare Center/Weapons Division through the SMART scholarship Program (SMART ID: 2021-17978).

\bibliographystyle{ieeetr}
\bibliography{library}

\end{document}